\documentclass[a4paper,12pt]{article}
\usepackage[margin=1in]{geometry}
\usepackage{graphicx}

\usepackage{amssymb}
\usepackage{color}
\usepackage{multirow}
\usepackage{amsmath}
\usepackage{breqn}
\usepackage{bm}
\usepackage{ascmac}
\usepackage{hyperref}
\usepackage{wrapfig}
\usepackage{url}
\usepackage{color}
\usepackage[modulo]{lineno}
\usepackage{threeparttable}
\usepackage{float}
\usepackage{siunitx}
\usepackage{subfig}
\usepackage{comment}
\usepackage[blocks]{authblk}

\title{Proposal of the KOTO~II experiment}
\date{December 17, 2024, Two institutions are added.\\December 9, 2024, submitted to J-PARC PAC}

\author{John Fry}
\author{Evgueni Goudzovski}
\author{Cristina Lazzeroni}
\author{Thomas Reddel}
\author{Angela Romano}
\author{Jack Sanders}
\author{Adam Tomczak}

\affil{School of Physics and Astronomy, University of Birmingham, Edgbaston, Birmingham, B15~2TT, United Kingdom}

\author{Gabriella Carini}
\author{Guang Yang}
\affil{Instrumentation Department, Brookhaven National Laboratory, Upton, NY 11973, USA}

\author{Jianming Bian}
\affil{Department of Physics and Astronomy, University of California Irvine, Irvine, California 92697, USA}

\author{Tom\'a\v{s} Husek}

\affil{Institute of Particle and Nuclear Physics, Faculty of Mathematics and Physics, Charles University, V Hole\v{s}ovi\v{c}k\'ach 2, 180 00 Prague, Czech Republic}
\affil{School of Physics and Astronomy, University of Birmingham, Edgbaston, Birmingham, B15~2TT, United Kingdom}

\author{Emile Augustine}
\author{Benjamin Stillwell}
\author{Joseph Redeker}
\author{Yau Wah}
\author{Maresa Wynd}
\affil{Enrico Fermi Institute, University of Chicago}

\author{Siavash~Neshatpour}
\affil{Universit\'e Claude Bernard Lyon 1, CNRS/IN2P3, Institut de Physique des 2 Infinis de Lyon, 69622 Villeurbanne, France}

\author{Farvah~Mahmoudi}
\affil{Universit\'e Claude Bernard Lyon 1, CNRS/IN2P3, Institut de Physique des 2 Infinis de Lyon, 69622 Villeurbanne, France}
\affil{Theoretical Physics Department, CERN, 1211 Geneva 23, Switzerland}
\affil{Institut Universitaire de France, 75005 Paris, France}

\author{Diego Mart\'inez Santos}
\author{Veronika Chobanova}
\author{Claire Prouve}
\author{John~Wendel}
\affil{Ferrol Industrial Campus, Universidade da Coru\~{n}a, Dr. V\'azquez Cabrera, s/n, 15403, A Coru\~{n}a, Spain}

\author{Alexander Glazov}
\author{Armine Rostomyan}
\affil{Deutsches Elektronen-Synchrotron DESY, Notkestr. 85, 22607 Hamburg, Germany}

\author{Laura Bandiera}
\author{Alberto Gianoli}
\affil{INFN Sezione di Ferrara, 44122 Ferrara, Italy}

\author{Marco Romagnoni}
\affil{Dipartimento di Fisica e Scienze della Terra, Università di Ferrara and INFN Sezione di Ferrara, 44122 Ferrara, Italy}

\author{Antonella~Antonelli}
\author{Matthew~Moulson}
\author{Ivano~Sarra}
\author{Mattia~Soldani}
\author{Joel~Swallow}
\affil{INFN Laboratori Nazionali di Frascati, 00044 Frascati, Italy}

\author{Victoria~Martin}
\author{Matthew~Needham}
\affil{School of Physics and Astronomy, University of Edinburgh, Edinburgh, EH9~3FD, United Kingdom}

\author{Dan~Protopopescu}
\affil{School of Physics and Astronomy, University of Glasgow, Glasgow, \mbox{G12 8QQ}, United Kingdom}

\author{Takeshi~Komatsubara}
\author{GeiYoub~Lim}
\author{Tadashi~Nomura}
\author{Koji~Shiomi}
\author{Hiroaki~Watanabe}
\affil{High Energy Accelerator Research Organization (KEK), Institute of Particle and Nuclear Studies, Tsukuba, Ibaraki 305-0801, Japan}
\affil{J-PARC Center, Tokai, Ibaraki 319-1195, Japan}

\author{Abhishek~Iyer}
\affil{Department of Physics, Indian Institute of Technology Delhi, Hauz Khas, New Delhi, Delhi 110016, India}

\author{Eun-Joo~Kim}
\affil{Jeonbuk National University, Division of Science Education, Jeonju 54896, Republic of Korea}

\author{Letizia~Peruzzo}
\author{Rainer~Wanke}
\affil{Physics Institute and PRISMA$^+$ Cluster of Excellence, Johannes Gutenberg University Mainz, Mainz, Germany}

\author{Jung~Keun~Ahn}
\affil{Department of Physics, Korea University, Seoul 02841, Korea}

\author{John~Bourke~Dainton}
\author{Roger~William~Lewis~Jones}
\author{Karim~Massri}
\author{Artur~Shaikhiev}
\affil{Physics Department, Lancaster University, LA1 4YB, United Kingdom}

\author{Martin~Gorbahn}
\affil{Department of Mathematical Sciences, University of Liverpool, Liverpool, L69 3BX, United Kingdom}

\author{David~Hutchcroft}
\affil{Oliver Lodge Laboratory, University of Liverpool, Liverpool, L69 3BX, United Kingdom}

\author{Stefano~De~Capua}
\affil{Department of Physics and Astronomy, The University of Manchester, Manchester, M13~9PL, United Kingdom}

\author{Babette D\"obrich}
\author{Samet Lezki}
\affil{Max-Planck-Institut f\"ur Physik (Werner-Heisenberg-Institut), 85748 Garching, Germany}

\author{Giancarlo~D'Ambrosio}
\affil{INFN Sezione di Napoli, 80126 Napoli, Italy}

\author{Chieh~Lin}
\affil{Department of Physics, National Changhua University of Education, Changhua 50007, Taiwan}

\author{Toru~Matsumura}
\affil{National Defense Academy of Japan, Depertment of Applied Physics, Yokosuka, Kanagawa 239-8686, Japan}

\author{Yu-Chen~Tung}
\affil{Department of Physics, National Kaohsiung Normal University, Kaohsiung 824, Taiwan.}

\author{Yee~B.~Hsiung}
\author{Tong~Wu}
\author{Yi-Ting~Su}
\affil{Department of Physics, National Taiwan University, Taipei 106, Taiwan}

\author{Mario~Gonzalez}
\author{Mei~Homma}
\author{Mai~Katayama}
\author{Yuto~Kawata}
\author{Katsushige~Kotera}
\author{Hajime~Nanjo\footnote{Contact person: Hajime Nanjo \\E-mail: \texttt{nanjo@champ.hep.sci.osaka-u.ac.jp}}}
\author{Daiki~Ogawa}
\author{Keita~Ono}
\author{Ryota~Shiraishi}
\affil{Osaka University, Department of Physics, 1-1 Machikaneyama, Toyonaka, Osaka 560-0043, Japan}

\author{Luigi~Montalto}
\author{Daniele~Rinaldi}
\affil{SIMAU Department and ICRYS, Universit\`a Politecnica delle Marche, 60131 Ancona, Italy}

\author{Francesco~Brizioli}
\author{Monica~Pepe}
\affil{INFN Sezione di Perugia, 06123 Perugia, Italy}

\author{Giuseppina~Anzivino}
\affil{Dipartimento di Fisica e Geologia, Università di Perugia and INFN Sezione di Perugia, 06123 Perugia, Italy}

\author{Jacopo~Pinzino}
\affil{INFN Sezione di Pisa, 56127 Pisa, Italy}

\author{Gianluca~Lamanna}
\author{Marco~Sozzi}
\affil{Dipartimento di Fisica, Università di Pisa and INFN Sezione di Pisa, 56127 Pisa, Italy}

\author{Sanghoon~Lim}
\author{Chong~Kim}
\affil{Pusan National University, Department of Physics, Busan 46241, Republic of Korea}

\author{Michal~Kreps}
\affil{Department of Physics, University of Warwick, Coventry, CV4 7AL, United Kingdom}

\author{Yasuhisa~Tajima}
\author{Hiroshi~Yoshida}
\affil{Faculty of Science, Yamagata University, 1-4-12 Kojirakawa,Yamagata 990-8560, Japan}

\begin{document}
\maketitle
\clearpage

\tableofcontents
\clearpage

\newcommand{\memo}[1]{\textcolor{red}{\textbf{#1}}}
\newcommand{\ie}{\textit{i}.\textit{e}.}

\newcommand{\kl}{K_L}
\newcommand{\klpionn}{K_L \to \pi^0 \nu \overline{\nu}}
\newcommand{\klpioll}{K_L \to \pi^0 \ell^+ \ell^-}
\newcommand{\klpioee}{K_L \to \pi^0 e^+ e^-}
\newcommand{\klpiomm}{K_L \to \pi^0 \mu^+ \mu^-}

\newcommand{\kpinn}{K \to \pi \nu \overline{\nu}}
\newcommand{\klgg}{K_L \to 2\gamma}
\newcommand{\klpiopio}{K_L \to \pi^0 \pi^0}
\newcommand{\klpiopiopio}{K_L \to \pi^0 \pi^0 \pi^0}
\newcommand{\klppm}{K_L \to \pi^+ \pi^- \pi^0}

\newcommand{\kpluspnn}{K^{+} \to \pi^+ \nu \overline{\nu}}

\newcommand{\pt}{p_{\mathrm{T}}}
\newcommand{\zvtx}{z_{\mathrm{vtx}}}

\newcommand{\kpien}{K_L \to \pi^\pm e^\mp \nu}
\newcommand{\kpimun}{K_L \to \pi^\pm \mu^\mp \nu}

\newcounter{secnumdepthsave}
\setcounter{secnumdepthsave}{\value{secnumdepth}}
\setcounter{secnumdepth}{4}

\section*{Preface}

We propose the KOTO~II experiment at J-PARC, a next-generation experiment to measure the branching ratio of the $\klpionn$ decay. The experiment is a successor to the KOTO experiment, which is searching for the $\klpionn$ decay with a single event sensitivity below $10^{-10}$.
Based on the experience in KOTO,  the KOTO~II experiment aims to achieve a single event sensitivity below $10^{-12}$, which is much smaller than the Standard Model prediction for the $\klpionn$ branching ratio of $3\times 10^{-11}$, allowing searches for new physics beyond the standard model as well as the first discovery of the $\klpionn$ decay with significance exceeding $5\sigma$. As the only experiment in the world dedicated to rare kaon decays, KOTO II will be indispensable in the quest for a complete understanding of flavor dynamics in the quark sector. Moreover, by combining efforts from kaon community worldwide, we plan to further develop the KOTO II detector and expand the physics reach of the experiment to include measurements of the branching ratio of the $K_L\to\pi^0\ell^+\ell^-$ decays, studies of other $K_L$ decays, and searches for dark photons, axions, and axion-like particles.
KOTO II will therefore obtain a comprehensive understanding of $K_L$ decays, providing further constraints on new physics scenarios when combined with existing $K^+$ results.

\section{Physics motivation}
\label{chap:physics}


For 75~years, experimental studies of kaon decays have played a unique role in propelling the development of the Standard Model. As in other branches of flavor physics, the continuing experimental interest in the kaon sector derives from the possibility of conducting precision measurements, particularly of suppressed or rare processes, which may reveal the effects of new physics with mass-scale sensitivity exceeding that which can be explored directly, e.g., at the LHC or a next-generation 
collider. 

Because of the relatively small number of kaon decay modes and the relatively simple final states, combined with the relative ease of producing intense kaon beams, kaon decay experiments are in many ways the quintessential intensity-frontier experiments. 

While some rare kaon decays can also be studied at colliders, namely at LHCb, the golden modes $K\to\pi\nu\bar\nu$ require specifically designed, dedicated experiments due to the presence of final-state neutrinos.
Since the kaon physics program at CERN will terminate in 2026, KOTO~II represents the only facility worldwide dedicated to rare kaon decays, and specifically, to these golden channels.

Measurements of quantities well predicted by the Standard Model (SM), like ${\cal B}(K\to\pi\nu\bar\nu)$, offer model-independent standard candles that can constrain any ``beyond SM (BSM)'' scenario, present or future.
KOTO~II will reach unprecedented levels of precision.
The status of BSM models in the future is hard to predict, but KOTO~II measurements will be durable standards against which many of those models will be judged.

\subsection{The $\klpionn$ decay}

\begin{figure}[tbph]
\begin{center}
\includegraphics[width=0.25\textwidth]{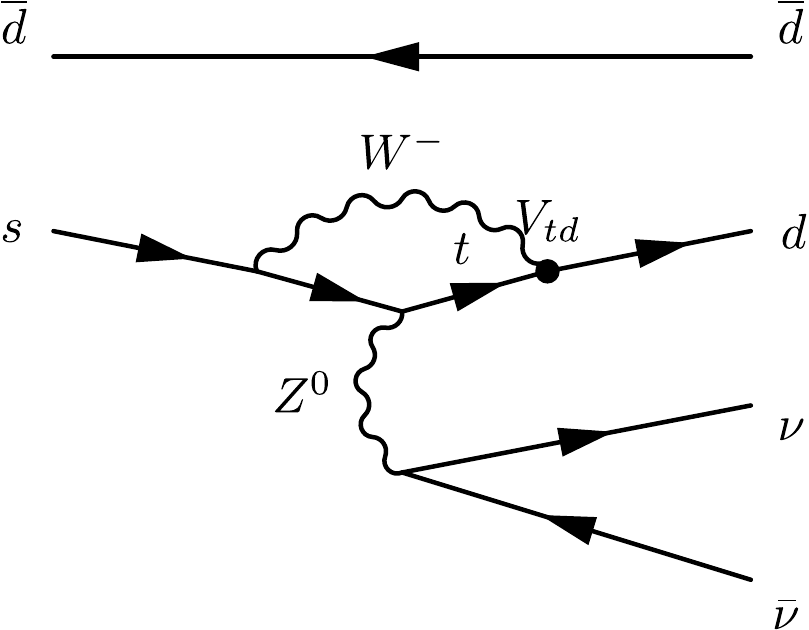}
\hspace*{3mm}
\includegraphics[width=0.25\textwidth]{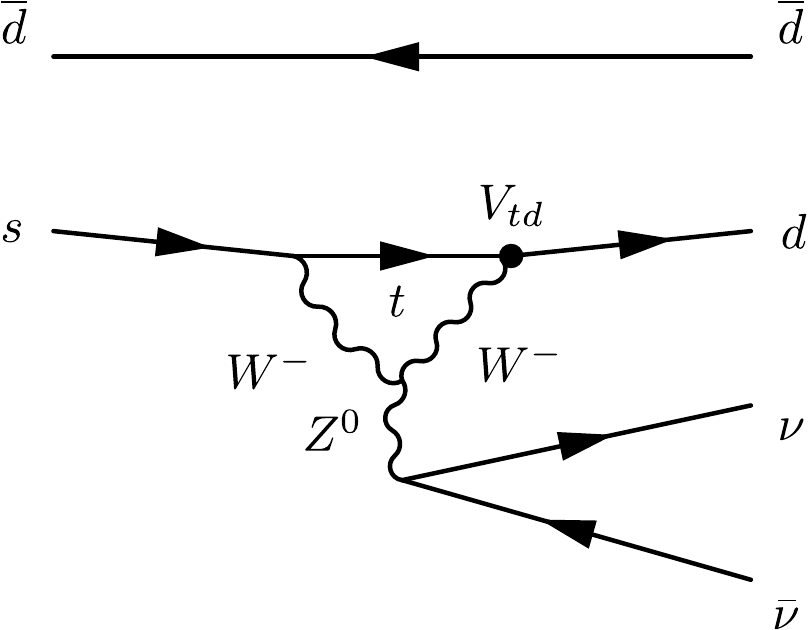} 
\vspace*{3mm}
 \includegraphics[width=0.25\textwidth]{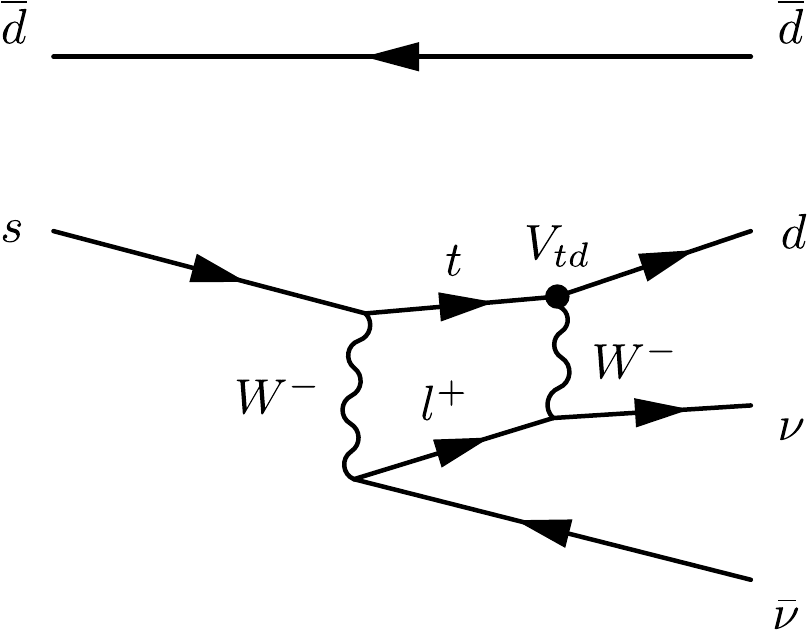}
\caption{
Feynman diagrams for the $\klpionn$ decay in the Standard Model. }
\label{fdiagram}
\end{center}
\end{figure}

The rare kaon decay $\klpionn$ provides a unique opportunity to search for physics beyond the SM.
The decay proceeds by a flavor-changing neutral current (FCNC) from a strange to a down quark ($s\to d$ transition) through loop effects expressed by the electroweak penguin and box diagrams shown in Fig.~\ref{fdiagram}. The $s\to d$ transition is most strongly suppressed in the SM among other FCNC transitions (such as $b\to d$ and $b\to s$) due to the Glashow-Iliopoulos-Maiani (GIM) mechanism and the hierarchical structure of the Cabibbo-Kobayashi-Maskawa (CKM) matrix.
On the other hand, the flavor structure on new physics is not in general expected to exhibit the CKM hierarchies.
The contribution of new physics in the loop could be observed as the deviation of the branching ratio from the SM prediction,
even if the energy scale of new physics is more than 100 TeV, which can not be reached by the LHC~\cite{ref:Zepto}.

The SM prediction of the branching ratio (BR) on the $\klpionn$ decay
~\cite{Brod:2021hsj,DAmbrosio:2022kvb,Buras:2022qip}
is $(2.94\pm 0.15)\times10^{-11}$~\cite{Buras:2022qip}.
The uncertainties are dominated by the parametric uncertainties on the CKM elements, and the theoretical uncertainties 
are only 2\%~\cite{Brod:2021hsj}, because the decay is entirely governed by short-distance physics involving the top quark.
The long-distance interaction mediated by photons does not exist, and the hadronic matrix element can be precisely estimated 
by $K \rightarrow\pi e\nu$ data.  Therefore, the small effects of new physics stand out against the SM contribution.

\begin{figure}[tbph]
 \begin{center}
  \includegraphics[width=12cm]{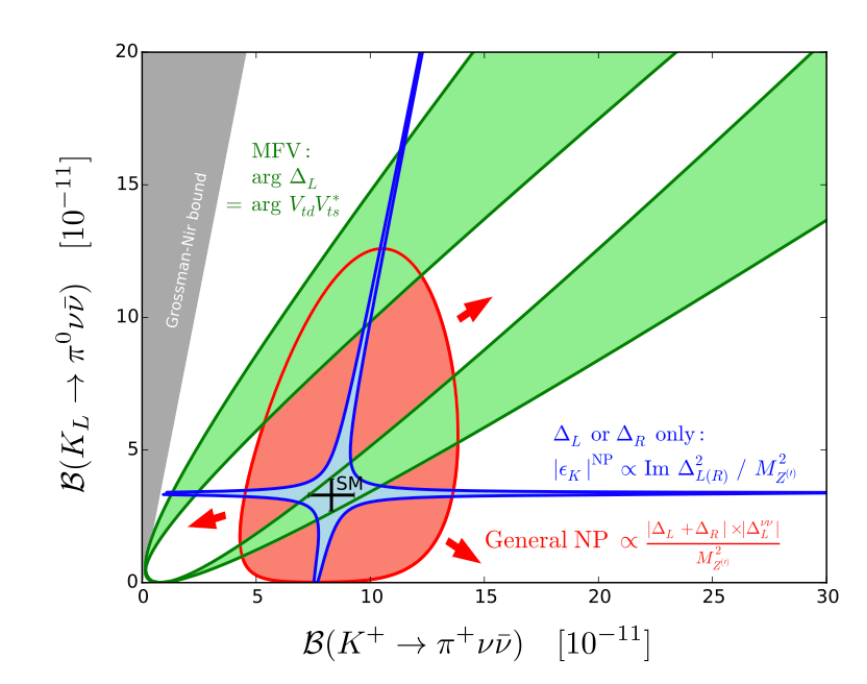}
  \end{center}
    \caption{Correlation between BR($\klpionn$) and BR($\kpluspnn$) for various new physics models.
    The blue region shows the correlation coming from the constraint by the $K$-$\overline{K}$ mixing parameter $\epsilon_{K}$ if
    only left-handed or right-handed couplings are present. The green region shows the correlation for models having
    a CKM-like structure of flavor interactions.  The red region shows the lack of correlation for models with general left-handed and
     right-handed couplings~\cite{ref:newphysics}. }
  \label{newphysics}
\end{figure}

In addition, the $\klpionn$ decay is sensitive to new physics that violates CP symmetry, 
because the $K_{L}$ is mostly $CP$-odd, while $\pi^{0}\nu\bar{\nu}$ is $CP$-even.
On the other hand, the charged-mode decay, $\kpluspnn$, has contributions from both the $CP$-violating and $CP$-conserving processes.
The $\klpionn$ and $\kpluspnn$ decay branching ratios are connected under isospin symmetry, leading to a model-independent bound~\cite{ref:Grossman} ${\cal B}(K_L\to\pi^0\nu\bar\nu)< 4.3\times {\cal B}(K^+\to\pi^+\nu\bar\nu)$. Figure~\ref{newphysics}, reproduced from \cite{ref:newphysics}, shows how the BRs of the $K_L\to\pi^0\nu\bar\nu$ and $K^+\to\pi^+\nu\bar\nu$ would be affected due to new physics effects. In some scenarios, these two BRs have a strong correlation, allowing the new physics scenario to be  distinguished by measurement of both branching ratios.

The BRs of the $\kpinn$ decays are related to other flavor observables, such as lepton-flavor universality violation in the $B$ sector. Most theoretical models put forward to explain such phenomena have strong couplings to the third generation fermions
which cause significant effects on the BRs of the $\kpinn$ decays through couplings to $\tau$ neutrinos~\cite{ref:Banomalies}.
Together with observables in the $B$ sector, measurements of BRs of the $\kpinn$ decays provide crucial information for the investigation of the flavor structure in new physics. 

\begin{figure}[p]
\begin{center}
\includegraphics[width=0.47\textwidth]{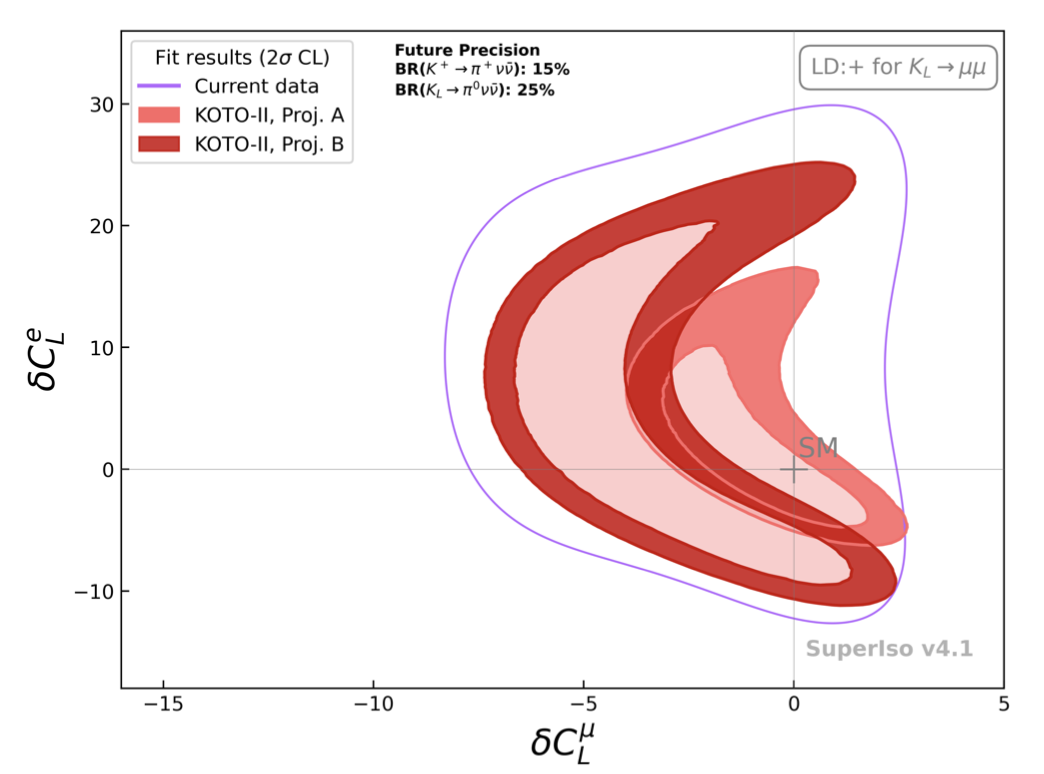}
\hspace*{3mm}
\includegraphics[width=0.47\textwidth]{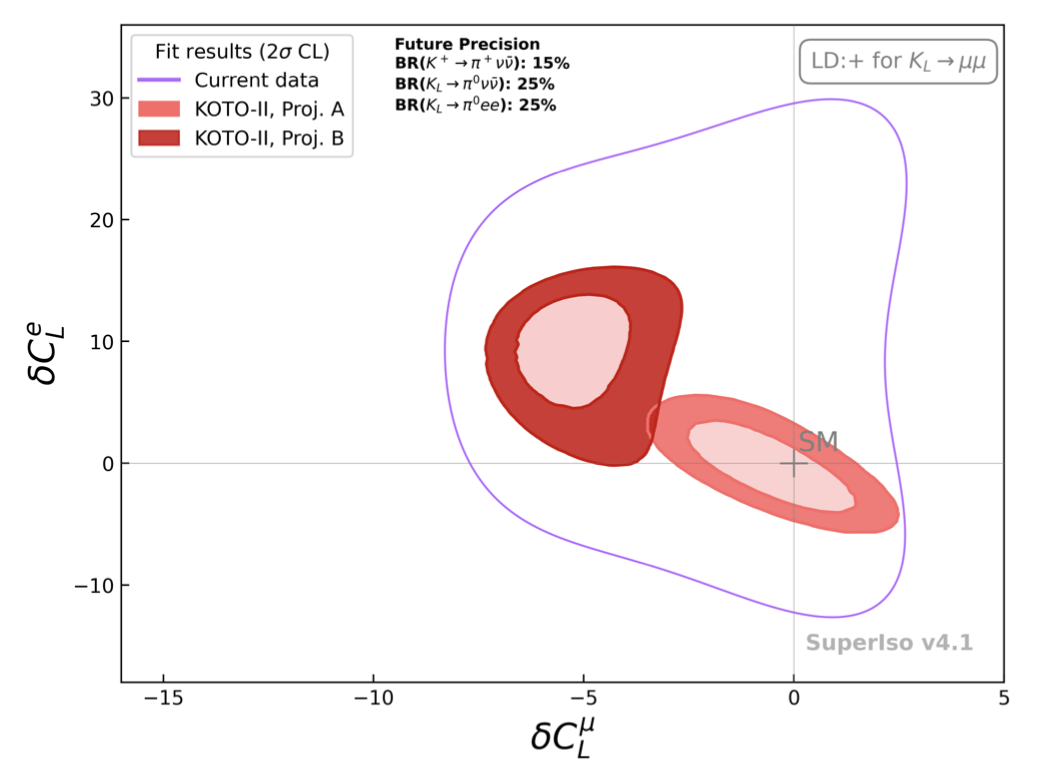} 
\caption{BSM parameter space for Wilson coefficients in scenarios with LFU violation where the NP effects for electrons are different from the those for muons and taus~\cite{DAmbrosio:2022kvb,DAmbrosio:2024ewg}. Left: Impact on allowed parameter space from measurements of the golden channels $K\to\pi\nu\bar\nu$ from NA62 and KOTO~II with the expected final precision. Right: Impact on the parameter space by the inclusion in the fits of a measurement of the $K_L\to\pi^0 e^+e^-$ branching ratio with 25\% precision.}
\label{newphysics1}
\end{center}
\end{figure}

The rare kaon decays investigated by KOTO~II offer the possibility to search for BSM physics with a global fit technique, for example, in the context of lepton-flavor universality (LFU) tests. In the SM, the three lepton flavors ($e$, $\mu$, and $\tau$) have exactly the same gauge interactions and are distinguished only through their couplings to the Higgs field and hence the charged lepton masses. BSM models, on the other hand, do not necessarily conform to the lepton-flavor universality hypothesis and may thereby induce subtle differences between the different generations that cannot
be attributed to the different masses. Among the most sensitive probes of these differences are rare kaon decays with electrons, muons, or neutrinos in the final state. 
For BSM scenarios with LFU violation where the NP effects
for electrons are different from the those for muons and taus~\cite{DAmbrosio:2022kvb,DAmbrosio:2024ewg}, the effect of KOTO~II measurements is shown in Figure~\ref{newphysics1}. The figure illustrates the extent to which
the parameter space can be reduced by projections corresponding to the final precisions of NA62 and KOTO~II for $K_L\to\pi^0\nu\bar\nu$ and $K_L\to\pi^0 \ell^+\ell^-$, and therefore highlights the impact of KOTO~II.

\begin{figure}[p]
\begin{center}
\includegraphics[width=10cm]{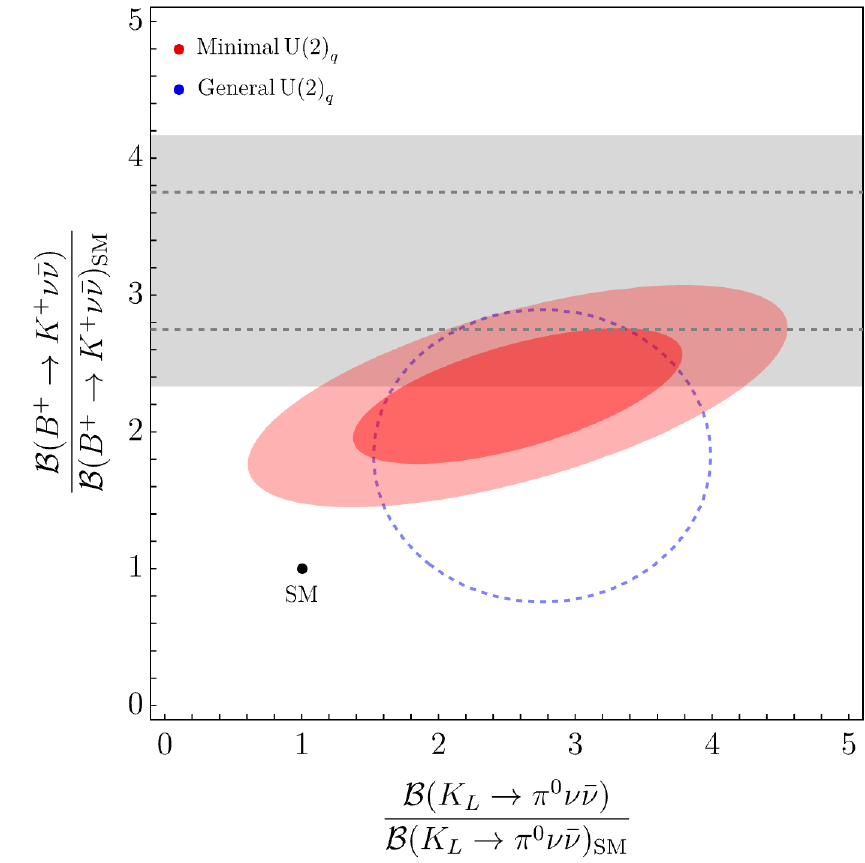}
\end{center}
\vspace{-6mm}
\caption{Correlation between $B^+\to K^+\nu\bar\nu$ and $K_L\to\pi^0\nu\bar\nu$ decay rates (normalized to their SM expected values) in several NP scenarios~\cite{Allwicher:2024ncl}. The red areas denote the parameter regions favored at $1\sigma$ and $2\sigma$ from a global fit in the limit of minimal $U(2)_q$ breaking. Dashed and dotted blue curves are $1\sigma$ and $2\sigma$ regions from a global fit where breaking is a free parameter. The gray bands indicate the current experimental constraints.}
\label{newphysics2}
\end{figure}

Specific models can also be considered to exemplify the
impact of KOTO~II on BSM searches and the interplay with other experiments, notably in $B$ physics. Recently, correlations with the measurements of $K^+\to\pi^+\nu\bar\nu$ by NA62 and evidence for $B^+\to K^+\nu\bar\nu$ by Belle~II have been highlighted, in the context of a new physics scenario aligned to the third generation, with an approximate $U(2)^5$ flavor symmetry acting on the light families~\cite{Allwicher:2024ncl}. The slight excess observed in both channels supports the hypothesis of non-standard TeV dynamics of this type, as also hinted at by other $B$-meson decays, and predicts a similar enhancement for $K_L\to\pi^0\nu\bar\nu$, as can be seen in Figure~\ref{newphysics2}.


\subsection{$\klpioll$ decays}

The ultra-rare $K_L\to\pi^0\ell^+\ell^-$ decays ($\ell=e,\mu$) are theoretically clean golden modes in kaon physics, 
allowing for direct exploration of NP contributions in the $s\to d\ell\ell$ short-distance interaction (to be compared to the $b\to s\ell\ell$ transition). As shown in Figure~\ref{newphysics1}, measurement of the $K_L\to\pi^0e^+e^-$ decay rate to the 25\% precision would provide crucial complementary information for interpretation of the $K\to\pi\nu\bar\nu$ measurements in terms of NP models with LFU violation.

The SM description of the $K_L\to\pi^0\ell^+\ell^-$ decays is provided in Refs.~\cite{DAmbrosio:1998gur,Isidori:2004rb,Mescia:2006jd}. The branching ratios depend on the CKM parameter $\lambda_t=V_{ts}^*V_{td}$ as follows:
\begin{eqnarray}
{\cal B}_{\rm SM}(K_L\to\pi^0e^+e^-) &\!=\!&
\left(15.7|a_S|^2~\pm~6.2|a_S|
\left(\frac{{\rm Im}~\lambda_t}{10^{-4}}\right)~+~2.4
\left(\frac{{\rm Im}~\lambda_t}{10^{-4}}\right)^2
\right)
\times 10^{-12},\nonumber \\
{\cal B}_{\rm SM}(K_L\to\pi^0\mu^+\mu^-) &\!=\!& 
\left(3.7|a_S|^2~\pm~1.6|a_S|
\left(\frac{{\rm Im}~\lambda_t}{10^{-4}}\right)~+~1.0
\left(\frac{{\rm Im}~\lambda_t}{10^{-4}}\right)^2
~+~5.2
\right)
\times 10^{-12}.
\nonumber
\end{eqnarray}
In the above expressions, the first three terms represent the indirect CPV contribution due to $K_S$--$K_L$ mixing, the interference of the indirect and direct CPV contributions (of unknown sign), and the direct CPV contribution determined by short-distance dynamics, respectively. The fourth term in the $K_L\to\pi^0\mu^+\mu^-$ case accounts for the long-distance CP-conserving component due to two-photon intermediate states, which is negligible in the helicity-suppressed $K_L\to\pi^0 e^+e^-$ case. The parameter $a_S$ describes the decay form factor, and has been measured with $K_S\to\pi^0\ell^+\ell^-$ decays to be $|a_S|=1.2\pm0.2$ by the NA48/1 experiment at CERN~\cite{NA481:2003cfm,NA481:2004nbc}.
While the value of $|a_S|$ can be determined from the $K_S\to\pi^0\ell^+\ell^-$ branching ratios, measuring the differential decay distributions as a function of the di-lepton invariant mass would give the form factor parameters, including, in principle, the sign of $a_S$. The LHCb experiment is planning to measure $a_S$ using the $K_S\to\pi^0\mu^+\mu^-$ decay; the prospects for this measurement are discussed in Ref.~\cite{AlvesJunior:2018ldo}.
The sign and value of $a_S$ can be in principle also determined theoretically, since the calculation is stable, with no large cancellations.



The SM expectation for the branching ratios is~\cite{Mescia:2006jd}
\begin{eqnarray}
{\cal B}_{\rm SM}(K_L\to\pi^0 e^+e^-) & = & 3.54^{+0.98}_{-0.85}~ \left(1.56^{+0.62}_{-0.49}\right)\times 10^{-11}, \nonumber\\
{\cal B}_{\rm SM}(K_L\to\pi^0\mu^+\mu^-) & = & 1.41^{+0.28}_{-0.26}~ \left(0.95^{+0.22}_{-0.21}\right)\times 10^{-11}, \nonumber
\end{eqnarray}
where the two sets of values correspond to constructive (destructive) interference between the direct and indirect CP-violating contributions. Beyond the SM, the decay rates can be enhanced significantly in the presence of large new CP-violating phases, in a manner correlated with the effects in $K_L\to\pi^0\nu\bar\nu$ and $\varepsilon^\prime/\varepsilon$~\cite{Aebischer:2022vky}.
These decays also provide unique access to short-distance BSM effects in the photon coupling via the tau loop~\cite{isidori-seminar}.

Numerically in the SM framework, in the assumption of constructive interference (which is preferred theoretically), using ${\rm Im}~\lambda_t=1.4\times 10^{-4}$ and $|a_S|=1.2$, the sensitivities to the input parameters are 
\begin{displaymath}
\frac{\delta{\cal B}(K_L\to\pi^0e^+e^-)}{{\cal B}(K_L\to\pi^0e^+e^-)}=
0.53 \cdot \frac{\delta\,{\rm Im}~\lambda_t}{{\rm Im}~\lambda_t} ,
\quad
\frac{\delta{\cal B}(K_L\to\pi^0\mu^+\mu^-)}{{\cal B}(K_L\to\pi^0\mu^+\mu^-)} =
0.44 \cdot \frac{\delta\,{\rm Im}~\lambda_t}{{\rm Im}~\lambda_t} 
\end{displaymath}
and
\begin{displaymath}
\frac{\delta{\cal B}(K_L\to\pi^0e^+e^-)}{{\cal B}(K_L\to\pi^0e^+e^-)} =
1.48 \cdot \frac{\delta|a_S|}{|a_S|},
\quad
\frac{\delta{\cal B}(K_L\to\pi^0\mu^+\mu^-)}{{\cal B}(K_L\to\pi^0\mu^+\mu^-)} =
0.88\cdot \frac{\delta |a_S|}{|a_S|}.
\end{displaymath}

Experimentally, the most stringent upper limits (at 90\% CL) on the branching ratios have been obtained by the KTeV experiment~\cite{KTeV:2003sls,KTEV:2000ngj}:
\begin{displaymath}
{\cal B}(K_L\to\pi^0 e^+e^-) < 28\times 10^{-11}, \quad {\cal B}(K_L\to\pi^0 \mu^+\mu^-) < 38\times 10^{-11}.
\end{displaymath}



\subsection{Other decays}

Given that the setup and detector will be optimized for the most challenging decay channels, KOTO~II will have the potential to also address searches for other rare decays and new particles, as outlined in this section.

\subsubsection{Rare kaon decays}
\begin{itemize}

\item \textbf{$K_L\to\pi^0\gamma \gamma$}: 
This decay is particularly important for determining the amplitude of CP-conserving components in the $K_L\to\pi^0 e^+ e^-$ process, 
where both CP-violating and CP-conserving contributions compete within the decay amplitude.
Previously, this decay was measured by Fermilab E832 and CERN NA48; their results differ by three standard deviations. Providing an improved  ${\cal B}(K_L\to\pi^0\gamma\gamma)$ measurement would require observing at least $\mathcal{O}(10^4)$ events, which is beyond the current KOTO capabilities but feasible with KOTO~II in a few years of data-taking.
\item \textbf{$K_L \to \pi^0 \pi^0 \nu \bar{\nu}$}:
This decay process with predominately clean short-distance contributions has a predicted branching ratio that is proportional to the square of the CKM matrix element \( \rho \), making it significant for testing the Standard Model. However, the expected value of  ${\cal B}(K_L\to\pi^0\pi^0\nu\bar\nu)$ is only \( (1.4 \pm 0.4) \times 10^{-13} \). This decay has previously been studied using the E391a data set, which yielded an upper limit of $8.1\times 10^{-7}$, with the precision of the result limited by the available statistics. With the expected increase in \( K_L \) yields, KOTO~II is anticipated to tighten the existing upper limit or even observe this rare decay.
\end{itemize}


\subsubsection{Searches for new particles}
\begin{itemize}
\item \textbf{Dark Photons}:
The dark photon, $\bar\gamma$, is a hypothetical gauge boson associated with a new abelian gauge symmetry, $U(1)_D$, under which states in the dark sector are charged. Depending on whether the $U(1)_D$ symmetry is exact or broken, the dark photon can be either massive or massless. The KOTO~II detector, with its hermetic veto system, is highly effective for probing for dark photons, particularly in cases where dark photons are invisible in their subsequent decays. These potential studies include $K_L\to\pi^0 \bar{\gamma}$, $K_L\to\pi^0 \gamma \bar{\gamma}$, and $K_L \to\pi^+\pi^- \bar{\gamma}$~\cite{Antel:2023hkf}.
\item \textbf{Axion-Like Particles}: KOTO~II is also sensitive to probe for the existence of axion-like particles (ALPs, $a$). Axions are theoretical particles originally proposed to resolve the strong CP problem in quantum chromodynamics, and ALPs are generalizations of axions that could be present in kaon decays via $s \to da$ transitions. Possible kaon decay channels involving ALP production include $K_L\to \pi^0 a$ and $K_L\to \pi^0 \pi^0 a$, with subsequent $a\to\gamma\gamma$ or $a\to\ell^+\ell^-$ decays.
The KOTO~II sensitivity for ALP decays $a\to\gamma\gamma$ has been studied, both for ALPs produced in the target (i.e., in beam-dump mode) and from kaon decays, demonstrating that
KOTO~II can probe new regions in the parameter space for ALP models, particularly for ALPs from kaon decays~\cite{Afik:2023mhj}.
The high-precision charged particle tracker in KOTO~II will enable detailed studies of channels with $a\to\ell^+\ell^-$ decays as well.
\end{itemize}

\section{History of the $\klpionn$ search}
\label{chap:history}
\begin{figure}[htbp]
 \begin{center}
  \includegraphics[width=12cm]{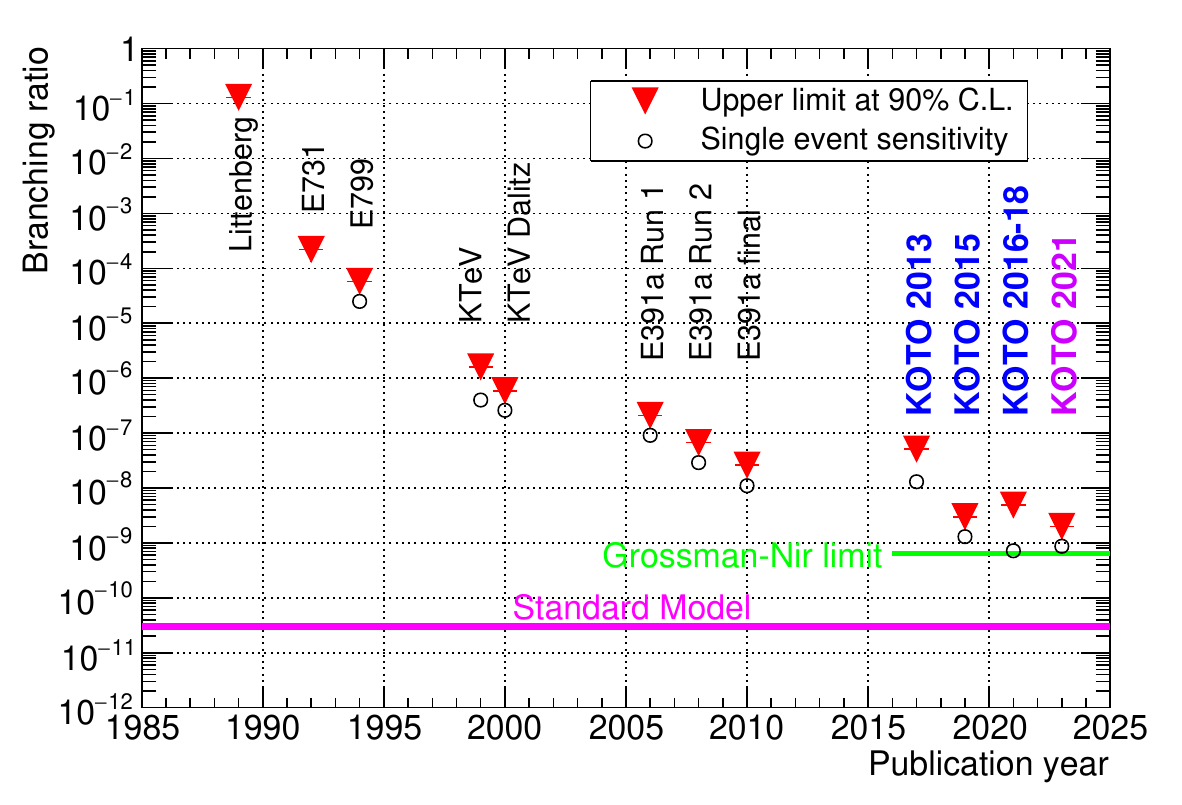}
  \end{center}
    \caption{Evolution of the $\klpionn$ search. Data points used in the figure are quoted from \cite{Littenberg:1989ix, Graham:1992pk,E799:1994amx,KTeV:1998taf,E799-IIKTeV:1999iym, E391a:2006fxm, E391a:2007qcj, E391a:2009jdb, KOTO:2016vwr, KOTO2015, KOTO2016-18, KOTO:2024zbl}. The Grossman-Nir limit in the figure is calculated from the measured $\kpluspnn$ branching ratio by the CERN NA62 experiment, published in 2021~\cite{NA62:2021zjw}.}
  \label{klpionn-evo}
\end{figure}

Figure~\ref{klpionn-evo} shows the evolution of the experimental search for the $\klpionn$ decay.
Starting from the re-analysis of the early $K_L \to 2\pi^0$ experiment~\cite{Littenberg:1989ix}, initial searches were conducted at FNAL~\cite{Graham:1992pk,E799:1994amx,KTeV:1998taf,E799-IIKTeV:1999iym}, KEK~\cite{E391a:2006fxm, E391a:2007qcj, E391a:2009jdb}, and J-PARC~\cite{KOTO:2016vwr, KOTO2015, KOTO2016-18, KOTO:2024zbl}.
The E391a experiment at KEK was the first experiment dedicated to the $\klpionn$ search, and the KOTO experiment at J-PARC is its successor.
Despite the technical difficulty of searching for an ultra rare decay that has only neutral particles in its initial and final states and has missing particles, the experimental sensitivity now reaches the level of $\sim 10^{-9}$ on the branching ratio.

The KOTO experiment will continue taking data for several more years, aiming to search for possible enhancements of the $\klpionn$ branching ratio from new physics effects. 
However, considering the realistically available beam power and beam time, the  sensitivity reachable will be better than $10^{-10}$ but worse than the branching ratio predicted by the SM, $3\times 10^{-11}$.
The expected number of background events will be $O(1)$, so the potential for signal observation will still be limited. It is therefore desirable to consider a next-generation experiment that is sensitive enough to observe a number of $\klpionn$ events that is sufficient to claim the discovery of the decay and to measure its branching ratio.

An idea for such an experiment, then called KOTO step-2, was described in the original KOTO proposal in 2006~\cite{KOTOproposal}.
Since then, we have gained a wealth of experience in the KOTO experiment and are ready to consider and design the next-generation experiment more realistically.
We are eagerly considering 
the realization of KOTO~II, aiming to run in the 2030s.

\section{Strategy}
\label{chap:concept}
\begin{figure}[htb]
\begin{minipage}{0.5\linewidth}
\includegraphics[width=\linewidth]{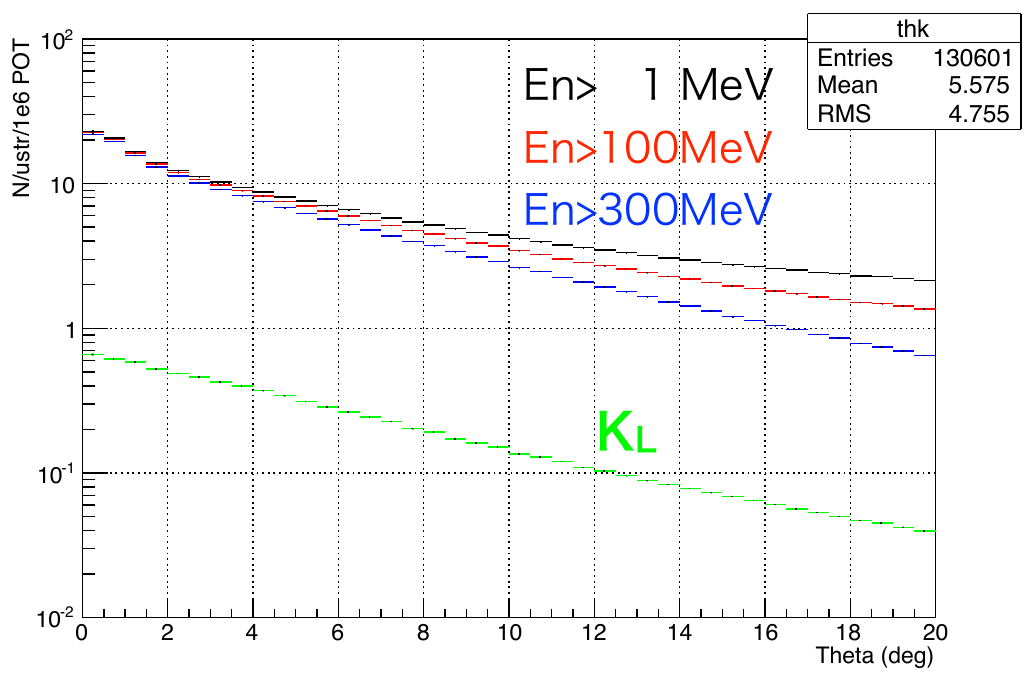}
\end{minipage}
\begin{minipage}{0.5\linewidth}
\includegraphics[width=\linewidth]{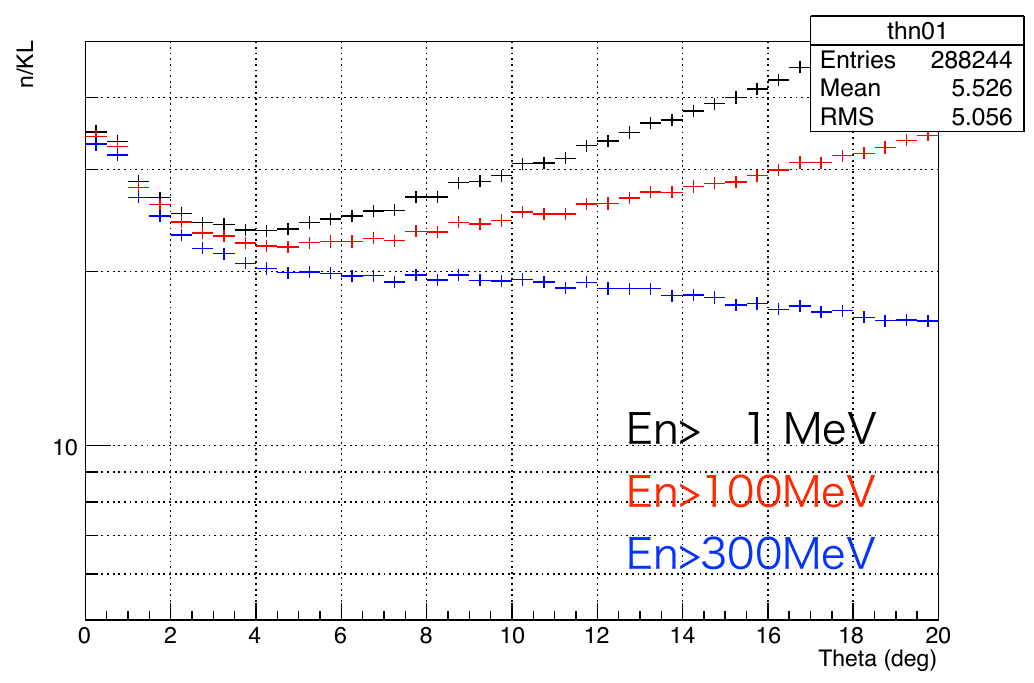}
\end{minipage} 
\caption{\label{fig:vsangle}Simulated $K_L$ and neutron yields (left) and their ratio (right) as functions of the production angle~\cite{ref:kaon2019}. 
The yields were evaluated at 1~m downstream of the target, 
normalized by the solid angle ($\mu$str).
Black, red, and blue points indicate the results when selecting neutrons with their energies of more than 1, 100, and 300~MeV, respectively.}
\end{figure}

\begin{figure}[htb]
\centering
\includegraphics[width=0.8\linewidth]{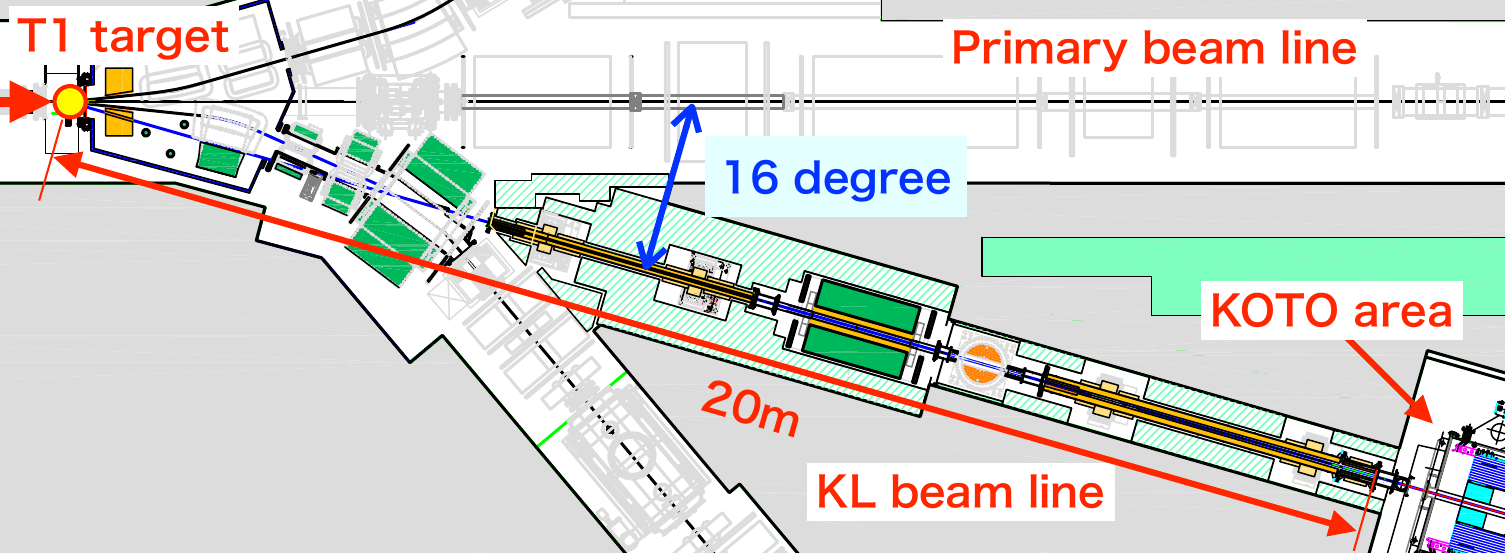}
\caption{\label{fig:kotobl}Schematic drawing of the $K_L$ beam line for the KOTO experiment in the current Hadron Experimental Facility.}
\end{figure}

\begin{figure}[htb]
\centering
\includegraphics[width=\linewidth]{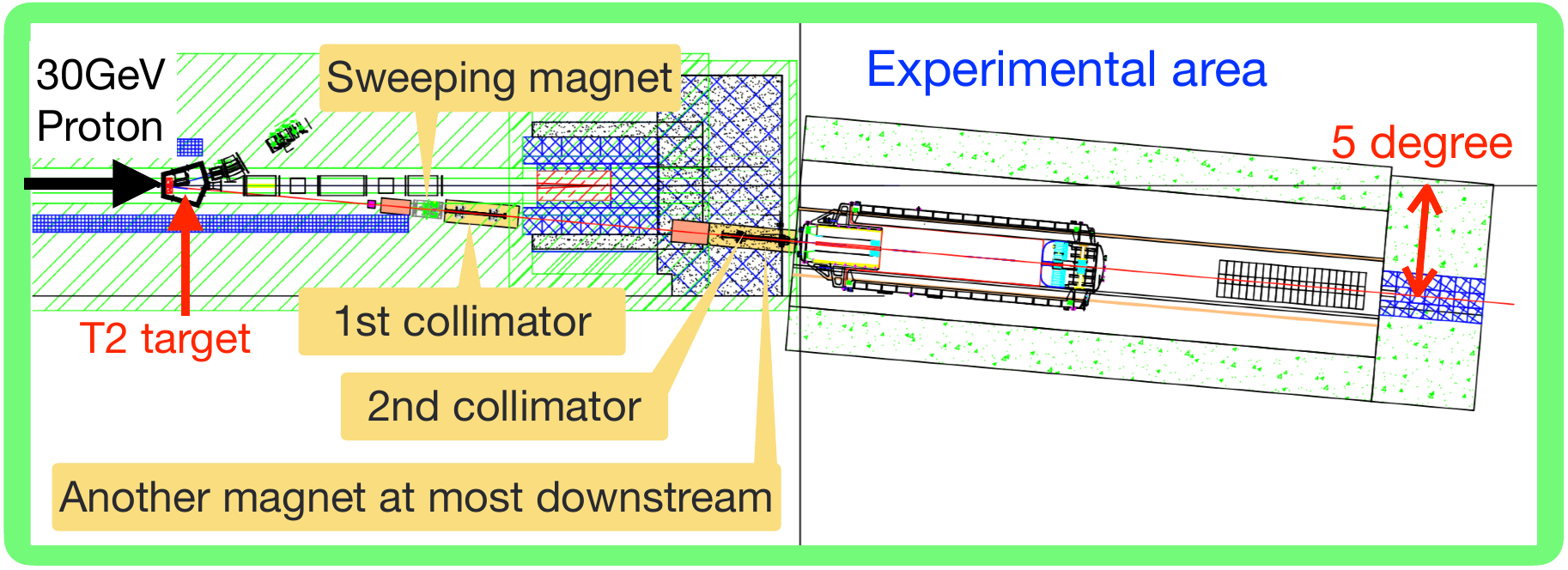}
\caption{\label{fig:koto2bl}Schematic drawing of the KOTO~II beam line in the Extended Hadron Experimental Facility.
The experimental area is located behind the beam dump. Distance between the T2 target and the experimental area is assumed to be 43~m in the present studies.
A modeled cylindrical detector with 20~m in length and 3~m in diameter is described in the experimental area as a reference.}
\end{figure}

To achieve high experimental sensitivity for the $\klpionn$ measurement, the $\kl$ flux, the detection acceptance for the signal, and the signal-to-background ratio must be maximized.
The $\kl$ flux is determined by the production angle and the solid angle of the secondary neutral beam.
Other important parameters are the achievable intensity of the primary proton beam and the target properties (material, thickness, etc.).
The production angle is defined as the angle between the directions of the primary proton beam and secondary neutral beam. 
Figure~\ref{fig:vsangle} shows the $\kl$ and neutron yields and the neutron-to-$\kl$ flux ratio as functions of the production angle when a 102-mm-long gold target is used.
%
For KOTO~II, a production angle of 5~degrees is chosen as an optimum balance between $\kl$ flux and smaller neutron fraction in the beam. 
In case of the KOTO experiment, the production angle is 16-degrees, as shown in Fig.~\ref{fig:kotobl}, 
which was chosen to utilize the T1 target, with the experimental area away from the primary beam.
%
In order to realize the 5-degree production while keeping the solid angle of the neutral beam as large as possible,
\ie, with the shortest beam line, a new experimental area behind the primary beam dump and a new target station close to the dump are necessary.
Figure~\ref{fig:koto2bl} shows a possible configuration
utilizing a new target (T2).

Here, we introduce the baseline design for the KOTO~II detector.
The acceptance of the signal is primarily determined by the detector geometry.
The basic detector configuration is the same as for the current KOTO experiment: a cylindrical detector system with an electromagnetic calorimeter to detect two photons from the $\pi^0$ decay at the downstream endcap. 
A longer $\kl$ decay region and a larger diameter of the calorimeter are being considered to obtain a larger acceptance.
The 5-degree production also provides a benefit in terms of the signal acceptance: the $\kl$ momentum spectrum is harder than it is in KOTO, so the two photons from $\pi^0$ decays are boosted more in the forward direction, allowing the acceptance gain from the longer decay region to be utilized.

The ability to claim discovery of the  signal decay and the precision of the branching ratio measurement depend on the signal-to-background ratio, as well as on the expected number of observed events.
The background level is affected by many factors such as the beam size at the detector, the flux of beam particles (neutrons, $\kl$) leaking outside the beam (beam-halo),
 the fraction of charged kaons in the neutral beam, and the detector performances.

In the following two sections, Sections~\ref{chap:beamline} and \ref{chap:detector}, a beam line model and a conceptual design of the detector are described.
Discussions of the sensitivity and background follow in Section~\ref{chap:sensitivity}, with parameterized detector performance.
\clearpage

\section{Beam Line}
\label{chap:beamline}

\subsection{Performance of the beam line}
Table~\ref{tab:beampar} summarizes the beam parameters for KOTO~II
and those in the current KOTO experiment. 
\begin{table}[htb]
\caption{\label{tab:beampar}Beam parameters for KOTO~II and the current KOTO experiment.}
\begin{center}
\begin{threeparttable}
\begin{tabular}{lll}
\hline
&KOTO~II\tnote{*}&KOTO\\
\hline
Beam power&100~kW&80~kW (100~kW in future)\\
Target&102-mm-long gold&60-mm-long gold\\
Production angle&5$^{\circ}$&16$^{\circ}$\\
Beam line length&43~m&20~m\\
Solid angle&4.8~$\mu$sr&7.8~$\mu$sr\\
\hline
\end{tabular}
\begin{tablenotes}\footnotesize
\item[*] 
Note the parameters for KOTO~II are tentative for this study.
\end{tablenotes}
\end{threeparttable}
\end{center}
\end{table}

\begin{figure}[htb]
\centering
\begin{minipage}{0.45\linewidth}
\includegraphics[width=\linewidth]{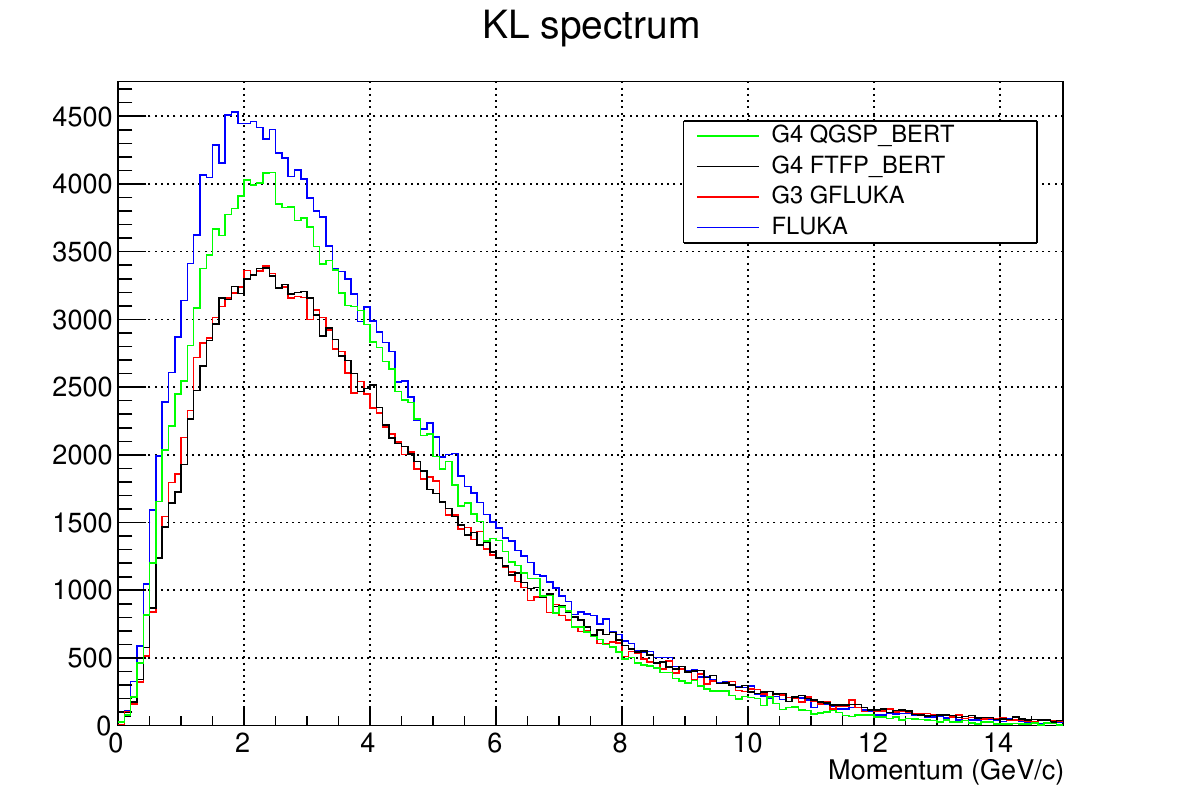}
\end{minipage}
\begin{minipage}{0.45\linewidth}
\includegraphics[width=\linewidth]{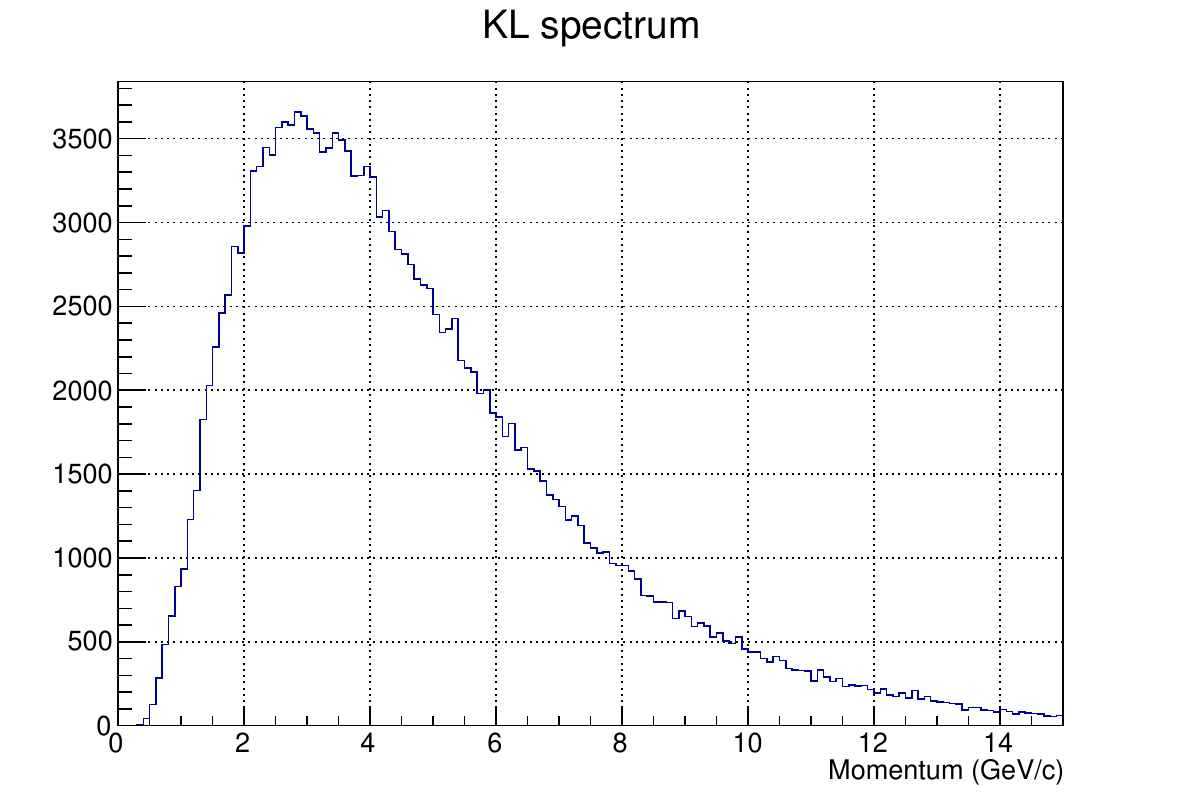}
\end{minipage}
\caption{\label{fig:klmom}
$\kl$ spectra at 1~m from the T2 target from the target simulation (left) and at the exit of the KL2 beam line at 43~m from the T2 target from the beam line simulation (right). In the left plot, the results obtained with various simulation packages are also shown, as well as the result by the GEANT3-based simulation (labeled ``G3 GFLUKA'') which is our default in this study.}
\end{figure}

To evaluate the performance of the beam line for KOTO~II (KL2 beam line), target and beam line simulations were conducted. The target in the study was chosen to be a simple cylindrical rod made of gold with a diameter of 10~mm and length of 102~mm, which corresponds to 1$\lambda_{I}$ (interaction length). The 30~GeV primary protons were incident on the target with a beam spot size ($\sigma$) of 1.6~mm in both the horizontal and vertical directions. No beam divergence was considered in the simulation. Secondary particles emitted at an angle of 5~degrees (within $\pm$0.3~degree) to the primary beam direction were recorded at 1~m downstream from the target 
to be used as input to the beam line simulation.
For the simulation of the particle production at the target, we used the GEANT3-based simulation as a default, and GEANT4 (10.5.1 with a physics list of QGSP\_BERT or FTFP\_BERT) and FLUKA (2020.0.3) for comparison, as shown in Fig.~\ref{fig:klmom} (left).
The resultant $\kl$ fluxes were found to agree with each other to within 30\%. GEANT3 provided 
the smallest $\kl$ yield
 and thus is considered to be a conservative choice in the discussion of the sensitivity.

In designing the KL2 beam line, we first follow the design strategy of the KOTO beam line (KL beam line)~\cite{ref:kotobl}.
The KL2 beam line consists of two stages of 5-m-long collimators, a photon absorber in the beam, and a sweeping magnet to sweep out charged particles from the target.
The photon absorber, made of 7-cm-thick lead, is located at 
4~m downstream of the target.
The first collimator, starting from 20~m from the target, defines the solid angle and shape of the neutral beam.
The solid angle is set to be 4.8~$\mu$str.
The second collimator, starting from 38~m from the target, eliminates the particles coming from the interactions at the photon absorber and the beam-defining edge of the first collimator. The bore shape of the second collimator is designed not to be seen from the target so that particles coming directly from the target do not hit the inner surface and thus do not generate particles leaking 
outside the beam.
The first sweeping magnet is located upstream of the first collimator in this study.
Although an additional sweeping magnet is needed at the end of the beam line in order to sweep out charged kaons which are produced by interactions of neutral particles with the collimators, its effect was evaluated independently from the baseline beam line design, as discussed later.

\begin{figure}[htb]
\centering
\begin{minipage}{0.45\linewidth}
\includegraphics[width=\linewidth]{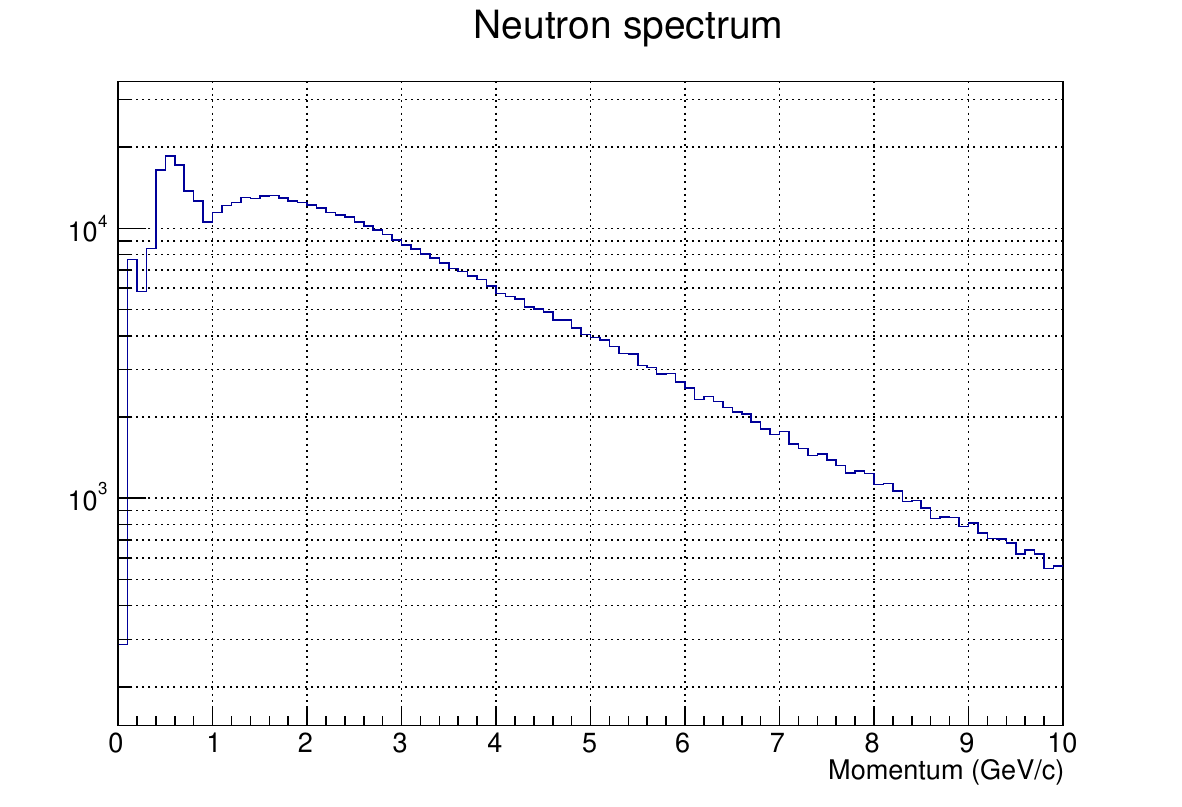}
\end{minipage}
\begin{minipage}{0.45\linewidth}
\includegraphics[width=\linewidth]{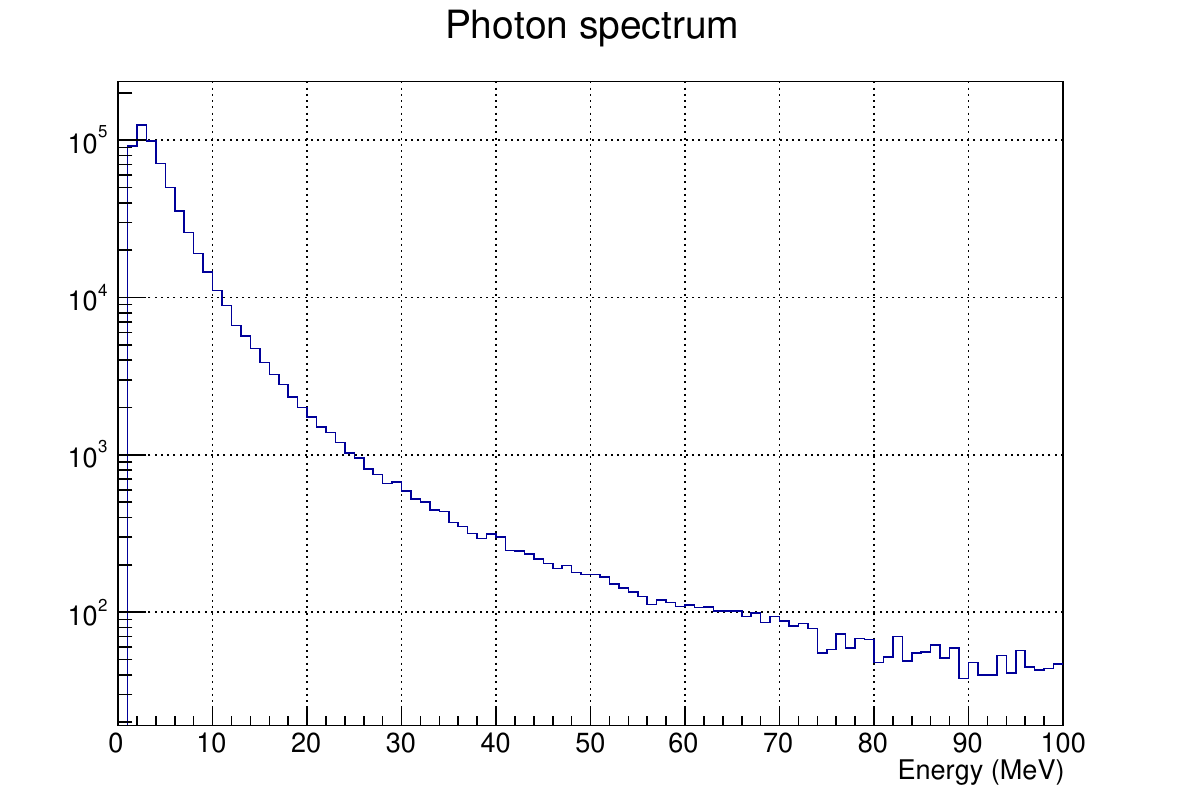}
\end{minipage}
\caption{\label{fig:ngspectra}Simulated neutron (left) and photon (right) spectra at the exit of the beam line.}
\end{figure}

Figures~\ref{fig:klmom} (right) and \ref{fig:ngspectra} show the simulated spectra of $\kl$, neutrons, and photons at the exit of the KL2 beam line, respectively.
The $\kl$ yield was evaluated to be $1.1\times 10^7$ per $2\times 10^{13}$ protons on the target (POT).
Note that the beam power of 100~kW corresponds to $2\times 10^{13}$~POT per second with 30~GeV protons.
The resultant $\kl$ flux per POT is 2.6 times higher than that of the current KOTO experiment.
The $\kl$ spectrum peaks at 3~GeV/c, while it is 1.4~GeV/c in the current KOTO experiment.
The simulated particle fluxes are summarized in Table~\ref{tab:yield}.
\begin{table}[htb]
\caption{\label{tab:yield}Expected particle yields estimated by the simulations.}
\begin{center}
\begin{threeparttable}
\begin{tabular}{cccc}
\hline
\multirow{2}{*}{Particle}&\multirow{2}{*}{Energy range}&Yield & On-spill rate\\
&&(per $2\times 10^{13}$~POT) & (MHz)\\
\hline
$\kl$&&$1.1\times 10^7$&24\\
\hline
\multirow{2}{*}{Photon}&$>$10~MeV&$5.3\times10^7$&110\\
&$>$100~MeV&$1.2\times10^7$&24\\
\hline
\multirow{2}{*}{Neutron}&$>$0.1~GeV&$3.1\times10^8$&660\\
&$>$1~GeV&$2.1\times10^8$&450\\
\hline
\end{tabular}
\begin{tablenotes}\footnotesize
\item[]
The beam power of 100~kW corresponds to $2\times 10^{13}$~POT/s with 30~GeV protons.
The on-spill rate means the instantaneous rate during the beam spill, assuming a 2-second beam spill every 4.2 seconds. 
\end{tablenotes}
\end{threeparttable}
\end{center}
\end{table}

\begin{figure}[htb]
\centering
\centering
\begin{minipage}{0.45\linewidth}
\includegraphics[width=\linewidth, height=\linewidth]{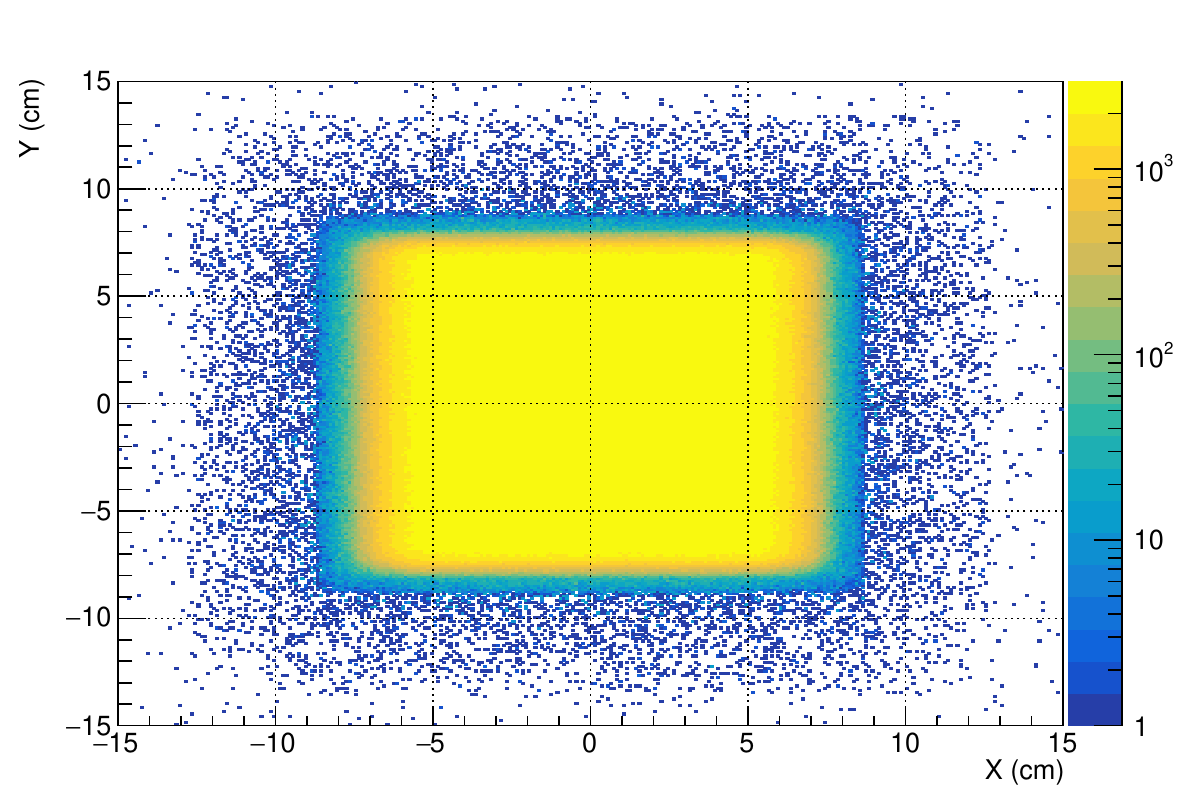}
\end{minipage}
\begin{minipage}{0.45\linewidth}
\includegraphics[width=\linewidth]{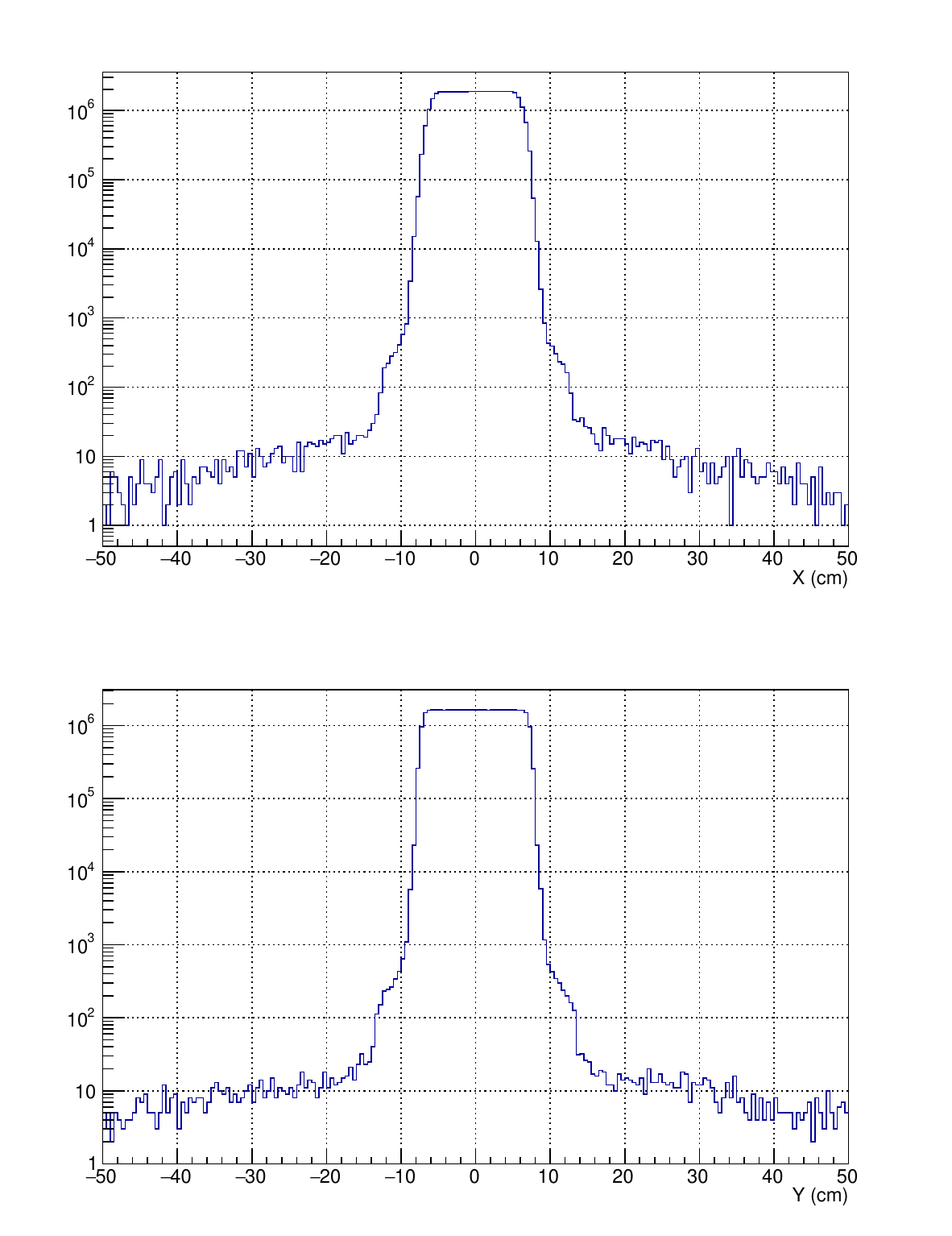}
\end{minipage}
\caption{\label{fig:nprof}Beam shape at the downstream endcap plane, represented by the neutron profile.
The left figure shows the distribution within $\pm 15$~cm of the beam center, and the right top (bottom) histogram indicates the horizontal (vertical) distribution within $\pm 50$~cm of the beam center.}
\end{figure}

Figure~\ref{fig:nprof} shows the neutron profile at the assumed calorimeter location, 64~m from the T2 target. As shown in the figure, the neutral beam is shaped so as to be a square at the calorimeter location.
The evaluation of neutron fluxes spreading into the beam halo region, called ``halo neutrons'', is important, since these are potential sources of backgrounds due to their interactions with the detector materials.
Here, we define the core and halo neutrons as those inside and outside the $\pm$10~cm region at the calorimeter, respectively.
The ratio of the halo neutron yield to the core yield was found to be $1.8 \times 10^{-4}$.

\subsubsection{Charged kaons in the neutral beam}
Contamination from charged kaons in the neutral beam line is harmful, since $K^\pm$ 
decays such as $K^\pm \to \pi^0 e^\pm \nu$
 in the detector region can mimic the signal, as pointed out in the analysis of data from the KOTO experiment~\cite{KOTO2016-18}.
The major production point of charged kaons is the second collimator. 
Neutral particles ($\kl$s and neutrons) hit the inner surface of the collimator, interacting to produce charged kaons that can reach the end of the beam line. Charged pions from $\kl$ decays hitting the collimator can also produce charged kaons.
According to the beam line simulation, 
the ratio of the charged kaon to $\kl$ fluxes entering the decay region is $R(K^\pm/\kl)=4.1\times 10^{-6}$.
To evaluate the reduction by an additional sweeping magnet, 
we conducted another beam line simulation with a sweeping magnet that provides a magnetic field of 2~Tesla in 1.5~m~long at the end of the beam line.
We confirmed that this can reduce the ratio to $R<1.1\times 10^{-6}$, which is limited by the simulation statistics.

\subsubsection{Discussion on the target length}
\begin{figure}[htb]
\centering
\includegraphics[width=0.7\linewidth]{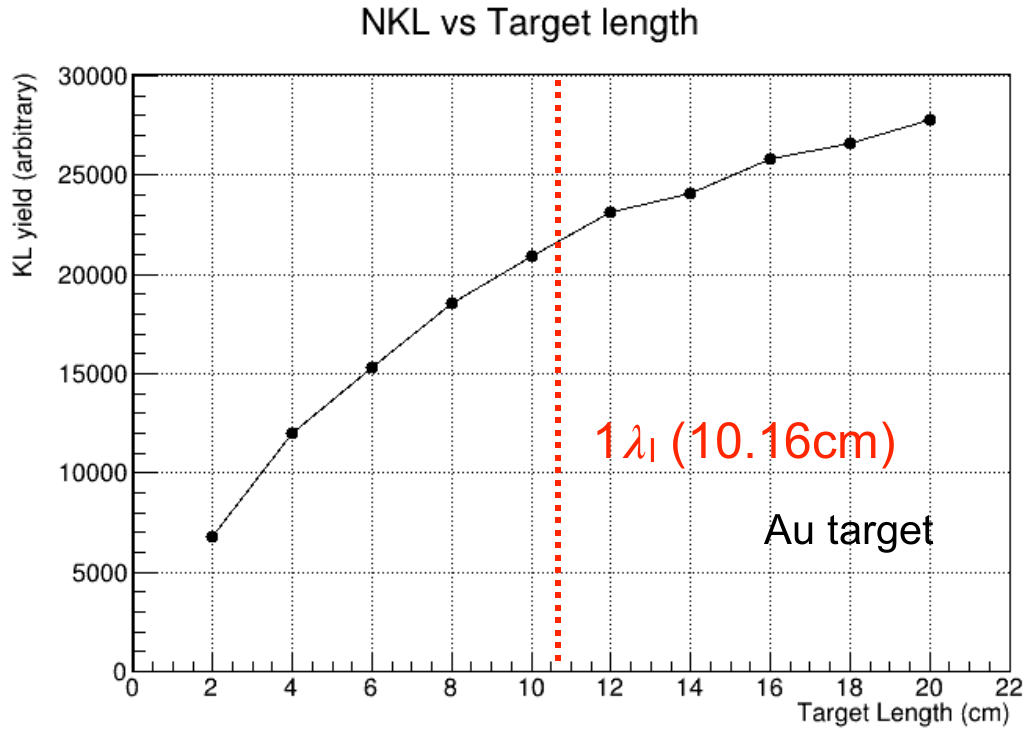}
\caption{\label{fig:tlen}$\kl$ yield as a function of the target length. The yield is evaluated at 1~m from the T2 target. A FLUKA-based simulation is used for this study.}
\end{figure}

In the target simulation,
the gold target was assumed to be 102~mm~long, while the T1 target used in the current Hadron Experimental Facility is 60~mm~long.
Although the use of a thicker target may pose some technical difficulties, there would be substantial advantages.
Figure~\ref{fig:tlen} indicates the relative $\kl$ yield as a function of the target length.
As can be seen, a 102-mm-long gold target provided 40\% more $\kl$ yield than a 60-mm-long gold target. In addition, the flux of photons emerging from a thicker target is reduced and the energy spectrum is softened. Thus, if a thinner target is used, the rates on the in-beam vetoes will be higher, and the thickness of the photon absorber may have to be increased in compensation. 
%
%

\subsubsection{Optimization of the photon absorber}
In order to keep the flux of photons in the neutral beam to levels that, even if challenging, can be realistically handled by the in-beam veto detectors, the beam-line has a 7-cm-thick photon absorber at 
$z=4$~m. 
Unfortunately, elastic and inelastic interactions of $K_L$s in the photon absorber reduce the kaon flux 
to 60\%, according to the beam-line simulation.
This loss of $K_L$ flux can be partially mitigated by the use of an aligned, high-$Z$, crystal-metal photon absorber. When an electromagnetic particle is incident on a crystal at a direction parallel to a crystal axis, the coherent superposition of the electric fields from the atoms in the lattice increases the probability of bremsstrahlung emission and pair production~\cite{Bak:1988bq,Kimball:1985np,Baryshevsky:1989wm}. The effects of coherent interactions increase with photon energy and for decreasing angle of incidence. For a tungsten crystal, the strong-field regime, in which coherent effects become particularly important, is reached for incident photons of energy of about 15 GeV; a good fraction of the high-energy photons incident on the absorber in the KOTO~II beam would be above this threshold.  

A series of exploratory tests was performed by the KLEVER and AXIAL collaborations at the CERN SPS in 2018 with a set of tungsten crystals and a tagged photon beam.
In particular, when the $\langle111\rangle$ axis of a commercial-quality tungsten crystal of 10-mm thickness was aligned with the beam, the effective radiation length for the conversion of the primary photon was seen to increase with incident photon energy over the range 30--100 GeV, ultimately reaching more than 3 times its nominal value at 100~GeV~\cite{Soldani:2022ekn}. This effect was maintained over an angular acceptance of at least a few mrad, which is consistent with the strong-field alignment angle of 1.75~mrad for tungsten, especially given the mosaicity of the commercial sample. 

The amount by which the KOTO~II photon absorber could be reduced in thickness while maintaining the required for photon conversion is amenable to simulation; simulations of this effect were shown to reproduce the results of~\cite{Soldani:2022ekn}. Notably, the ($\pm1.1$~mrad)$^2$ opening angle of the KOTO~II neutral beam is well within the angular acceptance for coherent interactions in a crystal such as tungsten. The use of an aligned crystal metal absorber requires only a static mechanical scheme to preserve the crystal alignment; a precise optical survey may be used, once the crystal axes have been measured, e.g., at a test beam. Alignment to within a few tenths of a mrad would be ideal but would be sufficient even to within 1~mrad. Since the alignment is technically straightforward
and could lead to a significant increase in the available $K_L$ flux, this option should be given further consideration.

\subsection{Activity in the experimental area behind the beam dump}
To realize the 5-degree production angle, the experimental area must be located behind the primary beam dump.
There is a concern that many particles from interactions in the beam dump can penetrate the shield and reach the experimental area, which causes a high rate of accidental hits in the detector.
To evaluate the flux in the experimental area, we performed a muon flux measurement behind the beam dump at the current facility~\cite{MATSUMURA2024169990}. Since the beam height is below ground level, we dug a hole for observation and measured a vertical distribution of the muon flux with a compact detector developed for this purpose.

\begin{figure}[htb]
\centering
\includegraphics[width=0.8\linewidth]{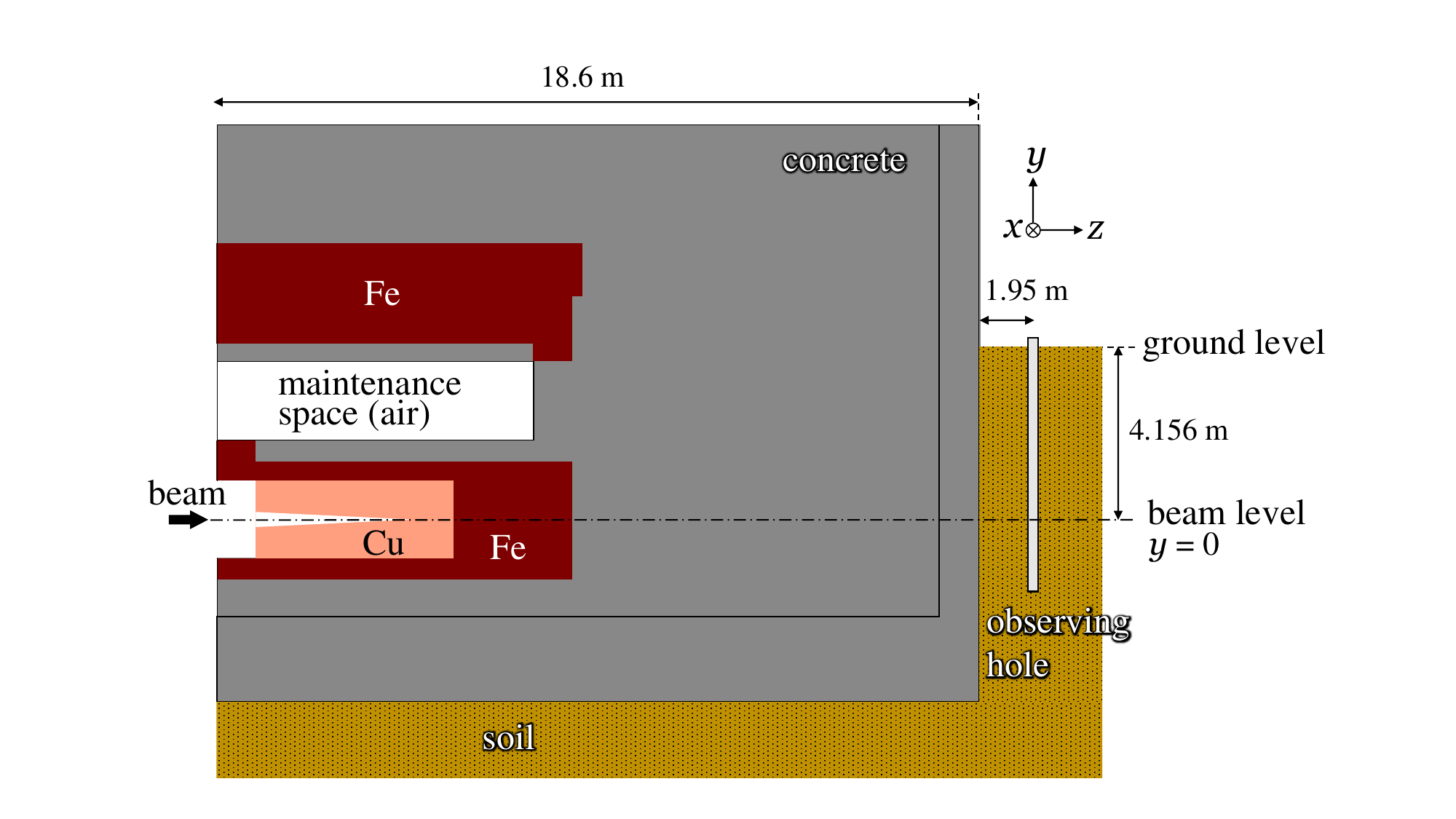}
\caption{Illustration of the side cross-sectional view of the current beam dump of the J-PARC Hadron Experimental Facility and the location of the observing hole.}\label{fig:bd-hole}
\end{figure}

The observing hole was located at 1.95~m downstream of the end wall of the beam dump as shown in Fig~\ref{fig:bd-hole}. 
The detector consisted of 12 plastic scintillator strips with a trapezoidal cross section, 
which were placed in a cylindrical shape. 
Each strip was 80~mm long, 14.7~mm wide on the short side, and 5~mm thick. 
The muon flux was determined by coincidence measurement of two facing strips. 
We measured the muon-coincidence rate by changing the vertical position of the detector at 0.5~m intervals in the observing hole. The beam power of the primary proton beam was 30~kW.
A detailed description of the experimental setup and conditions is found in Ref.\cite{MATSUMURA2024169990}.

\begin{figure}[htb]
\centering
\includegraphics[width=\linewidth]{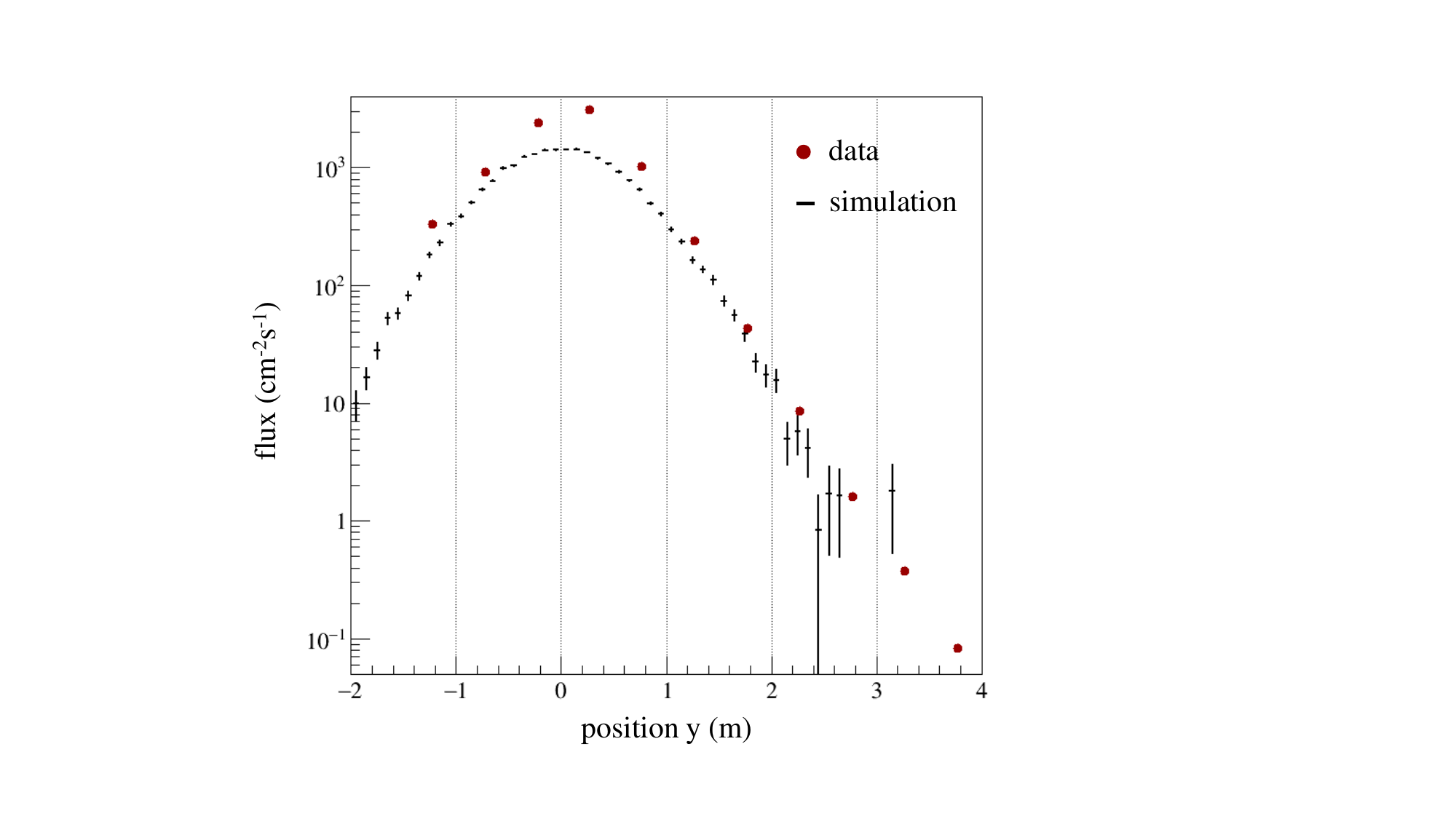}
\caption{Vertical position dependence of the muon flux at the observing hole. Black circles denote the measured muon fluxes with the 30 kW proton beam intensity, and red crosses show the muon flux calculated with the beam-dump simulation.}\label{fig:bd-flux}
\end{figure}

Figure.~\ref{fig:bd-flux} shows the vertical position dependence of the muon flux at the observing hole. 
The muon flux was highest near the beam level and decreased rapidly with distance from the beam axis. 
Compared to data, the GEANT4 simulation underestimates the muon flux by a factor of 2 near the beam level. 
In contrast, the simulation result agrees with the data to within 30\% at positions higher 
than $y=+1$~m. For the purposes of muon-flux estimation for the KOTO II experiment, 
which will be placed at the horizontally off-axis, 
we can conclude that the beam dump simulation can be used with reasonable precision.

%

%
\begin{figure}[htb]
\centering
\includegraphics[width=\linewidth]{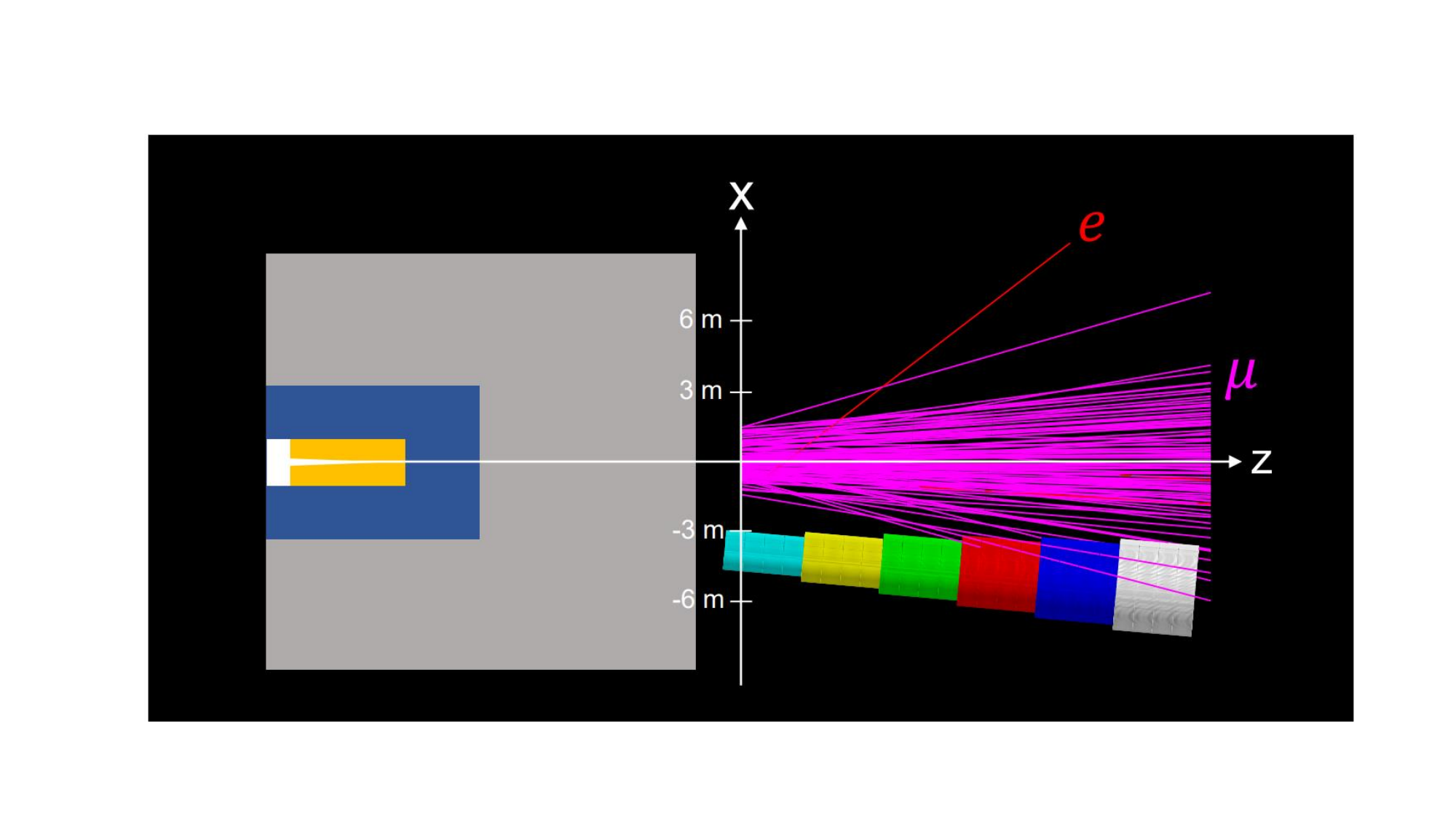}
\caption{Event display when 100 muons are generated from the location of the observing hole. The position of the observing hole is taken as the origin and the figure is shown as seen from the y-axis direction. Some of the muons hit the KOTO~II barrel detector.}\label{fig:bd-tracks}
\end{figure}

Figure~\ref{fig:bd-tracks} shows an event display that depicts 100 simulated muon events generated from the location of the observing hole.
It can be seen that some muons also hit the KOTO~II barrel detector, 
which is placed 5 degrees from the production target.
Assuming a 100 kW proton beam extracted with a 4.2 s spill cycle 
and with loss of 50\% at the target before entering the beam dump, 
the on-spill muon counting rate on the barrel detector was estimated to be 1.8~MHz.
These accidental hits result in a signal loss of 7.0\%, 
assuming a timing window of 40 ns for the barrel detector.

To reduce the muon flux further, part of the concrete shield must be replaced with steel.
Replacing the concrete with steel in a 6.7~m (horizontal) $\times$ 3~m (vertical) $\times$ 3~m (beam direction) volume reduces the flux by an order of magnitude, to 0.39~MHz, reducing the signal loss due to accidental muon hits to 1.6\%.

%

\clearpage

\section{Detector}
\label{chap:detector}
\subsection{Concept of detector}
The signature of the $K_L\to \pi^0\nu\overline{\nu}$ decay is
two photons from the $\pi^0$
decay without any other detectable particles.
In addition,
a large transverse momentum ($\pt$) of the $\pi^0$
is expected
due to the kinematics of the decay.

\begin{figure}[h]
 \centering
 \includegraphics[width=0.8\textwidth]{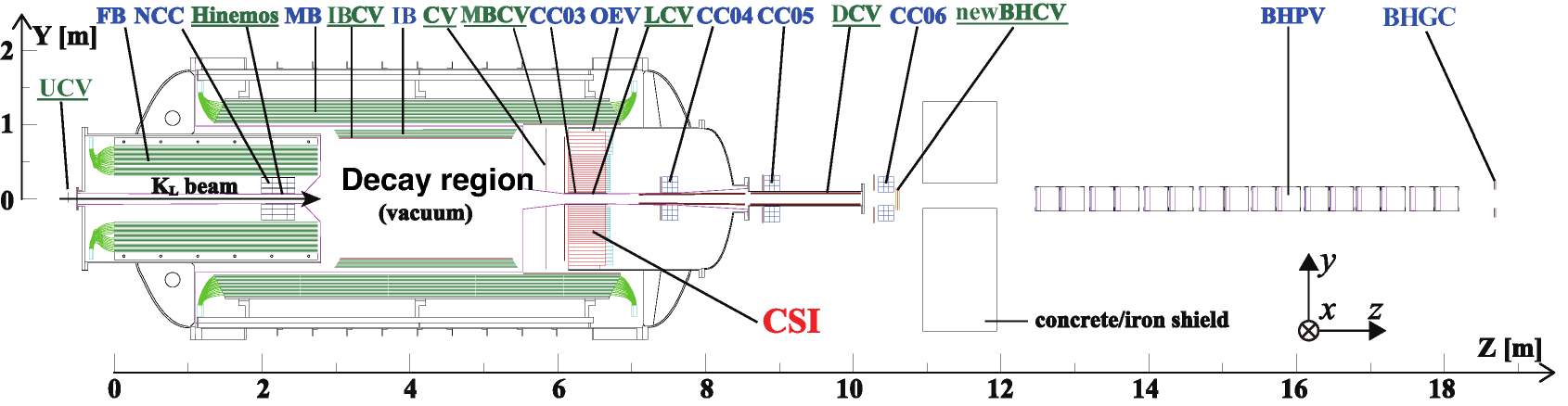}
 \caption{The KOTO detector. The beam enters from  the left.
 Detector components with their abbreviated names written in blue regular (in green regular and underlined) are photon (charged-particle) veto counters.
 \label{fig:KOTOdet2020}}
\end{figure}

The detector concept is the same as in the KOTO experiment.
The KOTO detector is shown in Fig.~\ref{fig:KOTOdet2020}.
The calorimeter ("CSI" in the figure) is used to detect two photons. 
The evacuated decay region upstream of the calorimeter is surrounded
by the hermetic veto-detector system to ensure that no other detectable particles are present.
The decay vertex of the $\pi^0$ is reconstructed on the beam axis by
assuming the nominal $\pi^0$ mass for the invariant mass of the two photons.
With the reconstructed decay vertex, the
$\pt$ of the $\pi^0$ can be calculated.
The diameter of the calorimeter is 2~m and the decay region is 3~m
in the KOTO experiment.
We have been using the signal region within 2~m in the decay region
from 3~m to 5~m in z for the $K_L\to\pi^0\nu\bar{\nu}$ search.

Backgrounds can be classified into three categories:
$K_L$ decays, $K^\pm$ decays, and halo-neutron backgrounds.
\begin{itemize}
 \item $K_L$ decays\\
       The $K_L$ decays with branching fractions
       larger than $10^{-4}$ are listed in Table~\ref{tab:KLdecay}.
       $\kpien$ (Ke3),
       $\kpimun$ (K$\mu$3),
       $K_L\to\pi^+\pi^-$,
       and
       $K_L\to 2\gamma$ have only two observable particles in the final state.
       The Ke3, K$\mu$3, and $K_L\to\pi^+\pi^-$ decays can be reduced by identifying the charged particles.
       The $K_L\to 2\gamma$ decay can be reduced by requiring large $\pt$
       for the reconstructed $\pi^0$,
       although a fake $\pi^0$ can be  reconstructed
       from the two clusters in the calorimeter.
       A $K_L$ which spreads out into the beam-halo region is called a ``halo $K_L$''.
       When such a halo $K_L$ decays into two photons, its
       reconstructed $\pt$ can be larger
       because the $K_L$ itself has $p_T$ and because 
       the vertex is reconstructed on the beam axis assuming the $\pi^0$ mass.
       This halo $K_L\to2\gamma$ background can be reduced with
       incident-angle information on the photons at the calorimeter.
       The other $K_L$ decays have more than two particles
       in the final state,
       and
       extra particles which are not used to reconstruct a $\pi^0$
       can be used to veto the events.
 \item $K^\pm$ decays\\
       $K^\pm$s are generated from the interaction of $K_L$s,
       neutrons, or $\pi^\pm$s
       in the collimator in the beam line.
       The second sweeping magnet
       near the entrance of the detector
       will reduce the contribution.
       Some $K^\pm$ can pass through
       the second magnet
       and decays as $K^\pm\to\pi^0 e^\pm\nu$ in the detector.
       This becomes a background if the $e^\pm$ is undetected.
       The kinematics of the $\pi^0$ is similar to
       $K_L\to\pi^0\nu\overline{\nu}$, and thus
       this decay is one of the more serious backgrounds.
       Detection of $e^\pm$s is one of the keys to reduce the background.
 \item Halo-neutron background\\
       Neutrons in the beam halo (halo neutrons) can interact with the detector material
       and produce $\pi^0$s or $\eta$s, which can decay into two photons
       with large branching fractions (98.8\% for $\pi^0$, 39.4\% for $\eta$).
       Minimizing 
       material near the beam is essential to reduce these backgrounds.
       The fully active detector will reduce the background by
       efficiently detecting
       other particles generated in $\pi^0$ or $\eta$ production.
       This background can be further suppressed if the calorimeter can provide information on the photon angle of incidence.
       Another type of halo-neutron background is
       ``hadron cluster background'':
       A halo neutron hits the calorimeter to produce a first hadronic shower,
       and another neutron in the shower travels inside the calorimeter, producing a second hadronic shower at some distance from the first one.
       These two hadronic clusters can also give a vertex inside the signal region with the $\pi^0$ mass hypothesis applied, mimicking the signal.
       
\end{itemize}
\begin{table}[h]
 \centering
 \caption{Properties of $K_L$ decays into the signal channel and background channels with the branching fraction larger than $10^{-4}$.}
 \label{tab:KLdecay}
 \begin{tabular}{llll}\hline
  Decay mode                 &branching fraction &$\pi^0$ maximum $\pt$
	   & key to reduce background \\
  \hline
  $\pi^0 \nu\overline{\nu}$  &$3\times 10^{-11}$ (in SM) &$230~\mathrm{MeV}/c$ &\\
  \hline
  $\pi^\pm e^\mp\nu$         &40.6\% & &charged particle ID\\
  $\pi^\pm \mu^\mp\nu$       &27.0\% & &charged particle ID\\
  $3\pi^0$                   &19.5\% & $139~\mathrm{MeV}/c$&extra-photon veto\\
  $\pi^+\pi^-\pi^0$          &12.5\% & $133~\mathrm{MeV}/c$&charged-particle veto\\
  $\pi^+\pi^-$               &$1.97\times 10^{-3}$  & &charged particle ID\\
  $2\pi^0$                   &$8.64\times 10^{-4}$  &$209~\mathrm{MeV}/c$&extra-photon veto\\
  $\pi^\pm e^\mp\nu\gamma$ &$3.79\times 10^{-3}$  & & extra-particle veto\\
  $\pi^\pm \mu^\mp\nu\gamma$ &$5.65\times 10^{-4}$  & &extra-particle veto\\
  $2\gamma$                  &$5.47\times 10^{-4}$  & &$\pt$ of reconstructed $\pi^0$ \\
  \hline
 \end{tabular}
\end{table}

\subsection{Conceptual detector for the base design}
\begin{figure}[h]
 \includegraphics[width=\textwidth]{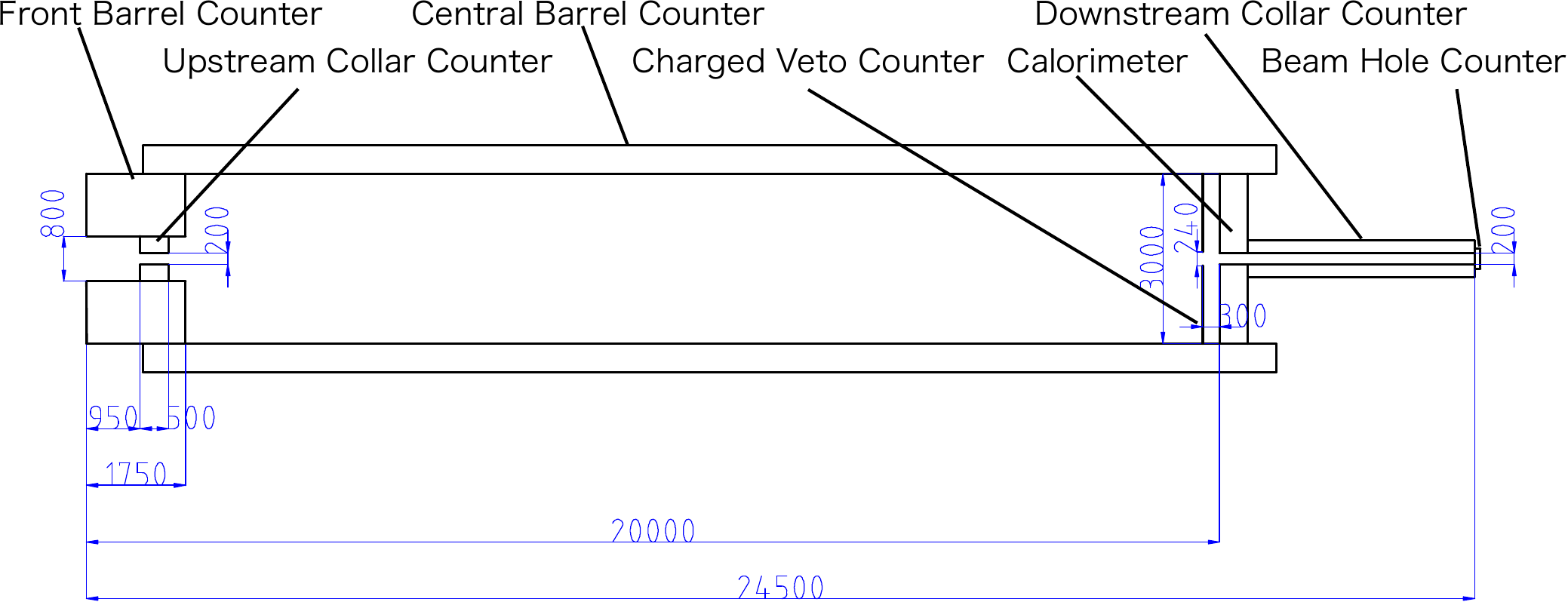}
 \caption{Conceptual KOTO II detector.
 The upstream edge of the Front Barrel Counter is 44~m from the T2 target.}
 \label{fig:conceptualDetector}
\end{figure}

A conceptual detector used in the baseline design of KOTO~II
is shown in Fig.~\ref{fig:conceptualDetector}.
We define the $z$ axis on the beam axis pointing downstream
with the origin at the upstream surface of the Front Barrel Counter,
which is 44~m from the T2 target (43-m long beam line and 1-m long space).
We use the following conceptual detector:
The diameter of the calorimeter is 3~m to gain the signal acceptance;
the beam hole in the calorimeter is 20~cm $\times$ 20~cm to accept
the 15~cm $\times$ 15~cm beam size; 
the $z$ position of the calorimeter is 20~m with
a larger decay volume to enhance the $K_L$ decay.
The larger decay volume is effective in KOTO II,
because the boost from the higher $K_L$ momentum collimates the final state photons for signal events, keeping them within the acceptance.

The Charged Veto Counter
is a charged-particle veto counter positioned 30~cm upstream of the calorimeter
to veto $K_L\to \pi^{\pm} e^{\mp}\nu$, $K_L\to \pi^{\pm} \mu^{\mp}\nu$,
or $K_L\to \pi^+\pi^-\pi^0$.
The beam hole at the Charged Veto Counter is 24~cm $\times$ 24~cm to avoid the production of 
$\pi^0$s or $\eta$s from neutron interactions.
The Barrel Counter is 20-m long
and is mainly used to veto $K_L\to 2\pi^0$ decays
by detecting the additional photons.
the Upstream Collar Counter is 0.5-m long
to veto $K_L\to3\pi^0$ decays  upstream of and inside this counter.
The Downstream Collar Counter is 4-m long and vetoes particles passing
through the beam hole in the calorimeter but at angles causing them to exit the beam region.
The Beam Hole Counter covers the in-beam region starting from $z = 24.5$~m
to veto particles escaping through the calorimeter beam hole.
The Barrel Counter, the Upstream Collar Counter,
the Central Barrel Counter,
and the Downstream Collar Counter,
act as both photon and charged-particle vetoes
for the conceptual design.
For the Beam Hole Counter,
we introduce two separate counters, 
a beam-hole charged-veto counter and 
a beam-hole photon-veto counter.
The fiducial signal region is defined as a 12-m region from 3~m to 15~m in $z$.

We use the conceptual detector in the following sections
to evaluate the signal acceptance, 
the background contributions, and the hit rates.

\subsection{Base design of detector}
\begin{figure}[h]
 \centering
 \subfloat[]{
 \includegraphics[width=0.53\textwidth,bb=50 150 550 430,clip]
 {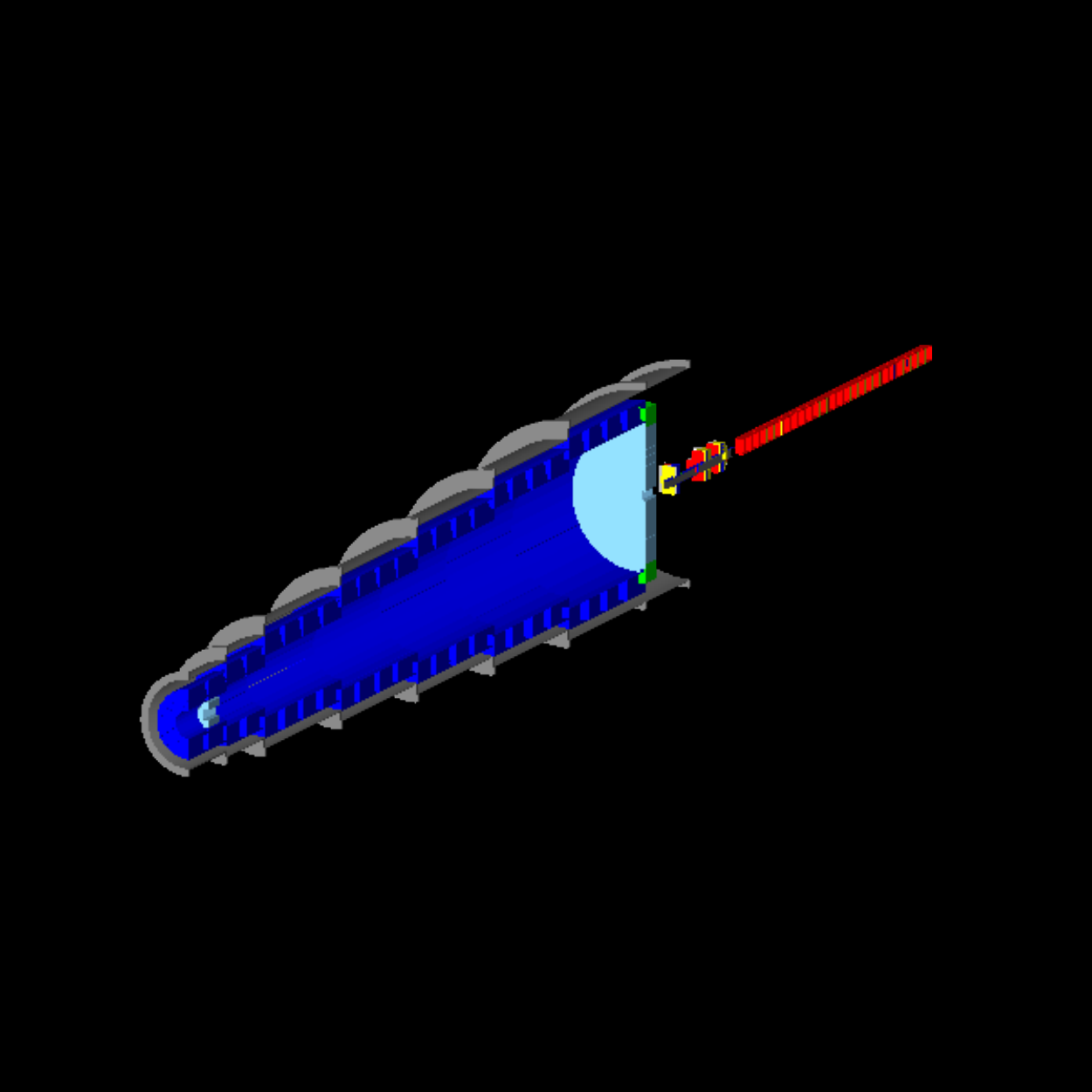}
 }
 \subfloat[]{
 \includegraphics[width=0.45\textwidth]
 {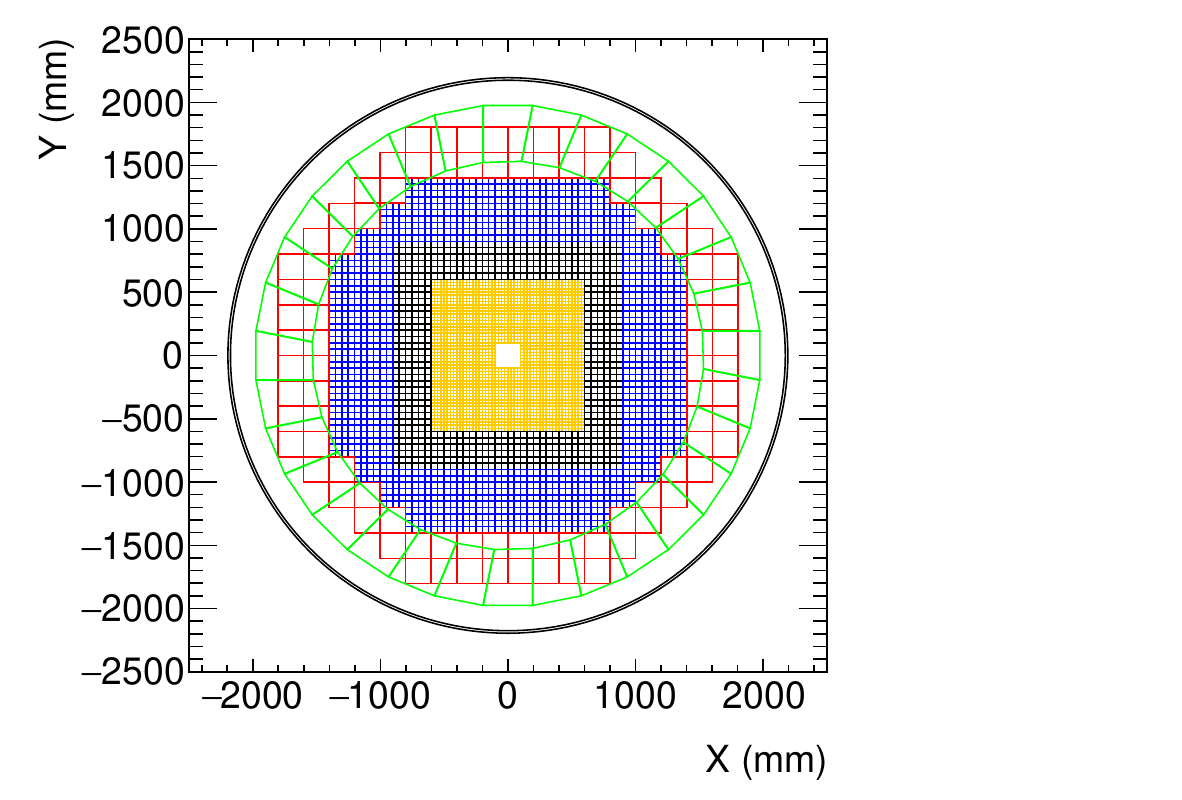}
 }
 \caption{(a) 3D cutaway view of a model for the base detector design. (b) Plane view of the base design of the calorimeter. 
 The calorimeter consists of 
 $\SI{2.5}{cm}\times \SI{2.5}{cm}\times \SI{50}{cm}$ undoped CsI crystals (the central orange region) and $\SI{5}{cm}\times \SI{5}{cm}\times \SI{50}{cm}$ undoped CsI crystals (the outer black and blue regions). 
 The calorimeter is surrounded by veto counters (red) with the size
 $\SI{20}{cm}\times \SI{20}{cm}\times \SI{50}{cm}$.
 The veto counter is a sandwich counter with
 1-mm-thick lead and  5-mm-thick plastic-scintillator plates.
 The modules of the Barrel Counter just upstream of the calorimeter are projected into the plane with green lines.
 The black region can be prepared with current existing crystals. New crystals are needed in the blue region.
 }
 \label{fig:baseDetector}
\end{figure}

A base design of the detector is explained 
for modeling the detector performance and for the cost evaluation.
A 3D cutaway view of the base design is shown in Fig.~\ref{fig:baseDetector}(a).
Most of the counters are in the vacuum tank. The region 
of core beam and its halo should be evacuated to $10^{-5}$~Pa
to reduce the interaction of beam particles to the residual gas,
because a $\pi^0$ or $\eta$ 
could be generated though the interaction and become a background source.

The Barrel Counter is a sandwich counter
with 1-mm thick lead and 5-mm thick plastic-scintillator plates.
The layers are stacked in the $z$ direction. 
The counter has a modular structure in $z$ and $\phi$ directions.
The module $z$-length is \SI{80}{cm}. The number of $\phi$ divisions is 20--32 depending on the $z$ position.
The module hit rate is reduced, and the $z$ coordinate can be obtained with the $z$ segmentation. 

The Upstream Collar Counter consists of 50-cm long undoped CsI crystals ($27X_{0}$). The lateral segmentation and the fast response of undoped CsI help its operation in the high rate.

The Charged Veto Counter located 30-cm upstream of the calorimeter is a plastic scintillator in $\SI{7}{cm}\times \SI{7}{cm}$ tiles for the base design.

The calorimeter shown in Fig.~\ref{fig:baseDetector} (b) 
consists of $\SI{2.5}{cm}\times \SI{2.5}{cm}\times \SI{50}{cm}$ undoped CsI crystals (the central orange region) and $\SI{5}{cm}\times \SI{5}{cm}\times \SI{50}{cm}$ undoped CsI crystals (the outer black and blue region). 
The orange and black regions can be prepared with current existing crystals. New crystals are needed in the blue region.
The calorimeter is surrounded by veto counters (red) with the size
 $\SI{20}{cm}\times \SI{20}{cm}\times \SI{50}{cm}$.
 The outer veto counter consists of lead and plastic-scintillator plates.

The Beam Hole Charged Veto counter consists of
three layers of thin MWPC gas-wire chambers (Thin Gap Chambers)
as in the KOTO experiment.
The wire interval is 1.8~mm, and the wire-to-cathode gap is 1.4~mm.
The cathode is a carbon-coated \SI{50}{\micro m}-thick foil.

The Beam Hole Photon Veto counter consists of
25 modules of lead and aerogel Cherenkov counters
as in KOTO.

\subsection{Modeling of detector response}
The interaction of particles in the calorimeter can be modeled
in terms of the {\bf energy / position resolutions},
and {\bf two-photon fusion probability}
(the probability to identify two incident photons nearby
as a single cluster).
The interaction of particles with respect to the veto performance 
is modeled by {\bf inefficiency} as a function of the particle type,
the incident energy, and the incident angle.
Time smearing for hits in the Central Barrel Counter is applied
for some studies.
Other energy or time smearing
is not applied for the veto counters.

The photon-detection inefficiency of the barrel counter
is estimated with a full-shower simulation 
with the counter in the base design.
The inefficiency of the in-beam detector
(the Beam Hole Counter in the conceptual detector)
is based on the performance of the current detector in KOTO.
The other modelings are the same as in the proposal~\cite{KOTOproposal}.

\subsubsection{Energy / position resolutions of the calorimeter}
The $\pt$ and $\zvtx$ resolutions of 
the reconstructed $\pi^0$ are affected by
the energy and position resolutions of the calorimeter.

\begin{figure}[h]
 \centering
 \includegraphics[bb=0 0 216 215,clip,width=0.45\textwidth]
 {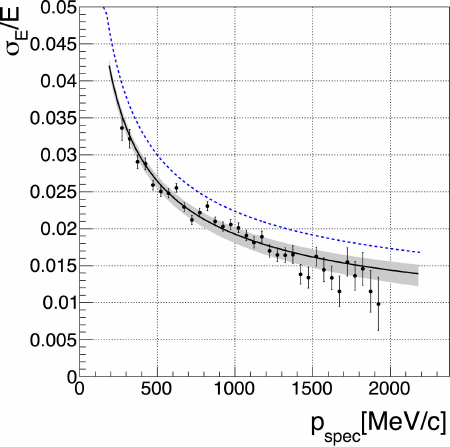}
 \includegraphics[bb=0 0 201 198,clip,width=0.45\textwidth]{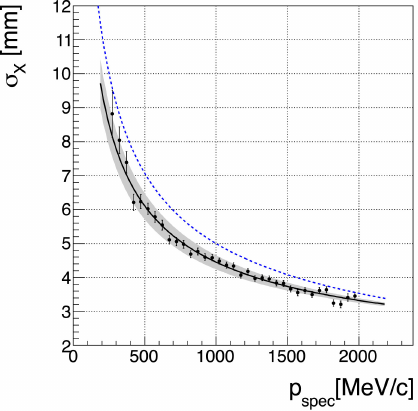}
 \caption{Energy (left) and position (right) resolutions
 for the central region of the calorimeter~\cite{Sato:2020kpq}.
 An electron sample was obtained in a special setup with a spectrometer and the calorimeter in the KOTO beam line. Resolutions are obtained as a function of electron momentum, and the electron momentum is treated as a photon energy.
 The points with error bars show the measured data with electrons,
 the solid line shows
 a fit with a function,
 and the filled area shows the combined statistical and systematic errors.
 The dashed line shows the model used in this study.
 }\label{fig:calreso}
\end{figure}

The energy resolution is modeled as follows:
\begin{align*}
 \frac{\sigma_E}{E}=&\left(1\oplus \frac{2}{\sqrt{E(\mathrm{GeV})}}\right)\%.
\end{align*}
The position resolution is modeled as follows:
\begin{align*}
 \sigma_x=&\frac{5}{\sqrt{E(\mathrm{GeV})}}\left(\mathrm{mm}\right).
\end{align*}
The comparisons between the model and actual measurements in the existing KOTO calorimeter are shown in Fig.~\ref{fig:calreso}~\cite{Sato:2020kpq}.
In the inner region of the KOTO calorimeter ($2.5\times2.5$~cm$^2$ cells), the model is conservative, in that it gives energy and position resolution worse than those actually measured.
\footnote{
The modeled energy resolution is also more conservative
than the actual measurements in the outer region.
The modeled position resolution 
is better by at most 3.2~mm 
than the actual resolution
in the outer region of the current calorimeter
for the incident energy smaller than 2~GeV.
5-cm-square CsI crystals 
are used in the outer region instead of 2.5-cm-square ones
in the inner region.}

\subsubsection{Two-photon fusion probability in the calorimeter}
Fusion of photon clusters in the calorimeter contributes to the $K_L\to 2\pi^0$ background.
Fusion is one of the mechanisms to miss a photon; missing two of the four photons from the decay causes the background.

\begin{figure}[h]
 \centering
 \includegraphics[bb=0 0 535 363,clip,width=0.5\textwidth]{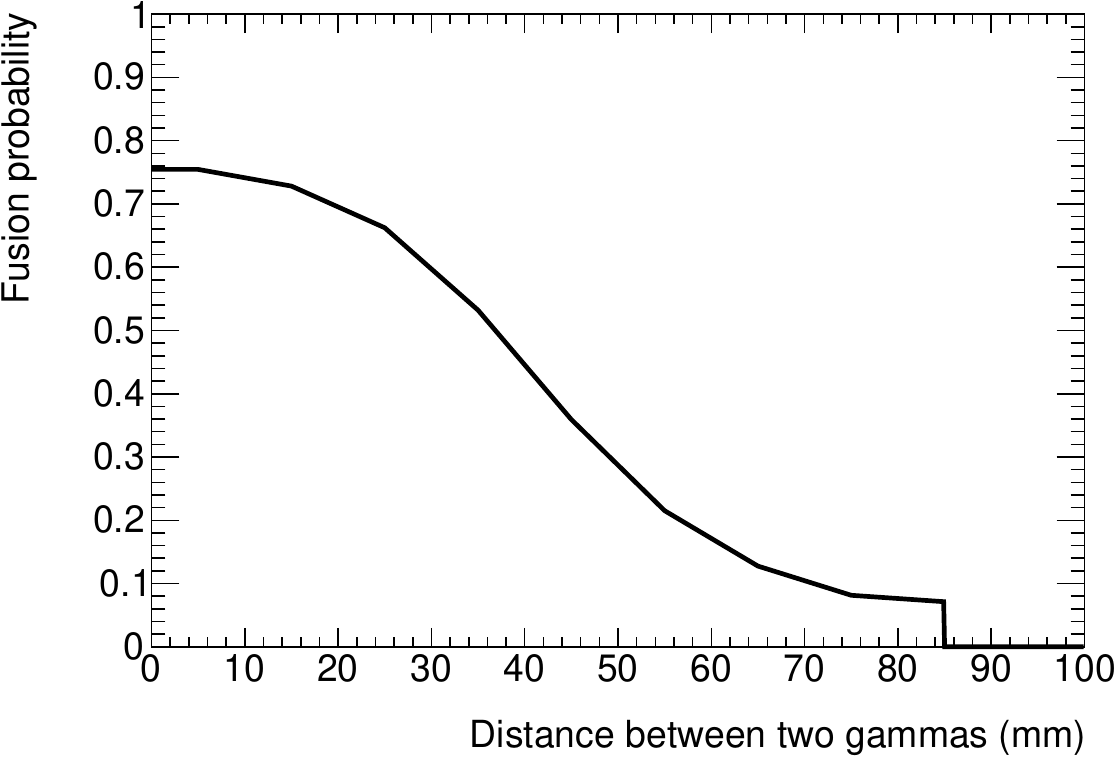}
 \caption{Model for fusion probability as a function of the distance between two photons on the calorimeter.\label{fig:fusionProb}}
\end{figure}

The model of the fusion probability is shown in Fig.~\ref{fig:fusionProb}
as a function of the distance between the positions of two photons incident on the calorimeter.
This model was prepared with a MC study using
the existing KOTO calorimeter.

\subsubsection{Inefficiency of the particle veto}
\label{det:inefficiency}

\begin{figure}[h]
 \centering
 \includegraphics[bb=0 0 540 363,clip,width=0.5\textwidth]{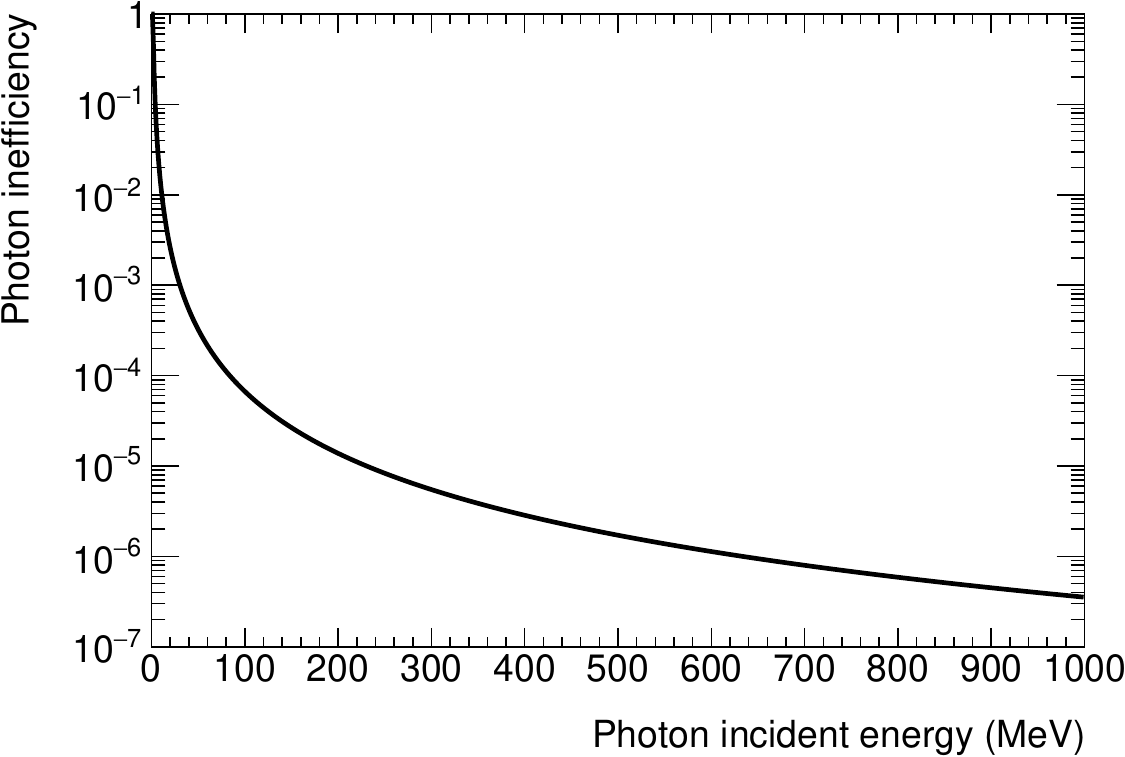}
 \caption{Model for photon inefficiency for the calorimeter.}\label{fig:csiIneffi}
\end{figure}

\subparagraph{Calorimeter photon inefficiency}
The photon inefficiency of the calorimeter contributes
to the $K_L\to2\pi^0$ background.
The modeled inefficiency is shown in Fig.~\ref{fig:csiIneffi},
which is the same as in the KOTO proposal and was obtained with a MC study.

\begin{figure}[h]
 \centering
 \subfloat[]{
 \includegraphics[page=1,width=0.48\textwidth]
 {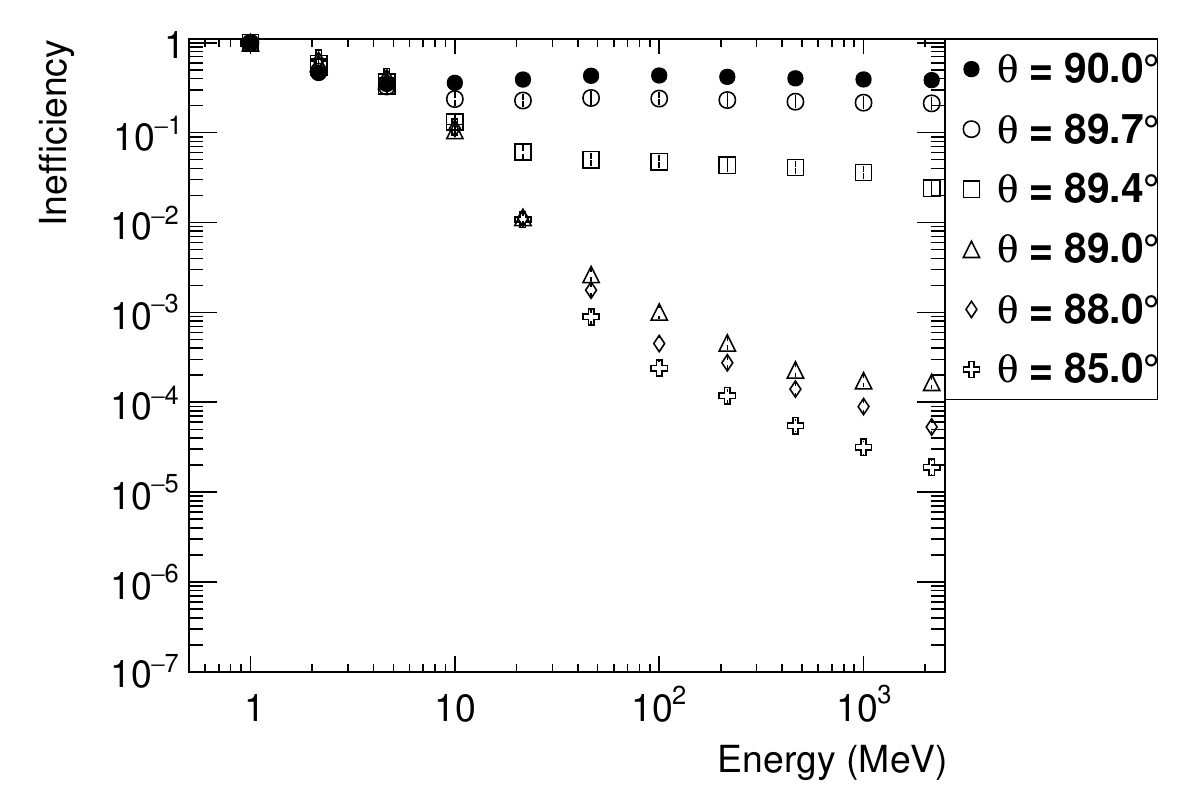}
 }
 \subfloat[]{
 \includegraphics[page=2,width=0.48\textwidth]
 {detector/figure/ineffiSB.pdf}
 }
 \caption{Photon inefficiency of the Barrel Counter
 for photons incident at angles of
 $85-90^{\circ}$ (a) and
 $5-75^{\circ}$ (b).
 The photon incident angle is the angle between the momentum vector and the z-axis.
 When the photon direction is perpendicular to the barrel detector surface,
 the incident angle is $90^\circ$.
 }
 \label{fig:ibIneffi}
\end{figure}

\subparagraph{Barrel photon inefficiency}
The photon inefficiency of the barrel counter
is shown in Fig.~\ref{fig:ibIneffi}.
The inefficiencies were obtained for a sandwich counter
composed of 1-mm-thick lead and 5-mm-thick plastic-scintillator plates.
These plates were stacked in the z direction.
The inefficiencies were prepared with full shower simulations
by applying a threshold for the deposited energy at 1 MeV.
The inefficiencies were given
as a function of the photon energy and the incident angle
measured from the z axis.

\subparagraph{Charged Veto Counter inefficiency for penetrating charged particles}
The inefficiency of the Charged Veto Counter contributes to
the backgrounds from 
the $\kpien$ and $\kpimun$ decays.
The two charged particles from these decays could make two clusters on the calorimeter,
which mimics the signal if these are not detected with the Charged Veto Counter.

In the conceptual design, reduction of $10^{-12}$ for
these backgrounds with the Charged Veto Counter is required.
For example, 
two planes of the Charged Veto Counter with $10^{-3}$ reduction
of a single charged particle per plane
will reduce the background
with the two charged particles by $10^{-12}$.
In KOTO,
we achieved the $10^{-5}$ reduction of a single charged particle with one plane~\cite{Naito:2015vrz}.
Therefore, the $10^{-3}$ reduction with one plane is achievable.


\subparagraph{Beam Hole Counter charged-particle inefficiency}
In KOTO, we are operating a MWPC-type gas-wire chamber (Thin Gap Chamber), with which we have achieved $5\times 10^{-3}$ inefficiency
for charged particles~\cite{Kamiji:2017deh}.
Based on this result,
we assume the same $5\times 10^{-3}$ inefficiency
for the beam-hole charged-veto counter
in the KOTO~II design.

\subparagraph{Beam Hole Counter photon inefficiency}
In KOTO, we are operating 16 modules of
a lead-aerogel Cherenkov counter
~\cite{Maeda:2014pga}
as the in-beam photon veto counter.
This detector is insensitive to beam neutrons,
because protons or charged pions generated from the neutron-interaction
tend to be slow and emit less Cherenkov radiation.
By taking three-consecutive coincident hits in the modules,
electromagnetic shower is efficiently detected,
because it develops in the forward direction and is laterally well collimated.

\begin{figure}[h]
 \centering
 \includegraphics[page=2,width=0.5\textwidth]{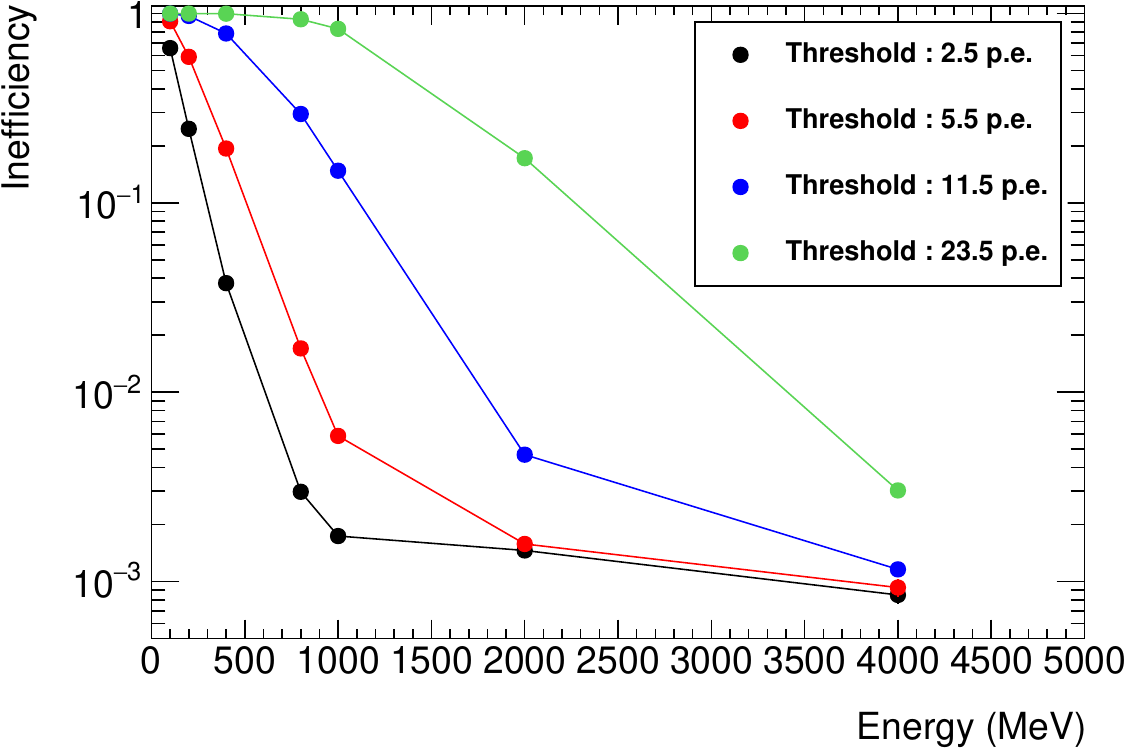}
 \caption{Photon inefficiency of the Beam Hole Counter.
 }\label{fig:bhpvinef}
\end{figure}

We assume 25 modules of such a counter for the 
beam-hole photon-veto counter.
The photon detection performance was studied with
a reliable full-shower simulation developed for KOTO.
Inefficiencies as a function of the incident-photon energy are
shown in Fig.~\ref{fig:bhpvinef} for several detection
thresholds on the number of observed photoelectrons.
For this report,
we use the threshold of 5.5 photoelectrons.

\subsubsection{Timing resolution of the Central Barrel Counter}
\label{sec:barrelTreso}
\begin{figure}[h]
 \centering
 \includegraphics[width=0.5\textwidth]{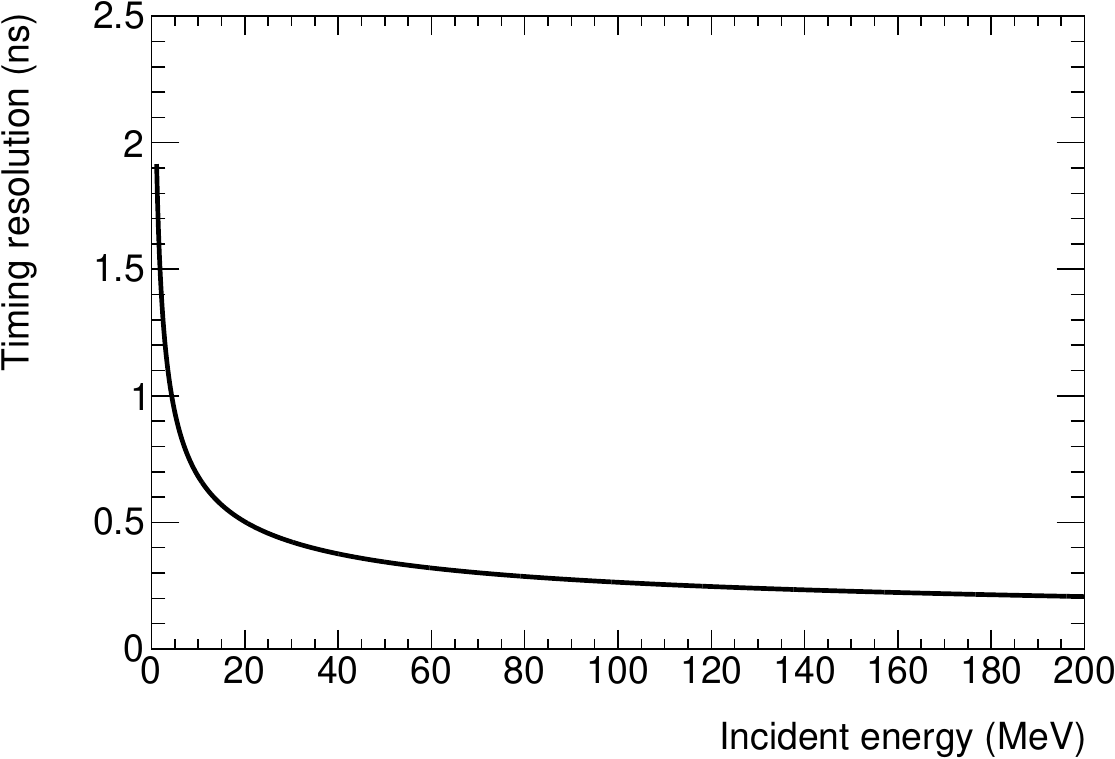}
 \caption{Assumed timing resolution of the Central Barrel Detector.
 }\label{fig:barrelTreso}
\end{figure}

We assume the timing resolution of the Central Barrel Counter
as shown in Fig.~\ref{fig:barrelTreso}
based on the study performed
for the new 
barrel photon counter installed in KOTO
~\cite{Murayama:2020mcp}.
The resolution is modeled as a function of the incident energy.
It is 2~ns for the incident energy of 1~MeV, for example.

\clearpage

\section{Sensitivity and Background Estimation}
\label{chap:sensitivity}

\subsection{Beam conditions}
We assume the beam conditions and running time as in Table~\ref{tab:beam}.
\begin{table}[h]
 \centering
 \caption{Assumed beam conditions and running time.}\label{tab:beam}
 \begin{tabular}{lll}\hline
  Beam power & 100~kW &(at 1-interaction-length T2 target) \\
  && ($1.1\times 10^7 K_L/2\times 10^{13}~\mathrm{POT}$)\\
  Repetition cycle &4.2~s  & \\
  Spill length &2~s  & \\
  Running time &$3\times 10^7$~s  & \\\hline
 \end{tabular}
\end{table}

\subsection{Reconstruction}
We evaluated yields of the signal and backgrounds
with Monte Carlo simulations.
The calorimeter response was simulated either with model responses
as explained in Section~\ref{chap:detector}
or with shower simulations in the calorimeter.
We assumed 50-cm-long CsI crystals for the calorimeter material in the shower simulations.

A shower is generated by a particle incident on the calorimeter.
A cluster is assembled from the energy deposits in multiple, adjacent calorimeter cells.
Assuming the incident particle to be a photon,
the energy and position are reconstructed for the cluster, which is treated as a photon in the analysis
regardless of the original particle species.

\begin{figure}[h]
 \centering
 \includegraphics[width=0.5\textwidth]{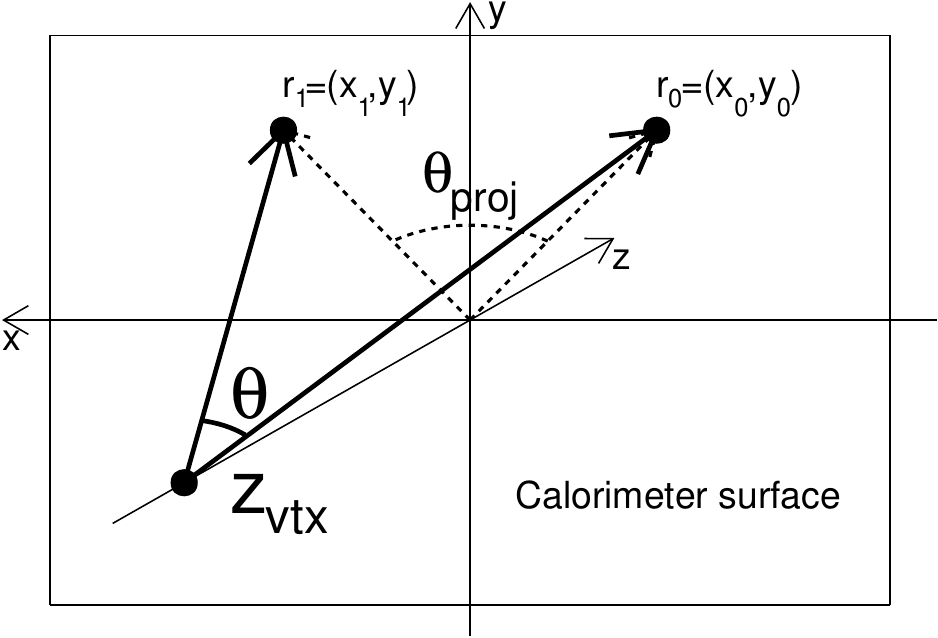}
 \caption{Geometrical relation in the vertex reconstruction.}\label{fig:zvert}
\end{figure}

A $\pi^0$ is reconstructed from the two photons on the calorimeter;
The opening angle
between the momentum directions of the two photons ($\theta$)
can be evaluated with the energies of the two photons ($E_0, E_1$)
from 4-momentum conservation:
\begin{align*}
 {p}_{\pi^0}=&{p}_0+{p}_1,\\
 m_{\pi^0}^2=& 2 E_0 E_1 (1-\cos\theta),
\end{align*}
where ${p}_{\pi^0}$ is the four-momentum of the $\pi^0$,
${p}_0$ and ${p}_1$ are the four-momenta
of the two photons, and $m_{\pi^0}$ is the nominal mass of the $\pi^0$.
The vertex position of the $\pi^0$ is assumed to be on the $z$ axis,
owing to the narrow beam, and
the $z$ vertex position ($\zvtx$) is calculated from the
geometrical relation among $\theta$ and hit positions
${\bf r}_0=(x_0, y_0)$, ${\bf r_1}=(x_1, y_1)$ on the calorimeter
as shown in Fig.~\ref{fig:zvert}.
The value of $\zvtx$ gives the momenta of two photons,
and the sum of the momenta gives the momentum of the $\pi^0$.
Accordingly, the the $\pi^0$ transverse momentum ($\pt$) is obtained.

\subsection{Event selection}\label{sec:cuts}
\begin{figure}[h]
 \centering
 \includegraphics[width=0.5\textwidth]{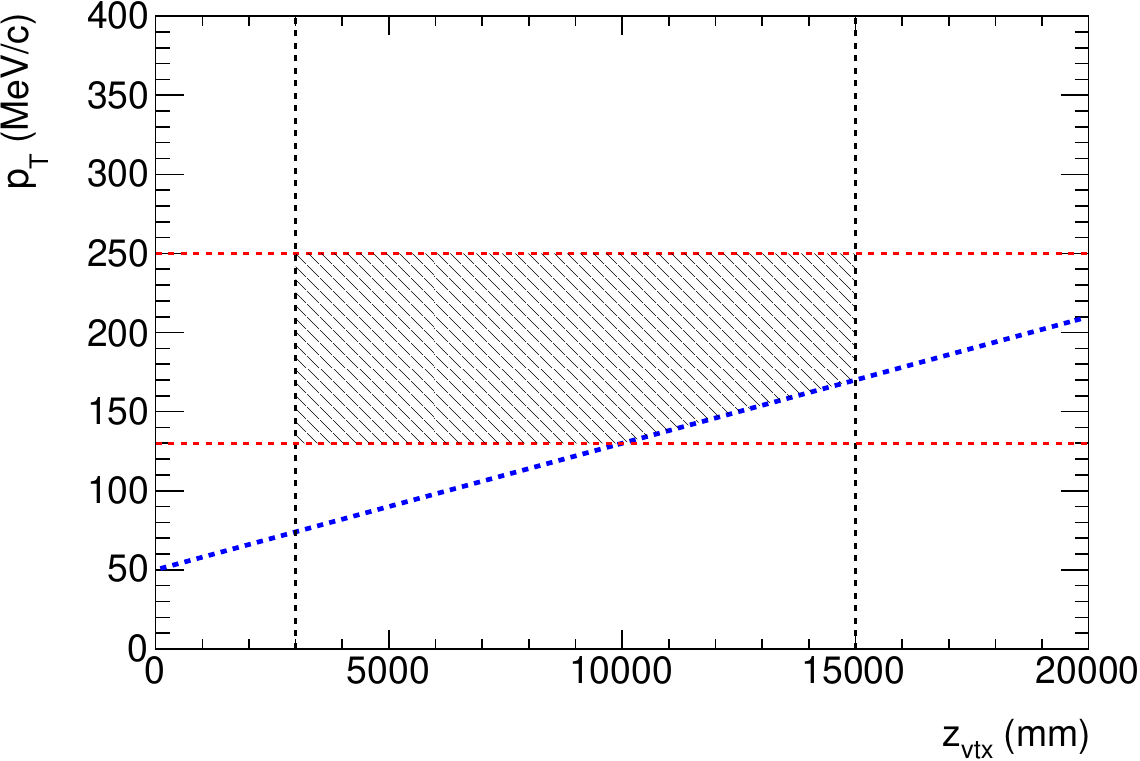}
 \caption{The shaded area shows the $\pt$ criteria in the $\zvtx$-$\pt$ plane.
 The blue dotted line
 shows the tighter $\pt$ criteria in the downstream region.}
 \label{fig:hexiagonal}
\end{figure}

We use the following event selection criteria
for events with two clusters in the calorimeter.
\begin{enumerate}
 \item Sum of two photon energies : $E_0+E_1>500~\mathrm{MeV}$.
 \item Calorimeter fiducial area :
       $\sqrt{x_0^2+y_0^2}<1350~\mathrm{mm}$,
       $\sqrt{x_1^2+y_1^2}<1350~\mathrm{mm}$.
 \item Calorimeter fiducial area :
       $\mathrm{max}(|x_0|, |y_0|)>175~\mathrm{mm}$,
       $\mathrm{max}(|x_1|, |y_1|)>175~\mathrm{mm}$.
 \item Photon energy :
       $E_0>100~\mathrm{MeV}$,
       $E_1>100~\mathrm{MeV}$.
 \item Distance between the two photons : $|{\bf r}_1-{\bf r}_0|>300~\mathrm{mm}$.
 \item Projection angle ($\theta_{\mathrm{proj}}$ as shown in Fig.~\ref{fig:zvert}) : $\theta_{\mathrm{proj}}\equiv \mathrm{acos}\left( \frac{{\bf r}_0\cdot{\bf r}_1}{|{\bf r}_0||{\bf r}_1|}\right)<150^{\circ}$.
 \item $\pi^0$ decay vertex : $3~\mathrm{m}<\zvtx <15~\mathrm{m}$.
 \item $\pi^0$ transverse momentum : $130~\mathrm{MeV}/c<\pt <250~\mathrm{MeV}/c$.
 \item Tighter $\pi^0$ $\pt$ criteria in the downstream region (Fig.~\ref{fig:hexiagonal}):\\ 
       $\frac{\pt}{(\mathrm{MeV}/c)}>\frac{\zvtx}{(\mathrm{mm})}\times 0.008+50$.
 \item Selection to reduce hadron cluster background:\\
       In order to reduce neutron clusters,
       cluster shape, pulse shape, and 
       depth information of the hits in the calorimeter
       are used as in the analysis of KOTO data.
       A signal selection efficiency of $0.9^3=0.73$ is assumed.
       The reduction of the background is discussed in Sec.~\ref{sec:hadronCluster}.
 \item Selection to reduce halo $K_L\to 2\gamma$  background:\\
       The photon angle-of-incidence information is used to reduce the halo $K_L\to 2\gamma$  background
       as in KOTO. A signal selection efficiency of $0.9$ is assumed.
       The reduction of the background is discussed in Sec.~\ref{sec:haloKL}.
\end{enumerate}
The first five criteria ensure the quality of the photon cluster.
The sum of the photon energies is useful to reduce a trigger bias,
because we plan to use the sum of the calorimeter energy for the trigger.
The edge region of the calorimeter is avoided in order to reduce the effects of energy leakage from the calorimeter.
The use of higher energy photons give better energy, time, and position resolution.
The requirement of a large distance between the two photons
reduces the overlap of the two clusters.

The next four criteria are kinematic selections.
The projection angle selection requires
no back-to-back configuration of the two photons, to reduce contamination from  $K_L\to2\gamma$.
Larger values of $\pi^0$ $\pt$ are required
to match the kinematics of the signal.
The tighter $\pt$ selection is required in the downstream region
because the reconstructed $\pt$ tends to be larger for decays near the calorimeter due to the  
worse $\pt$ resolution in this region.

The last two criteria make use of the particle-identification and reconstruction capabilities of the calorimeter
to discriminate between photon and neutron clusters, or between correct and incorrect photon angles of incidence.

\subsection{Signal yield}
The $\klpionn$ yield can be factorized into
the decay probability within the $z$ region
from 3~m to 15~m,
the geometrical acceptance to have 2 photons in the calorimeter,
and the cut acceptance described in the following subsections.
The signal losses
called  ``accidental loss'' and ``shower-leakage loss''
will be introduced and discussed in the later subsections.

\subsubsection{Decay probability and geometrical acceptance of two photons at the calorimeter}

\begin{figure}[h]
 \centering
 \subfloat[]{
 \includegraphics[page=1,width=0.45\textwidth]{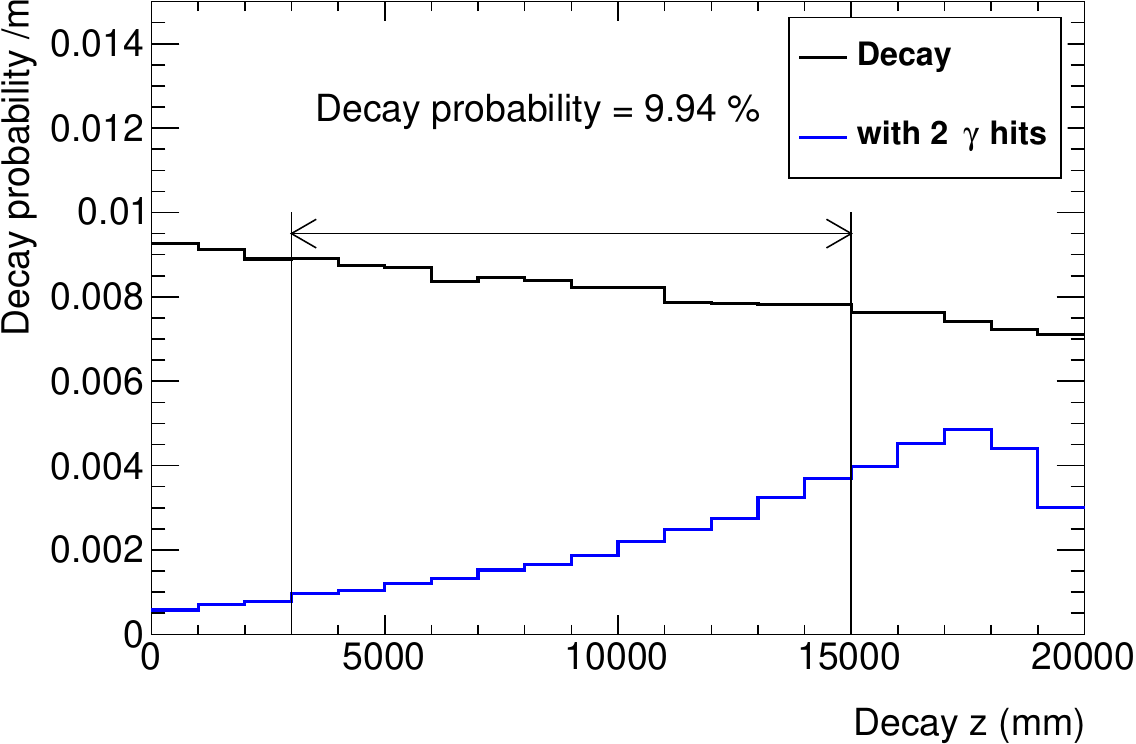}
 }
 \subfloat[]{
 \includegraphics[page=2,width=0.45\textwidth]{sensitivity/figure/decayGeom.pdf}
 }
 \caption{Decay probability (a) and geometrical acceptance (b).}
 \label{fig:decayGeom}
\end{figure}

The decay probability ($P^{\mathrm{truth}}_{\mathrm{decay}}$) is defined:
\begin{align*}
 P^{\mathrm{truth}}_{\mathrm{decay}}=&
 \frac{\text{Number of $K_L$s that decay in $3~\mathrm{m}<z<15~\mathrm{m}$}}
 {\text{Total number of $K_L$s at $z=-1$~m}}.
\end{align*}
It is evaluated by MC simulation as in Fig.~\ref{fig:decayGeom}(a)
to be $9.9$\%.

The geometrical acceptance ($A^{\mathrm{truth}}_{\mathrm{geom}}$) is defined:
\begin{align*}
 A^{\mathrm{truth}}_{\mathrm{geom}}=&
 \frac
 {\text{Number of $K_L$s with 2$\gamma$'s in the calorimeter that decayed in $3<z<15$~m}}
 {\text{Number of $K_L$s that decayed in $3~\mathrm{m}<z<15~\mathrm{m}$}}.
\end{align*}
It is also evaluated by MC simulation 
to be 24\%
as shown in Fig.~\ref{fig:decayGeom}(b).

The quantities
$P^{\mathrm{truth}}_{\mathrm{decay}}$ and 
$A^{\mathrm{truth}}_{\mathrm{geom}}$
relate to the true decay $z$ position. 
In the following sections, the reconstructed $\zvtx$ is used
to give a realistic evaluation.
$P^{\mathrm{truth}}_{\mathrm{decay}}\times A^{\mathrm{truth}}_{\mathrm{geom}}$
can be compared with $A_{2\gamma}$, which is defined in terms of the 
reconstructed $\zvtx$:
\begin{align*}
 A_{2\gamma}=&
 \frac{\text{Number of events with 2$\gamma$ hits with $3~\mathrm{m}<\zvtx <15~\mathrm{m}$}}
 {\text{Total number of $K_L$ at $z=-1$~m}}.
\end{align*}
The MC simulation gives
$A_{2\gamma} = 2.4\%$,
which is consistent with
$P^{\mathrm{truth}}_{\mathrm{decay}}\times A^{\mathrm{truth}}_{\mathrm{geom}}=2.4\%$.

\subsubsection{Cut acceptance}
\begin{figure}[h]
 \centering
 \includegraphics[page=21,width=0.5\textwidth]{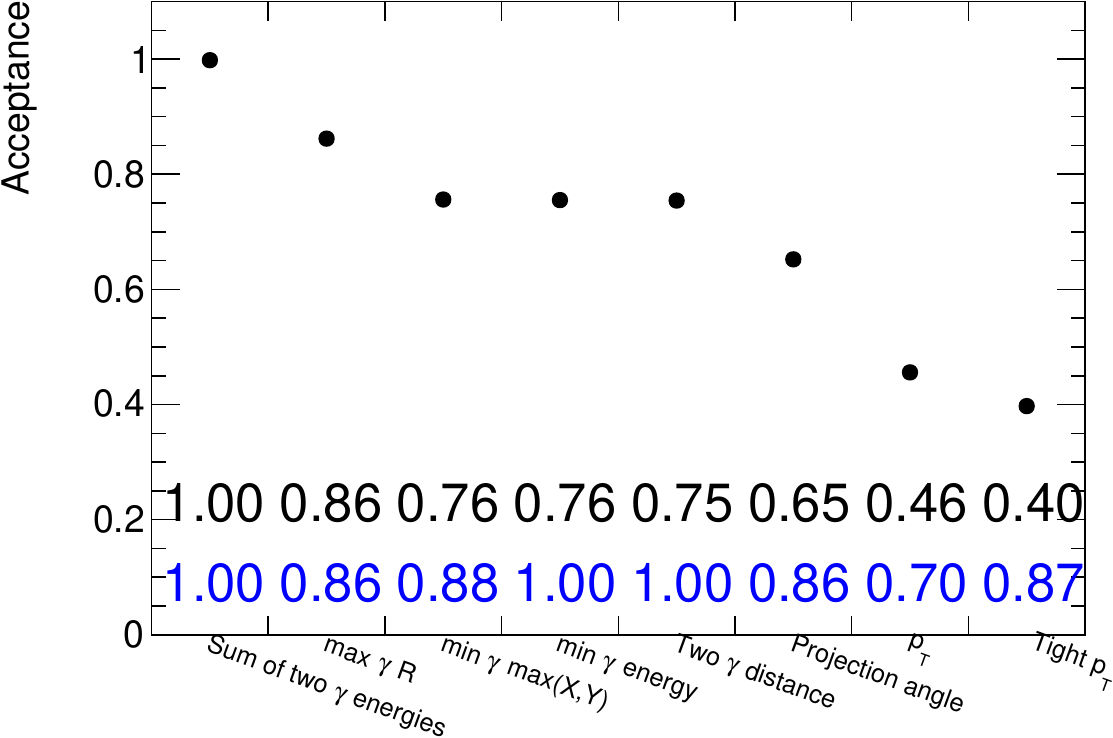}
 \caption{Cut acceptance. The black numbers in the figure show cumulative acceptances,
 the blue numbers show individual acceptances.
 }
 \label{fig:cutAccSummary}
\end{figure}

\begin{figure}[h]
 \centering
 \subfloat[]{
 \includegraphics[page=12,width=0.32\textwidth]{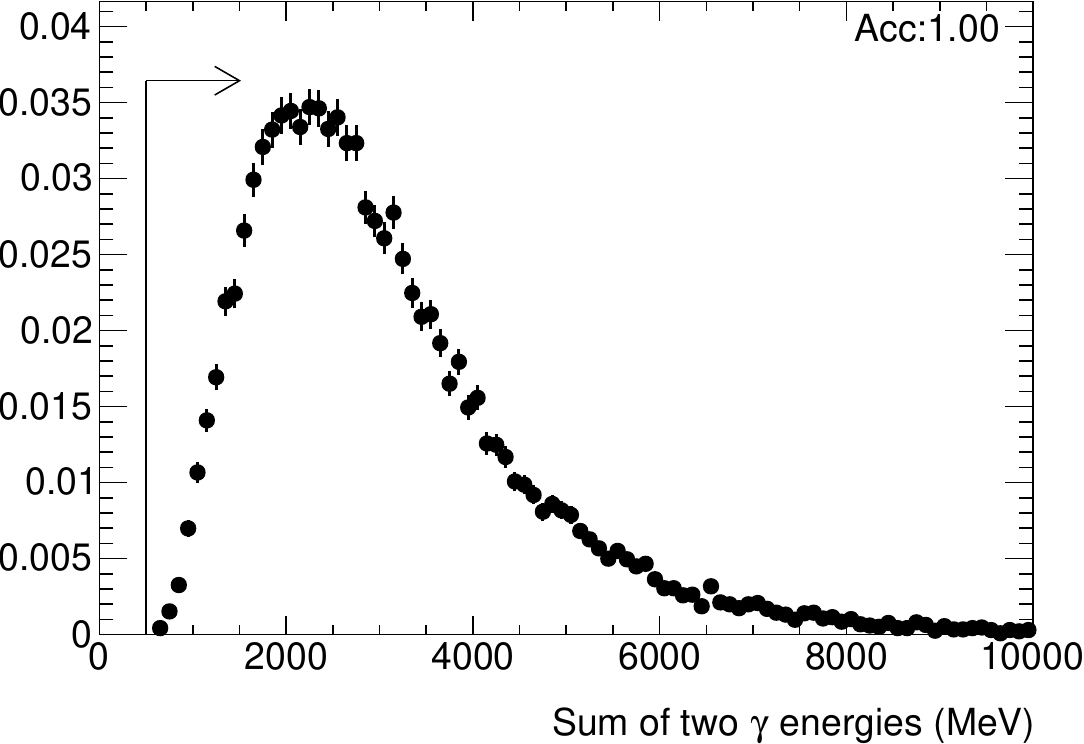}
 }
 \subfloat[]{
 \includegraphics[page=13,width=0.32\textwidth]{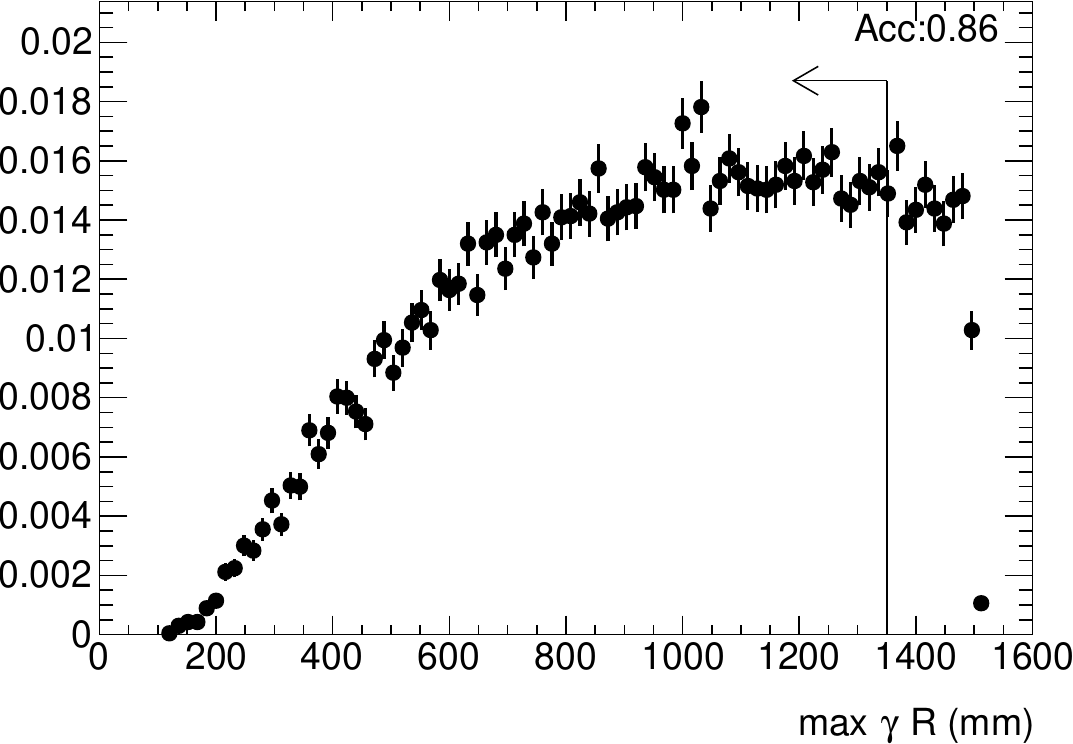}
 }
 \subfloat[]{
 \includegraphics[page=14,width=0.32\textwidth]{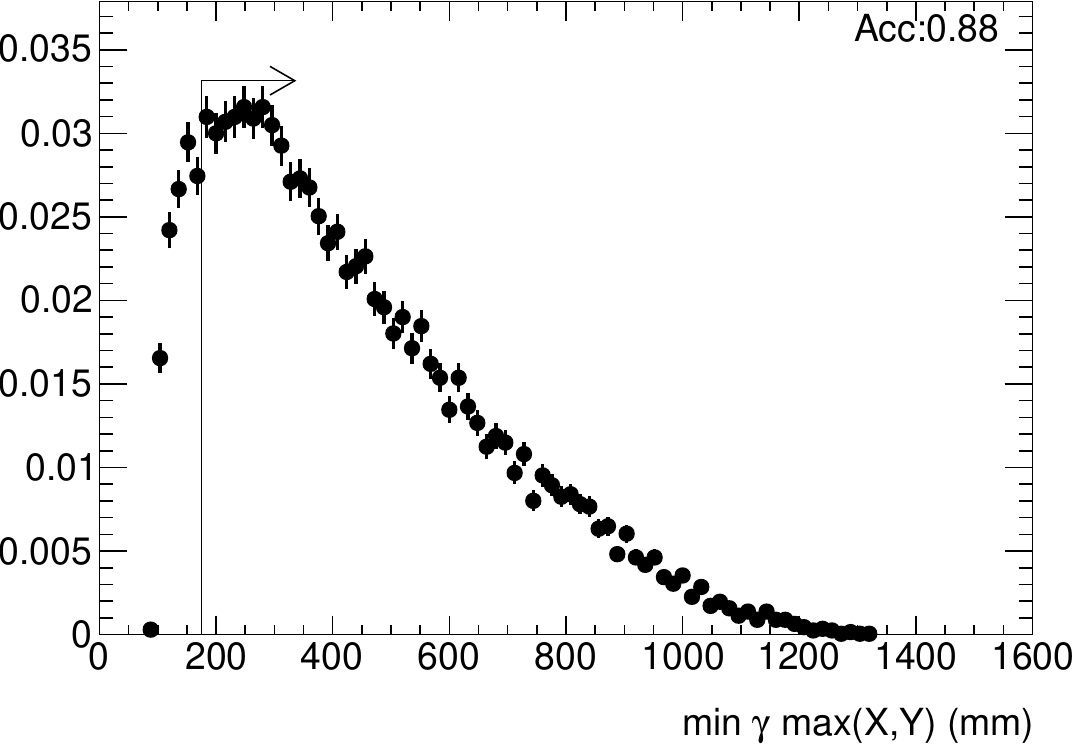}
 }
 
 \subfloat[]{
 \includegraphics[page=16,width=0.32\textwidth]{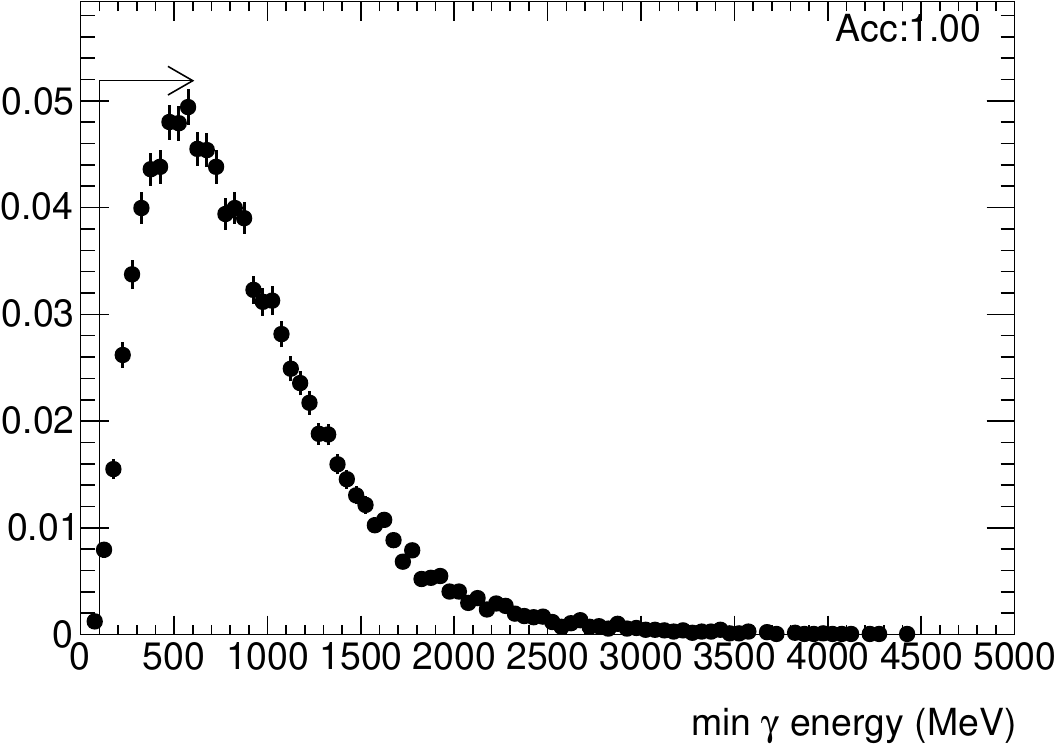}
 }
 \subfloat[]{
 \includegraphics[page=17,width=0.32\textwidth]{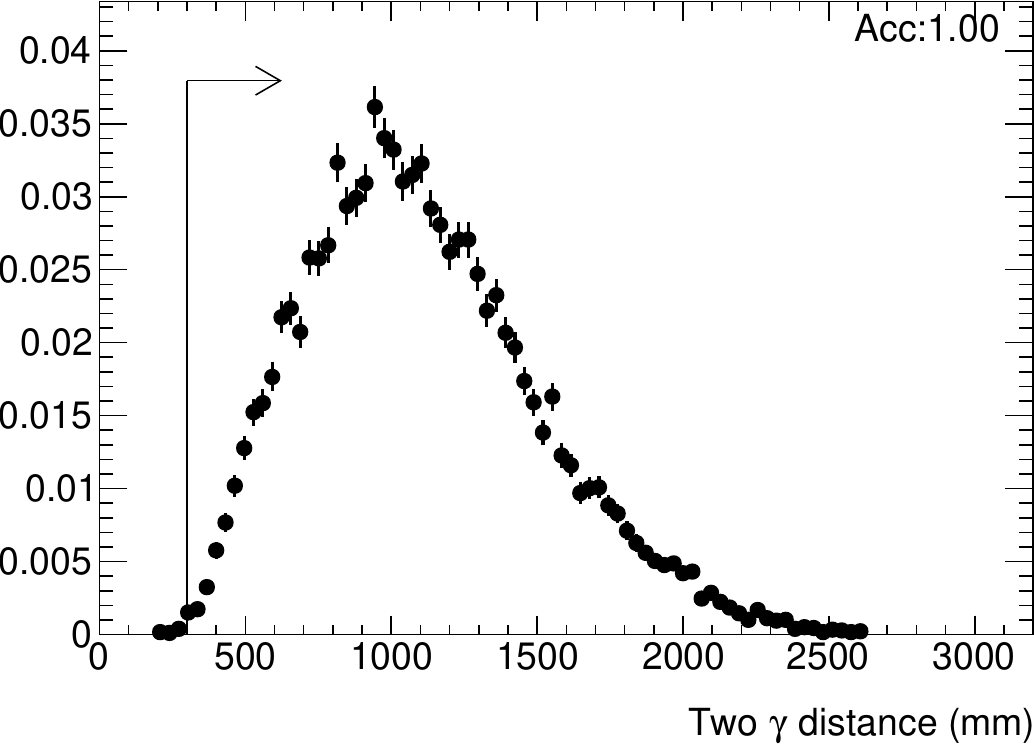}
 }
 \subfloat[]{
 \includegraphics[page=18,width=0.32\textwidth]{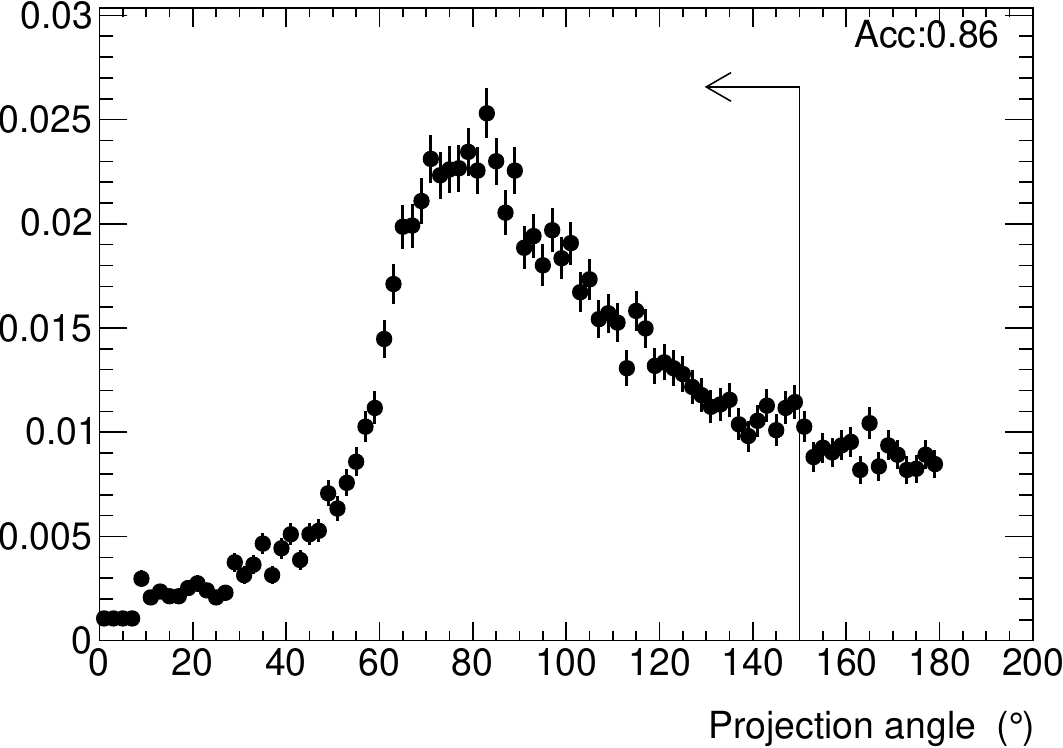}
 }
 
 \subfloat[]{
 \includegraphics[page=19,width=0.32\textwidth]{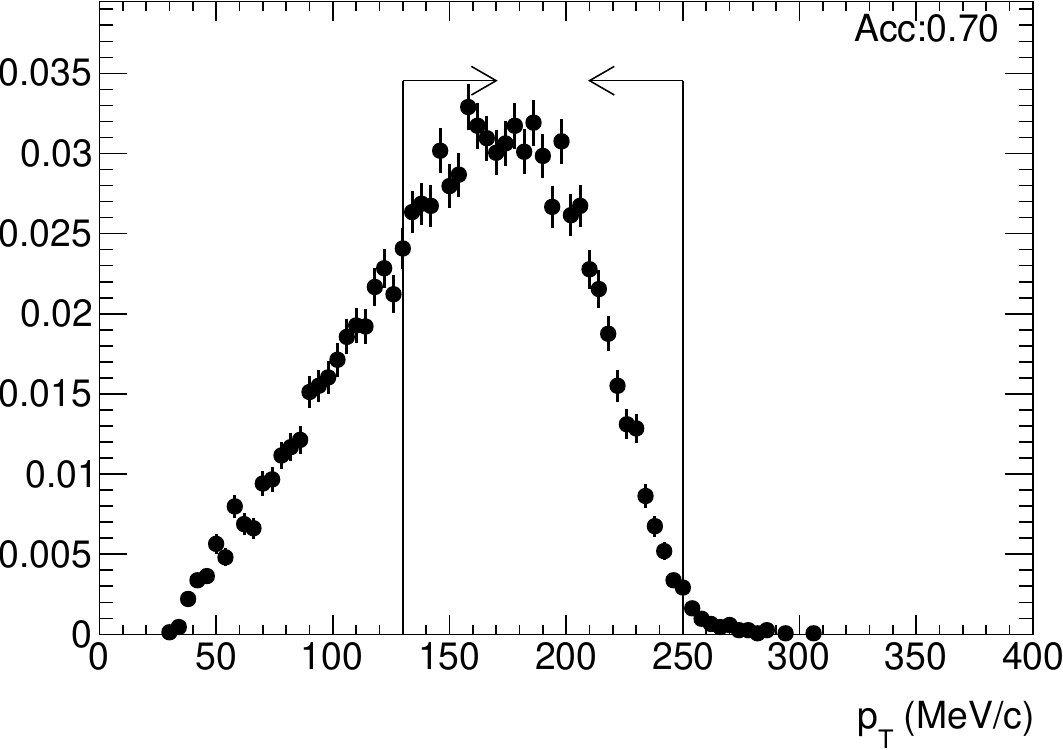}
 }
 \subfloat[]{
 \includegraphics[page=20,width=0.32\textwidth]{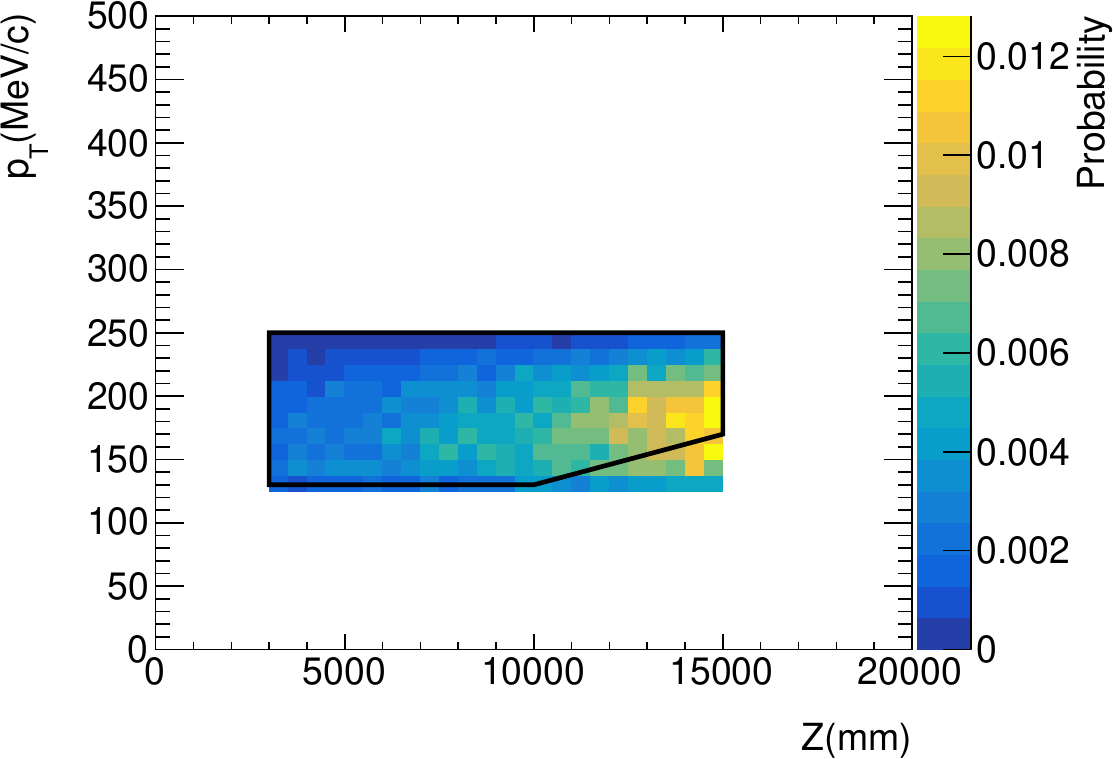}
 }
 \caption{Distributions of variables used in the event selections:
 (a) sum of two photon energies, (b) radial hit position,
(c) inner hit positions (d) minimum photon energy,
 (e) distance between two photons,
 (f) projection angle, (g) $\pt$, and (h) tighter $\pt$ selection in the downstream
 }
 \label{fig:cutAcc}
\end{figure}

The acceptances for cuts 1 through 6, 8, and 9\footnote{Cut 7, the $\zvtx$ selection, is already treated in the previous section.}
listed in Sec.~\ref{sec:cuts}
are summarized in Fig.~\ref{fig:cutAccSummary}.
The overall acceptance
after applying all of these cuts is 40\%.
The distributions of the cut variables are shown in Fig.~\ref{fig:cutAcc}.
The assumed acceptance for the additional cuts
to reduce the hadron-cluster and halo $K_L\to 2\gamma$ backgrounds is $0.9^4=66\%$.
Including all of the above, the overall cut acceptance is 26\%.

\subsubsection{Accidental loss}
\label{sec:accidentalLoss}
In order to veto background events,
we set a timing window (veto window)
to detect extra particles 
with respect to the two-photon hit timing at the calorimeter.
The width of the veto window is set to 40 ns for the Central Barrel Counter,
30 ns for the beam-hole charged-veto counter,
6 ns for the beam-hole photon-veto counter,
and 20 ns for the other counters
throughout this report.

When the $\klpionn$ signal is detected with the calorimeter,
another $K_L$ might decay accidentally
and its daughter particle may hit a counter at the same time.
Similarly, a photon or neutron in the beam might hit the Beam Hole Counter
at the same time.
These accidental hits will veto the signal if the hit timing is within the veto window.
We call this type of signal loss ``accidental loss''.

In the following, we first estimate the accidental losses from the detectors
other than the beam-hole counter; we then evaluate  the accidental loss from the Beam-Hole Counter.

\subparagraph{Detectors other than Beam-Hole Counter}


The hit rate for each detector and the veto width are summarized in Table~\ref{tab:rateWidth}. 
The detector rates were obtained with a $K_L$-decay simulation. 
In the simulation,
$K_L$s were shot from $z=-10~\mathrm{m}$, 
and a 4-m-long modeled collimator with a $\SI{20}{cm}\times \SI{20}{cm}$ square beam hole was located at $z$ from $-5$~m to $-1$~m.
The rate of $K_L$ decays with detector hits was found to be 5.1~MHz,
to which the main contribution is from $K_L$-decays in the region $-1~{\rm m} < z < 23~{\rm m}$.
We evaluated the accidental loss from hits on detectors other than the beam-hole counters to be 17.9\% in total.
This is a conservative number, because two or more different counters can have hits in coincidence
from the same $K_L$-decay, but
these hits are counted separately in the rates for different counters. 

%

\begin{figure}[h]
 \centering
 \includegraphics[page=40,width=0.5\textwidth]{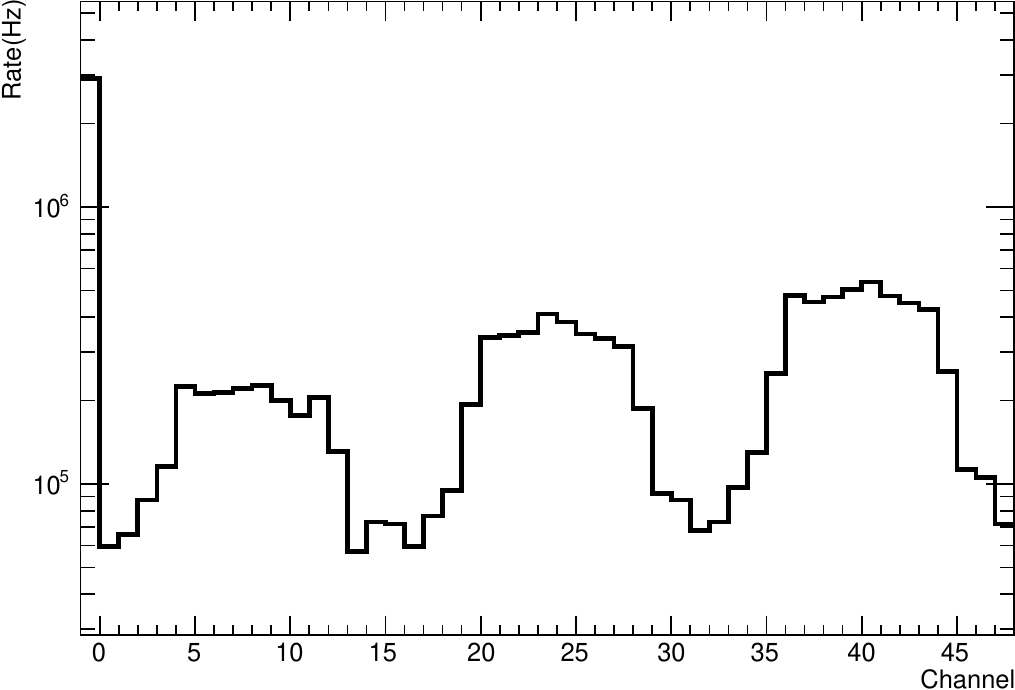}
 \caption{Expected hit rate for the KOTO-II beam-hole charged-particle veto counter
 with a 1/4 minimum-ionizing-particle peak threshold.
 It consists of three layers of a MWPC-type wire chamber with 16-channel readout for each layer.
 Histogram channels 0--47 correspond to the individual readout channels.
 Histogram channel $-1$ corresponds to the detector rate as determined by two layer-hits in coincidence out of three layers.}\label{fig:newBHCV}
\end{figure}

\subparagraph{Beam-hole charged-veto counter}
\label{sec:rateBHCV}
We evaluated the hit rate of the beam-hole charged-veto counter based on the current detector design in KOTO.
It consists of three layers of a MWPC-type wire chamber~\cite{Kamiji:2017deh}
with a small amount of material:
the thickness of the gas volume for each layer is 2.8 mm, and
the cathode plane is a 50-$\mu$m-thick graphite-coated polyimide film.
This design reduces the hit rate from neutral particles,
such as photons, neutrons, and $K_L$s.
A layer-hit is
defined as an energy deposit larger than
1/4 of the minimum-ionizing-particle peak.
A counter-hit is
defined as
two layer-hits in coincidence out of three layers, providing a charged-particle efficiency of better than 99.5\%
with a minimal contribution from neutral particles.
The width of the veto window is 30 ns
to cover the drift time of the ionized electrons in the chamber.
The particles simulated with the beam line simulation
were injected into the beam-hole charged-particle veto counter.
Figure~\ref{fig:newBHCV} shows the hit rate for each readout channel.
The counter-hit rate
with the two-out-of-three logic is 2.9~MHz
as shown in channel $-1$ in the figure.
The accidental loss with this counter is 8.3\% with a 30-ns veto window.

\subparagraph{Beam-hole photon-veto counter}
We evaluated the hit rate of the beam-hole photon-veto counter
based on the current detector design in KOTO,
which consists of 16 modules of
lead-aerogel Cherenkov counters~\cite{Maeda:2014pga}.
A high-energy photon generates an $e^+e^-$ pair in the lead plate,
and these generate Cherenkov light in the aerogel radiator.
The Cherenkov light is guided by mirrors to a PMT.
In this report, we use 25 modules for KOTO II. Details on the lead or aerogel thickness for each module are described in Sec.~\ref{sec:beamHolePhotonVetoCounter}.

\begin{figure}[h]
 \centering
 \includegraphics[page=28,width=0.5\textwidth]
 {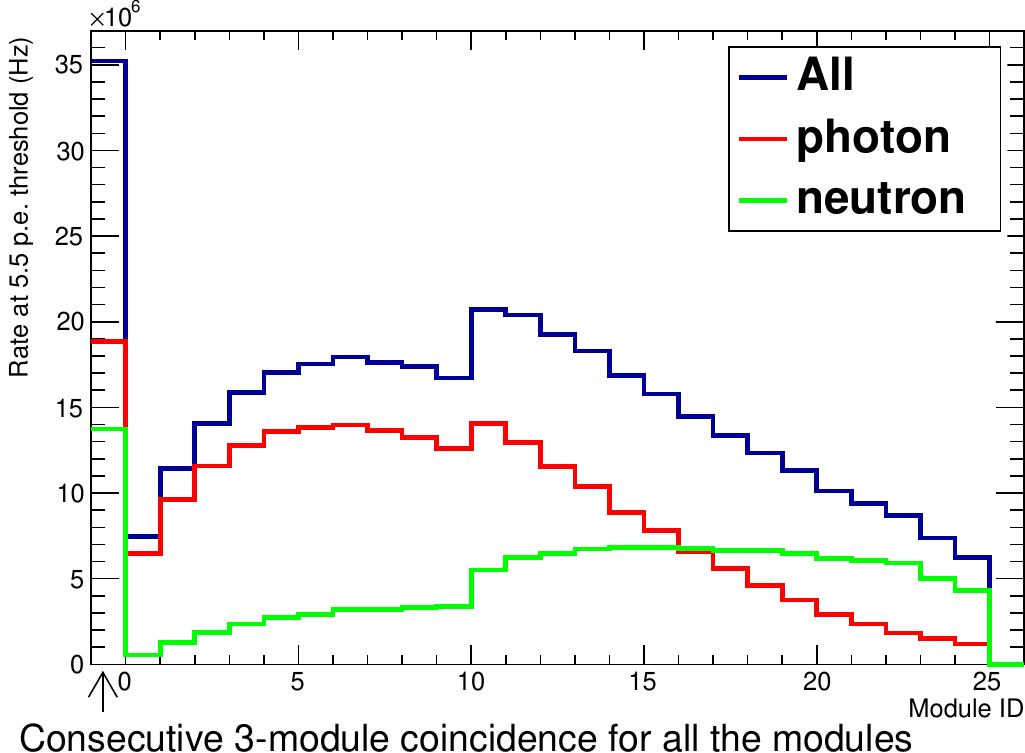}
 \caption{Expected hit rate for the beam-hole photon-veto counter for KOTO II.
 Histogram channels 0--24 show
 the module-hit rates with a 5.5-p.e.-threshold for each module. 
 Histogram channel $-1$ shows
 the counter-hit rate, defined as the rate of module-hits in coincidence on three consecutive modules. 
 }
 \label{fig:bhpvrate}
\end{figure}

The particles collected in the beam line simulation were injected into the beam-hole photon-veto counter with 25 modules.
A full-shower simulation with optical-photon tracking to the PMTs was performed,
and the observed number of photoelectrons was recorded.
The individual module-hit is defined with 
a 5.5-photoelectron (p.e.) threshold.
The counter-hit is defined by a coincidence of module-hits on three consecutive modules.
The module-hit rates and the counter-hit rate
are shown in Fig.~\ref{fig:bhpvrate}.
Of the counter rate of $35.2~\mathrm{MHz}$,
60\% comes from beam photons, and 30\% comes from the beam neutrons.
The accidental loss with this counter is 19\% with a 6-ns veto window.

\subparagraph{Conclusion on the accidental loss}
The rate ($r_i$), the veto width $(w_i)$, and
the individual loss $(1-\exp\left(- w_i r_i\right))$
for each detector are summarized
in Table~\ref{tab:rateWidth}.

The total accidental loss is evaluated:
\begin{align*}
 \text{Accidental loss}=&1-\exp\left(-\sum_i w_i r_i\right)\\
 =&39\%.
\end{align*}
\begin{table}[h]
 \centering
 \caption{Summary of rate, veto width, and individual accidental loss.}\label{tab:rateWidth}
 \begin{tabular}{llll}
  Detector&Rate(MHz) & Veto width (ns) & Individual loss (\%)\\ \hline
  Front Barrel Counter&0.18 &20 & 0.4\\
  Upstream Collar Counter&0.80 &20 & 1.6\\
  Central Barrel Counter&2.21 &40 & 8.5\\
  Calorimeter&3.45 &20 & 6.7\\
  Downstream Collar Counter&0.97 &20 & 1.9\\\hline
  Beam-hole charged-veto&2.9 &30 & 8.3\\
  Beam-hole photon-veto&35.2 &6 & 19\\
  \hline
 \end{tabular}
\end{table}

\subsubsection{Shower-leakage loss}\label{sec:backsplash}
When two photons from the $\klpionn$ decay are detected
in the calorimeter,
fragments from the shower can leak both downstream and upstream of the calorimeter,
and make hits on the other counters such as the Central Barrel Counter.
Such hits will veto the signal if the hit timing is within the veto window.
We call this signal loss ``shower-leakage loss''.
In particular,
we call the loss caused by
shower leakage in the upstream direction
``backsplash loss''.

\subparagraph{Downstream shower-leakage loss}
The downstream shower leakage causes shower-leakage loss in 
the Downstream Collar Counter and the Central Barrel Counter.

The Downstream Collar Counter is hit by shower leakage passing through
the 50-cm long (27 radiation-length) CsI crystal of the calorimeter.
This signal loss of 8\% is neglected, because
this effect can be mitigated with
hit-position information in the Downstream Collar Counter,
or an absorber upstream of the Downstream Collar Counter.
This is one of the design requirements for the Downstream Collar Counter.

For the Central Barrel Counter, shower leakage at the outer edge of the calorimeter makes
hits in the barrel counter,
because the barrel counter covers the side of the calorimeter.
This causes 3.4\% loss of the signal; however, we allow it because
the barrel coverage is also effective to reduce the background from $K_L\to2\pi^0$.

\subparagraph{Backsplash loss at the Charged Veto Counter}
The Charged Veto Counter covers
the upstream face of the calorimeter and would
suffer from backsplash.
Because the Charged Veto Counter is located 30-cm upstream of the calorimeter,
the timing of the Charged Veto Counter
defined with respect to the calorimeter timing differs by at least 2~ns between $K_L$-decay particles
and backsplash particles, with
the $K_L$ decay particles giving hits at least 1~ns earlier
than the calorimeter timing,
and the backsplash particles giving hits at least 1~ns later. 
We require good timing resolutions for
the Charged Veto Counter
and the calorimeter to discriminate between these types of hits
in order to avoid contributions to the backsplash loss.
This is one of the requirements on the detector design.
For example,
a timing resolution of 300~ps for both the calorimeter and the Charged Veto Counter
would work.
Reducing the material in the Charged Veto Counter will also
reduce the backsplash loss, because the backsplash particles are 95\% photons.

\begin{figure}[h]
 \centering
 \includegraphics[width=\textwidth]{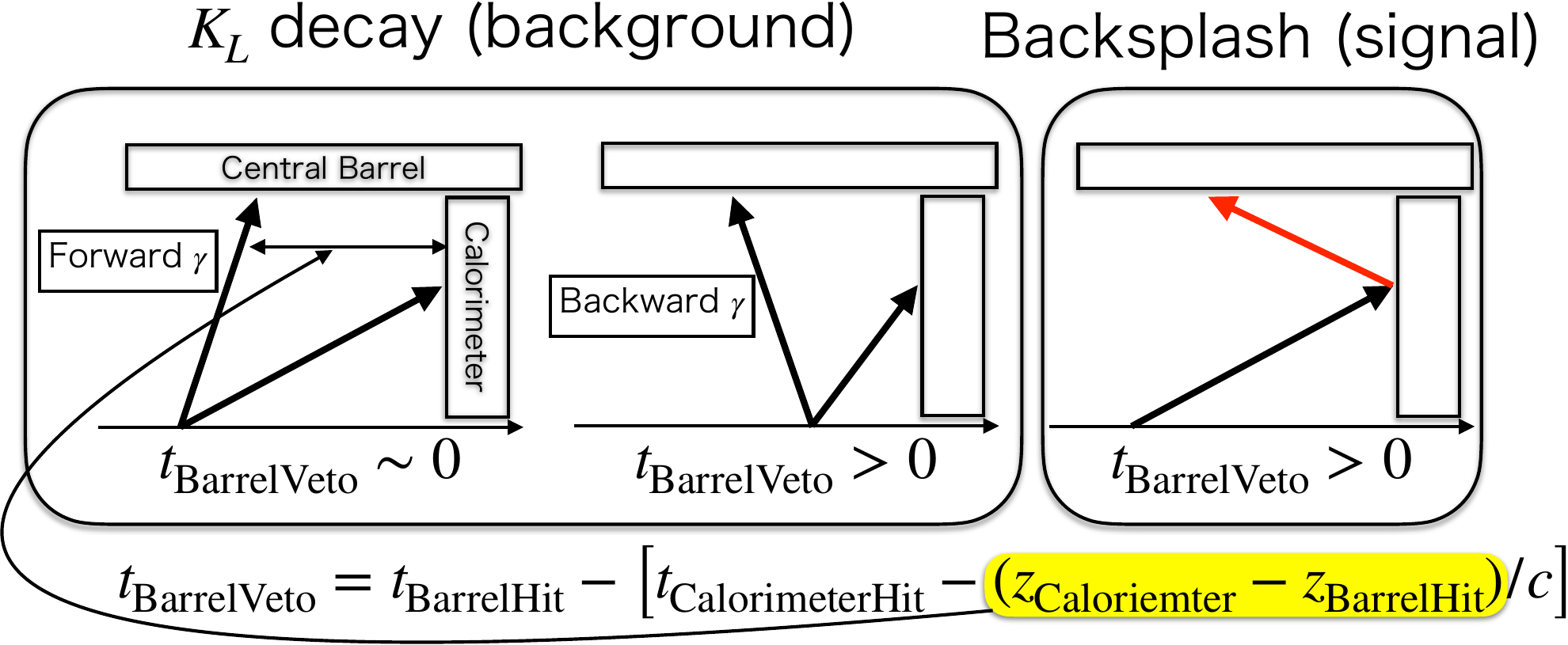}
 \caption{Configurations of barrel hits.}\label{fig:barrelTiming}
\end{figure}

\subparagraph{Backsplash loss and  barrel-timing definition}
The large coverage of the Central Barrel Counter gives rise to a significant contribution to the backsplash loss. First, we introduce a timing definition for the Central Barrel Counter ($t_{\mathrm{BarrelVeto}}$):
\begin{align*}
 t_{\mathrm{BarrelVeto}}=&
 t_{\mathrm{BarrelHit}}-
 \left[t_{\mathrm{CalorimeterHit}}
 -\frac{z_{\mathrm{Calorimeter}}-z_{\mathrm{BarrelHit}}}{c}
 \right].
\end{align*}
The concept of this definition is illustrated in Fig.~\ref{fig:barrelTiming}.
The calorimeter hit timing
($t_{\mathrm{CalorimeterHit}}$)
is corrected with
the expected time of flight
from the $z$ position of the barrel hit ($z_{\mathrm{BarrelHit}}$) and
the calorimeter $z$ position ($z_{\mathrm{Calorimeter}}$).
$t_{\mathrm{BarrelVeto}}$ is the relative timing of
the actual barrel hit ($t_{\mathrm{BarrelHit}}$) with respect to the corrected calorimeter timing.
This reduces the barrel timing fluctuations for $K_L$ decays,
and the veto window can be shortened.
The backward-going particle in the $K_L$ decay
shown
in the middle panel of Fig.~\ref{fig:barrelTiming} makes $t_{\mathrm{BarrelVeto}}$ larger, leading to a required veto window
from $-5$~ns to $35$~ns.
Backsplash particles also give large values of
$t_{\mathrm{BarrelVeto}}$, and a 40-ns veto-timing
requirement reduces the
backsplash loss for $t_{\mathrm{BarrelVeto}}>35~\mathrm{ns}$.

\begin{figure}[h]
 \centering
 \subfloat[]{
 \includegraphics[page=2,width=0.33\textwidth]{sensitivity/figure/anaBSL.pdf}
 }
 \subfloat[]{
 \includegraphics[page=1,width=0.33\textwidth]{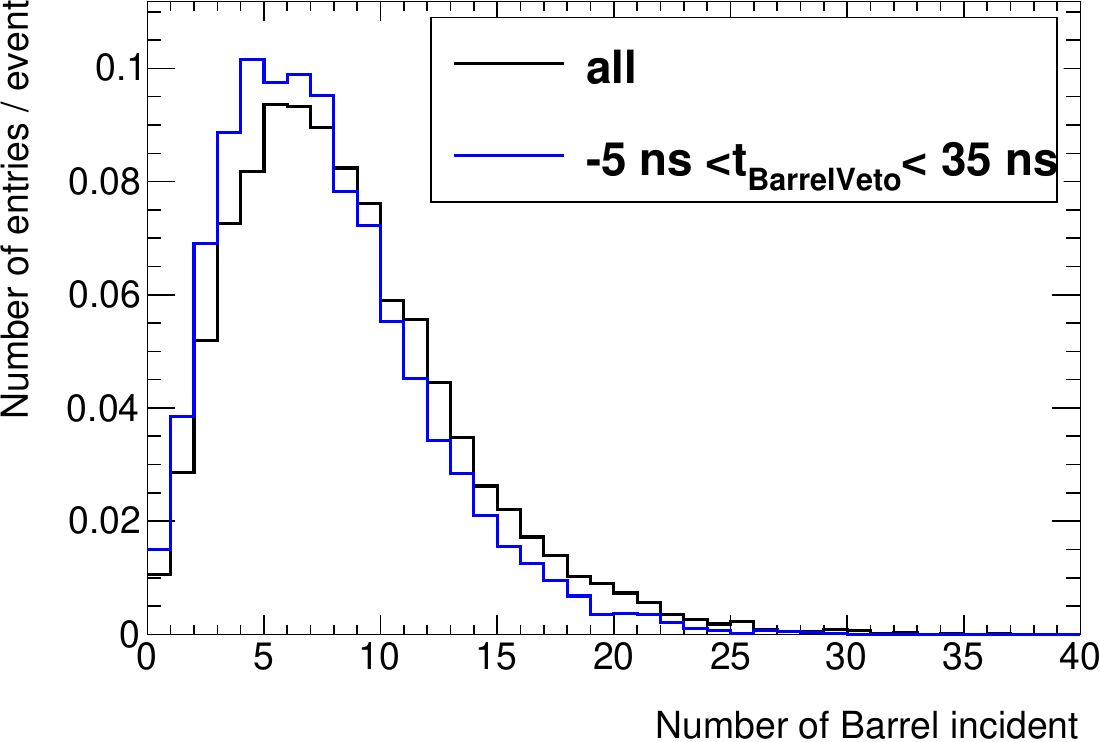}
 }
 \subfloat[]{
 \includegraphics[page=4,width=0.33\textwidth]{sensitivity/figure/anaBSL.pdf}
 }
 \caption{
 Distributions of 
 the arrival time on the barrel of particles from shower leakage (a),
 the multiplicity of shower-leakage particles incident on the barrel (b),
 and the maximum energy of the shower-leakage photons incident on the barrel
 within the veto window (c).
 The left-right arrow in (a) illustrates the veto window.
 }\label{fig:BShit}
\end{figure}

\subparagraph{Characteristics of backsplash}
The arrival time distribution of the shower-leakage particles is
shown in Fig.~\ref{fig:BShit}(a).
The incident-particle multiplicity is shown in Fig.~\ref{fig:BShit}(b).
The mean multiplicity is 8.0 without any timing requirement.
This large multiplicity
makes the reduction of the backsplash loss difficult,
because the detection of any one of these multiple particles
will kill the signal.
The timing requirement of the veto window reduces the mean multiplicity
to 7.1.

Most of the incident particles are low energy photons,
and the fraction of electrons or positrons is 4.8\%.
The distribution of the maximum photon energy 
in a event
within the veto window
is shown in Fig.~\ref{fig:BShit}(c).
The probability of containing one or more photons with energy larger than 1~MeV (3~MeV) is still 84\% (35\%).

Events survive the veto only when
none of the backsplash particles are detected.
The probability not to detect a particle is the inefficiency
($=1-\text{efficiency}$); therefore,
the event survival probability is obtained
by multiplying the inefficiencies for detection of each of the incident particles.
The inefficiency for 3-MeV photons is roughly 60\%, for example.
These contribute to the backsplash loss.
At this stage, the survival probability of a event
after vetoing the shower-leak particles is 57\%,
which can be improved as discussed in the following paragraph. 

\begin{figure}[h]
 \centering
 \includegraphics[page=2,width=0.5\textwidth]{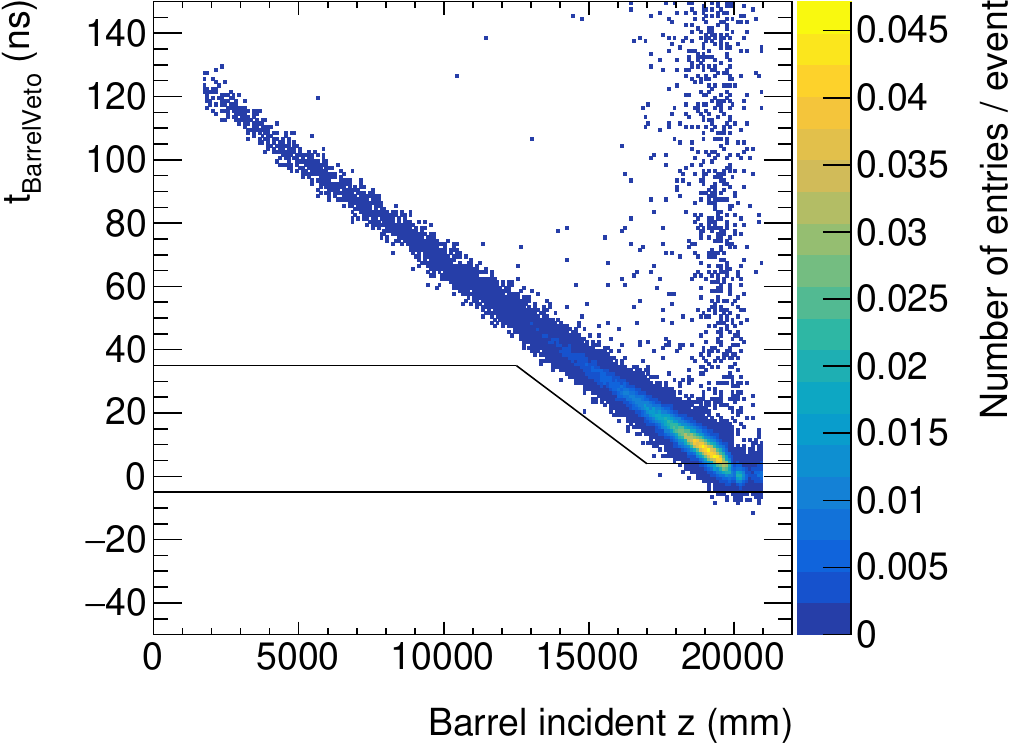}
 \caption{Arrival time of backsplash particles
 on the Central Barrel Counter
 as a function of the position of incidence in $z$. Time smearing is applied with time resolution
 depending on the incident energy.
 }\label{fig:BStiming}
\end{figure}

\subparagraph{Veto window depending on hit position}
Figure~\ref{fig:BStiming} shows the arrival time of the
backsplash particles on the Central Barrel Counter
as a function of the position of incidence in $z$.
The time smearing is applied with time resolution
depending on the incident energy
as discussed in Sec.\ref{sec:barrelTreso}.
The events with $t_{\mathrm{BarrelVeto}}$ at $z\sim 20000~\mathrm{mm}$
are generated by neutrons from the electromagnetic shower.
Events with smaller $z$ tend to have larger $t_{\mathrm{BarrelVeto}}$
due to the longer flight distance.
This clear correlation can be used to exclude the backsplash particles
from the veto to reduce the backsplash loss.
We loosen the veto criteria\footnote{
The loose veto criteria could reduce also the accidental loss
described in Sec.~\ref{sec:accidentalLoss}.
In this report, we do not change the accidental loss
obtained in Sec.~\ref{sec:accidentalLoss},
which can be improved with this loose veto.}
in the downstream region:
over the interval in $z$ from 12.5~m to 17~m, the 
upper limit of the veto time window decreases linearly from 35~ns to 4~ns, and
for $z>17~\mathrm{m}$, it is 4~ns.
The region within the two lines in the figure shows the 
new veto region, which gives a survival probability
of $91\%$, or equivalently, a backsplash loss of 9\%.
When an additional smearing in $z$ using a gaussian with $\sigma=500~{\rm mm}$ is applied,
the same survival probability is obtained.

\subsubsection{Signal yield including the signal losses}
\begin{figure}[h]
 \centering
 \includegraphics[page=8,width=0.5\textwidth]{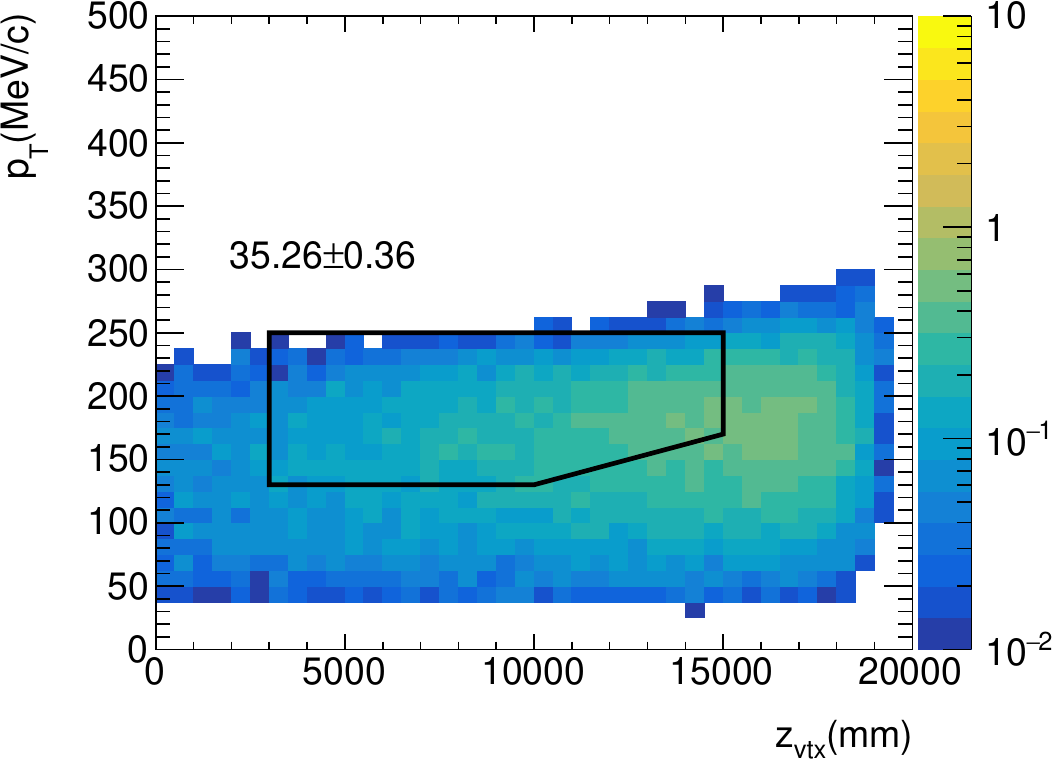}
 \caption{Distribution of $\klpionn$ events in the $\zvtx$-$\pt$ plane for
 for a running time of $3\times 10^7$~s.
 All cuts except for the $\pt$ and $\zvtx$ cuts are applied.
 }
 \label{fig:ptz_pi0nn}
\end{figure}

We evaluate the number of signal events ($S$) expected to be collected in $3\times 10^7$~s of  running time, assuming $\mathrm{BR}(\klpionn)=3\times 10^{-11}$:
\begin{align*}
 S=&
 \frac{(\text{beam power})\times (\text{running time})}
 {(\text{beam energy})}
 \times (\text{number of }K_L\text{/POT})\\
 &\times P_{\mathrm{decay}}
 \times A_{\mathrm{geom}}
 \times A_{\mathrm{cut}}
 \times (\text{1-accidental loss})
 \times (\text{1-backsplash loss}) \times \mathcal{B}_{\klpionn}\\
 =&
 \frac{
 (100~\mathrm{kW})\times (3\times 10^7~\mathrm{s})
 }
 {
 (30~\mathrm{GeV})
 }
 \times
 \frac{(1.1\times 10^7 K_L)}
 {(2\times 10^{13}~\mathrm{POT})}\\
 &\times 9.9\% \times 24\% \times 26\% \times (1-39\%) \times 91\%
 \times (3\times 10^{-11})\\
 =&35.
\end{align*}
Here, $P_{\mathrm{decay}}$ is the decay probability,
$A_{\mathrm{geom}}$ is the geometrical acceptance for the two photons
to enter the calorimeter,
$A_{\mathrm{cut}}$ is the cut acceptance, and
$\mathcal{B}_{\klpionn}$ is the branching fraction for $\klpionn$.
The distribution of the events in the $\zvtx$-$\pt$ plane is shown in Fig.~\ref{fig:ptz_pi0nn}.
The comparison of signal acceptance with KOTO~II and KOTO is shown in Table~\ref{tab:Factor_Comp_to_KOTO}. 

\begin{table}
\begin{center}
\caption{Comparison of the signal acceptance with KOTO~II and KOTO}
\label{tab:Factor_Comp_to_KOTO}
\begin{tabular}{lccc} \hline
     & KOTO~II & KOTO & Improvement factor \\ \hline 
  Number of $K_{L}$/POT   &  $\times$ 2.6     &        &  2.6 \\
 $P_{\mathrm{decay}}$ & 9.9\% &  3.3\% & 3.0 \\
 $A_{\mathrm{geom}}$  & 24\%  & 26\%   & 0.9 \\ 
 $A_{\mathrm{cut}}$   & 26\%  & 11\%   & 2.4 \\
 1-accidental loss    & 61\%  & 29\%   & 2.1 \\
 1-backsplash loss    & 91\%  & 56\%   & 1.6  \\ \hline
 Total                &     &      &   58   \\ \hline
\end{tabular}  
\end{center}
\end{table}

\subsection{Background estimation}

All estimated numbers of background events in this section refer to $3\times10^7$ seconds of running time.

\subsubsection{$\klpiopio$}
\begin{figure}[h]
 \centering
 \includegraphics[page=2,width=0.5\textwidth]{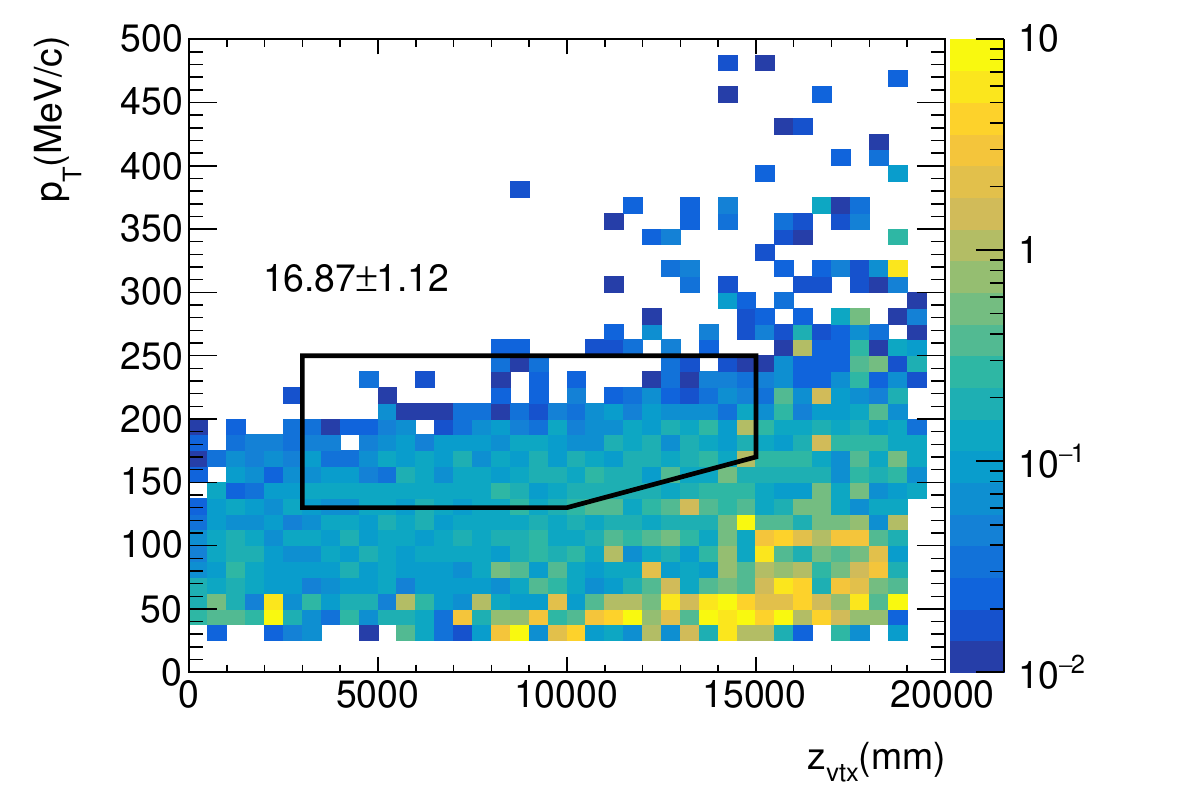}
 \caption{Distribution of  $\klpiopio$
 background events
 in the $\zvtx$-$\pt$ plane 
 for a running time of $3\times 10^7$~s.
 All cuts except for those on $\pt$ and $\zvtx$ are applied.
 }
 \label{fig:ptz_pi0pi0}
\end{figure}

$\klpiopio$ becomes a background
when two clusters are formed at the calorimeter and
the other photons are missed. The following cases can be distinguished:
\begin{enumerate}
 \item Fusion background: Three photons enter the calorimeter,
       and two of them are fused into a single cluster.
       The fourth photon is missed due to detector inefficiency.
 \item Even-pairing background: Two photons from a $\pi^0$-decay
       form clusters in the calorimeter. Two photons from
       the other $\pi^0$ are missed due to detector inefficiency.
 \item Odd-pairing background : One photon from one $\pi^0$ and
       one photon from the other $\pi^0$ form clusters in the calorimeter.
       The other two photons are missed due to detector inefficiency.
\end{enumerate}
We estimate the total number of background events from $\klpiopio$ 
to be 16.9, of which 1.6 events come from fusion, 11 events from the even-pairing background, and 
4.3 events from the odd-pairing background.
Fig.~\ref{fig:ptz_pi0pi0} shows the distribution of $\klpiopio$ background events
in the $\zvtx$-$\pt$ plane, with all cuts applied except for the $\zvtx$ and $\pt$ selections.
Among these 33.2 background events,
both of the two extra photons are missed in the Central Barrel Counter in 29 events, which gives the dominant contribution.

\begin{figure}[h]
 \centering
 \subfloat[]{
 \includegraphics[page=3,width=0.45\textwidth]{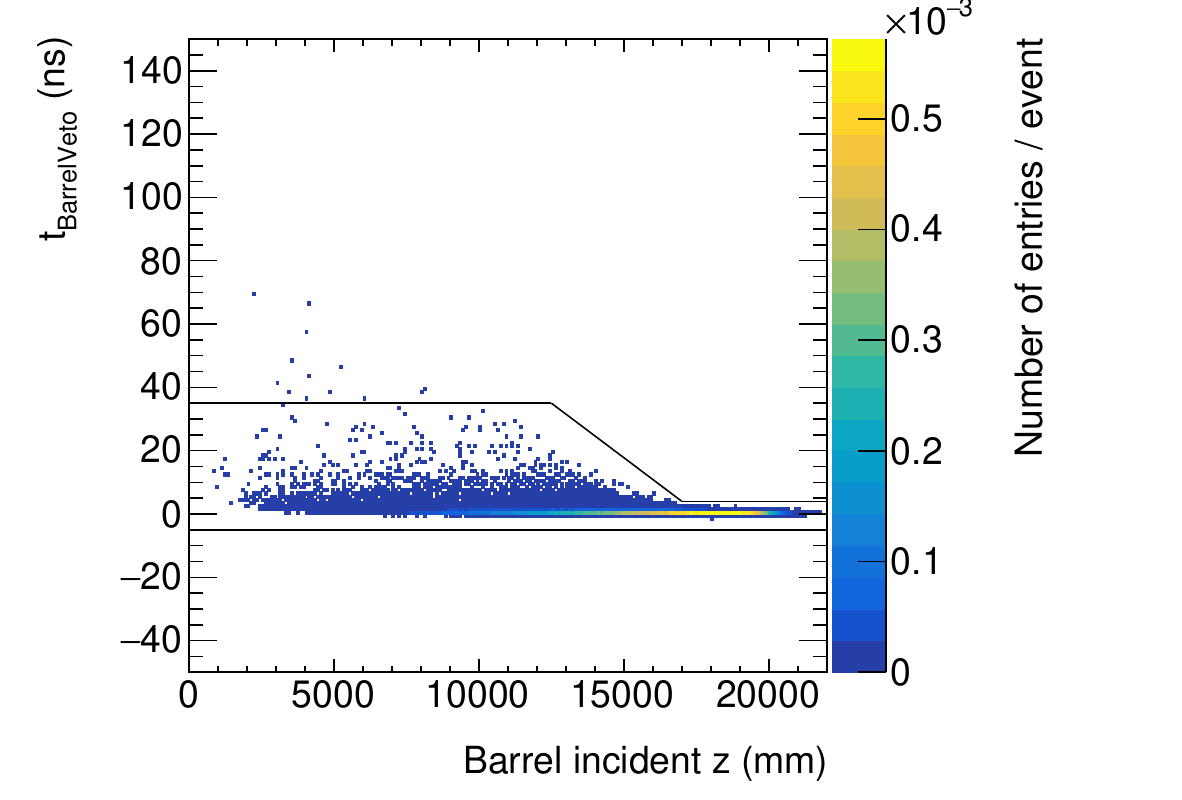}
 }
 \hspace{0.3cm}
 \subfloat[]{
 \includegraphics[page=6,width=0.45\textwidth]{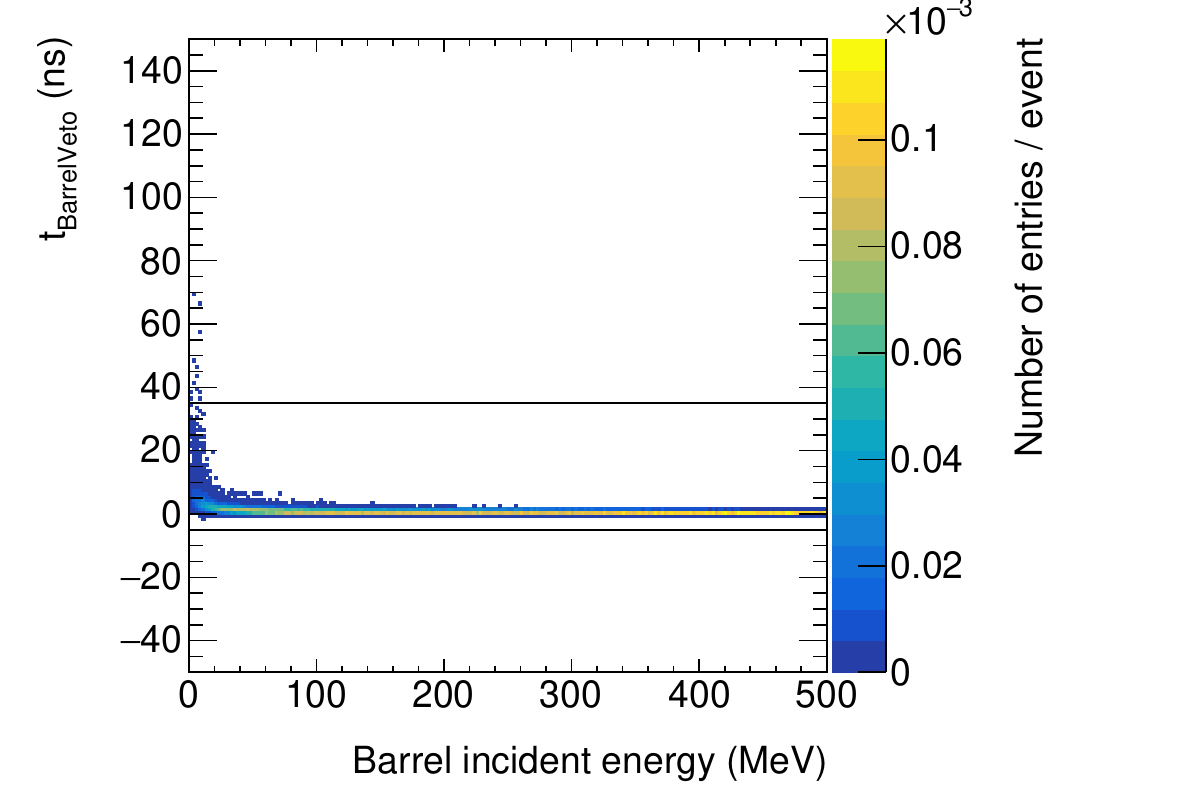}
 }
 \caption{Correlations between
 $t_{\mathrm{BarrelVeto}}$ and the position of incidence in $z$ (a), and between
 $t_{\mathrm{BarrelVeto}}$ and the energy of the incident photon (b),
 for photons from $\klpiopio$ arriving on the barrel, with all cuts imposed and time smearing with timing resolution
 depending on the incident photon  energy.
 }
 \label{fig:timing2pi0}
\end{figure}

For the barrel veto, the definition of the veto time $t_{\mathrm{BarrelVeto}}$,
and the veto window depending on the barrel hit position 
are introduced in Sec.~\ref{sec:backsplash}.
Fig.~\ref{fig:timing2pi0} shows the correlation between
$t_{\mathrm{BarrelVeto}}$ and position of incidence in $z$ on the barrel, with all cuts imposed.
Time smearing is applied with the time resolution
depending on the incident energy
as described in Sec.\ref{sec:barrelTreso}.

For photons incident on the barrel downstream of $z=15$~m,
the photon direction tends to be forward
because the signal region is upstream of 15~m.
Therefore, the timing fluctuation is small, and the time resolution
is better due to the higher energy photons.
The number of background events does not change by introducing the veto window with dependence on the barrel hit position.

For photons incident on the barrel upstream of 15~m,
the timing fluctuation is larger, and 
events with $t_{\mathrm{BarrelVeto}}>35~\mathrm{ns}$ exist.
For the events with larger $t_{\mathrm{BarrelVeto}}$,
the energy is small, 
as shown in Fig.~\ref{fig:timing2pi0} (b),
because of the backward photon.
These low energy photons are difficult to detect in any case, so missing them outside the veto window causes only a small increase in the $\klpiopio$ background. In fact, 
the expected number of events
with $t_{\mathrm{BarrelVeto}}>35~\mathrm{ns}$
is small (0.1 events).

If we define the veto window with the cut $t_{\mathrm{BarrelVeto}}<15~\mathrm{ns}$
instead of the baseline criterion $t_{\mathrm{BarrelVeto}}<35~\mathrm{ns}$,
the number of background events increases by only 3.4 events.

With an additional smearing in $z$ using a gaussian with $\sigma=500~{\rm mm}$ applied to the distribution in Fig.~\ref{fig:timing2pi0},
the same number of background events is obtained.

\subsubsection{$\klppm$}
\begin{figure}[h]
 \centering
 \includegraphics[page=20,width=0.5\textwidth]{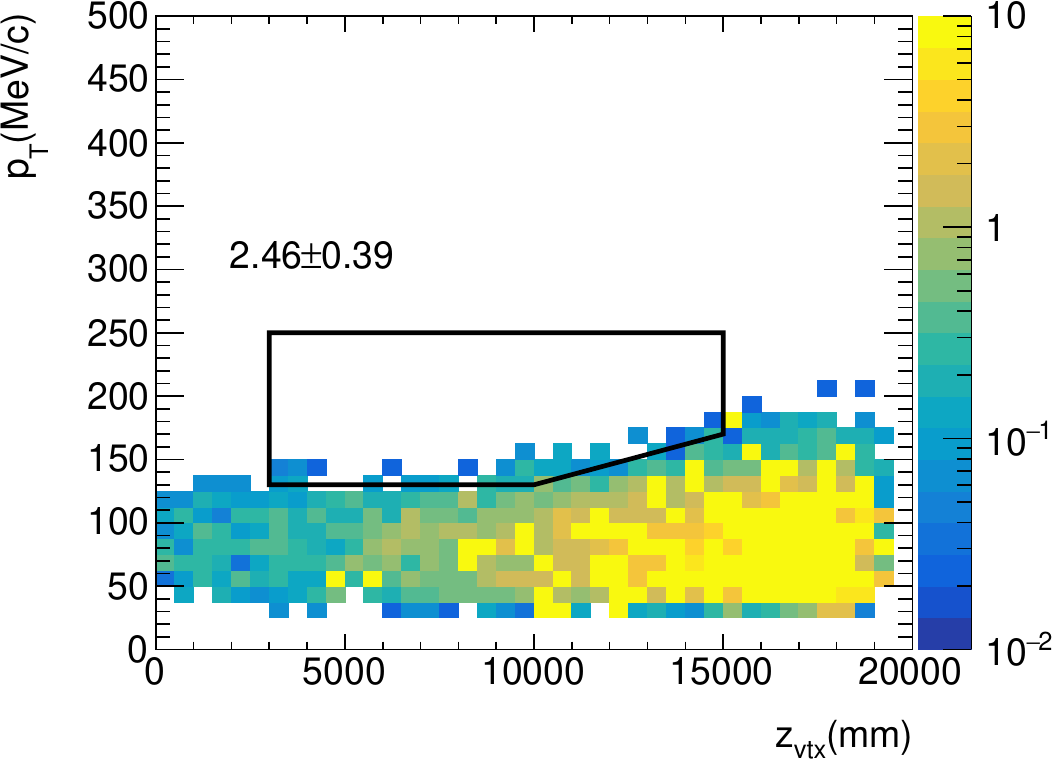}
 \caption{Distribution of  $\klppm$
 background events
 in the $\zvtx$-$\pt$ plane 
 for a running time of $3\times 10^7$~s.
 All cuts except for those on $\pt$ and $\zvtx$ are applied.}
 \label{fig:ptz_pipipi0}
\end{figure}

$\klppm$ becomes a background when the $\pi^+$ and $\pi^-$ are not detected.
The number of these background events
is evaluated to be 2.5
as shown in Fig.~\ref{fig:ptz_pipipi0},
where one charged pion is lost in the Charged Veto Counter,
and the other is lost in the Beam Hole Counter.
The maximum $\pt$ of the reconstructed $\pi^0$ is
limited by kinematics and the $\pt$ resolution.
The tighter $\pt$ selection in the downstream region makes the cut pentagonal in the $\zvtx$-$\pt$ plane
as shown in the figure.
This cut reduces the background
because the $\pt$ resolution is worse in the downstream region.

\subsubsection{$\kpien$ (Ke3)}
\begin{figure}[h]
 \centering
 \includegraphics[page=2,width=0.5\textwidth]{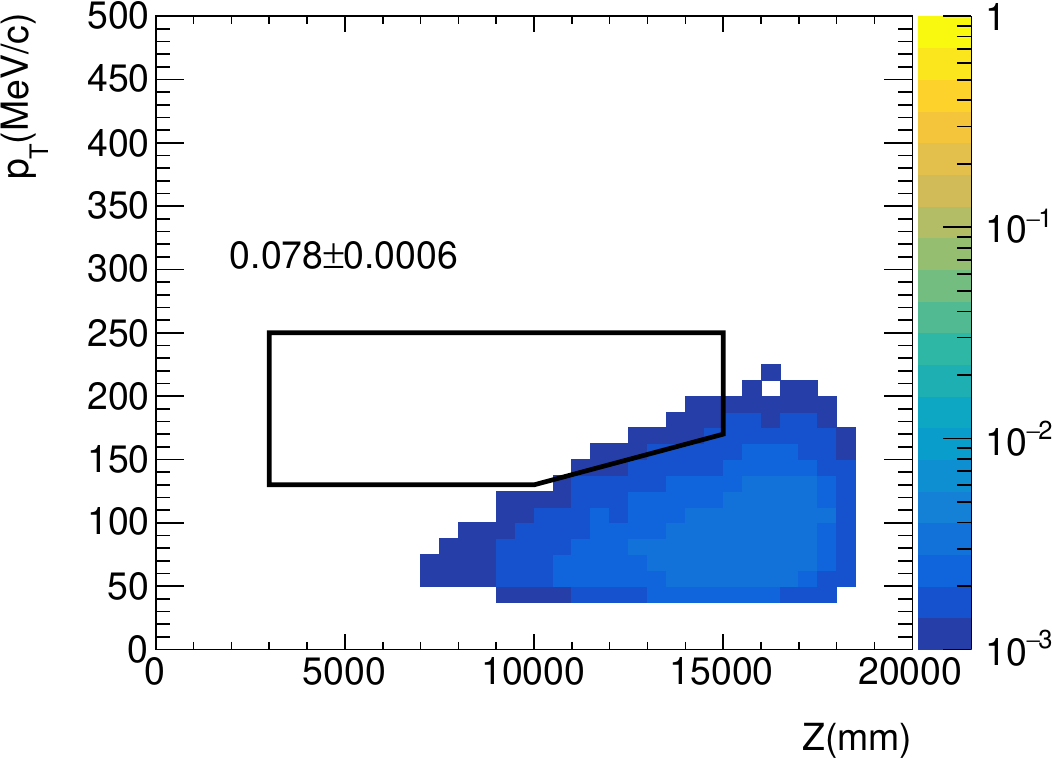}
 \caption{Distribution of Ke3
 background events
 in the $\zvtx$-$\pt$ plane 
 for a running time of $3\times 10^7$~s.
 All cuts except for those on $\pt$ and $\zvtx$ are applied.}\label{fig:ptz_ke3}
\end{figure}

Ke3 events become background when the electron and charged pion are not
identified with the Charged Veto Counter.
The number of background events is evaluated to be
0.08, assuming a $10^{-12}$ reduction with the Charged Veto Counter, 
as shown in Fig.~\ref{fig:ptz_ke3}.

\subsubsection{$\klgg$ for halo $K_L$} \label{sec:haloKL}
\begin{figure}[h]
 \centering
 \includegraphics[page=2,width=0.5\textwidth]{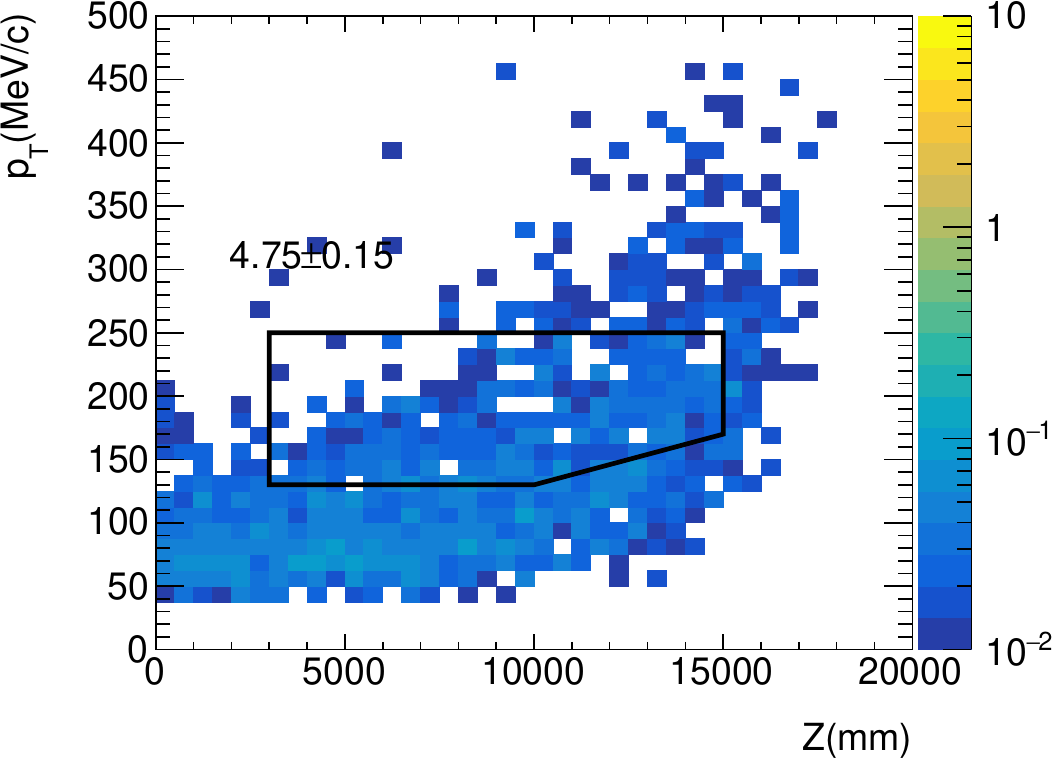}
 \caption{Distribution of halo $K_L\to2\gamma$ background events
 in the $\zvtx$-$\pt$ plane
 for a running time of $3\times 10^7$~s.
 All cuts other than the $\pt$ and $\zvtx$ cuts are applied.}
 \label{fig:ptz_haloK}
\end{figure}

$K_L$s in the beam can scatter off of the beam line components and enter the
beam halo region. When such a halo $K_L$ decays into two photons, an incorrectly large value for the reconstructed $\pt$ may be obtained 
due to the assumption of the vertex on the $z$-axis.
The decay vertex is incorrectly reconstructed
with the nominal pion mass assumption.
This incorrect vertex position gives incorrect values for the apparent angles of incidence of the photons on the calorimeter.
Therefore, the halo $K_L\to2\gamma$ background can be reduced with angle-of-incidence information at the calorimeter.
Another hypothetical vertex can be reconstructed with the nominal $K_L$ mass assumption,
which gives correct values for the angles of incidence of the photons.
In KOTO, by comparing the observed cluster shapes to 
those expected from the correct and incorrect vertex position hypotheses,
this background is reduced to 10\% of its original level,
while maintaining 90\% signal efficiency.
In this report, we assume that the background can be further reduced in KOTO~II
to 1\% of its original level,
because the higher energy photon in KOTO~II will give better resolution in the photon angle of incidence.
Further studies are in progress.
The number of $\klgg$ background events is evaluated to be 4.8,
as shown in Fig.~\ref{fig:ptz_haloK}.
For the halo $K_L$ generation in this simulation,
the core $K_L$ momentum spectrum and halo neutron directions were used.
Systematic uncertainties on the flux and momentum spectrum are under further study.

\subsubsection{$K^\pm\to\pi^0 e^\pm\nu$}
$K^\pm$s are generated in the interaction of $K_L$s,
neutrons, or $\pi^\pm$s
in the collimator in the beam line.
Here we assume that 
the second sweeping magnet
near the entrance of the detector
will reduce the flux to 10\% of its original value.

Higher momentum $K^\pm$s can survive
in the region downstream of the second magnet,
and $K^\pm\to\pi^0 e^\pm\nu$
decays can occur in the detector.
This becomes a background if the $e^\pm$ is undetected.
The kinematics of the $\pi^0$ is similar to that in
$\klpionn$ decays; therefore,
this is one of the more dangerous backgrounds.
Detection of  the $e^\pm$ is one of the keys to reduce this background.

\begin{figure}[h]
 \centering
 \includegraphics[page=2,width=0.5\textwidth]{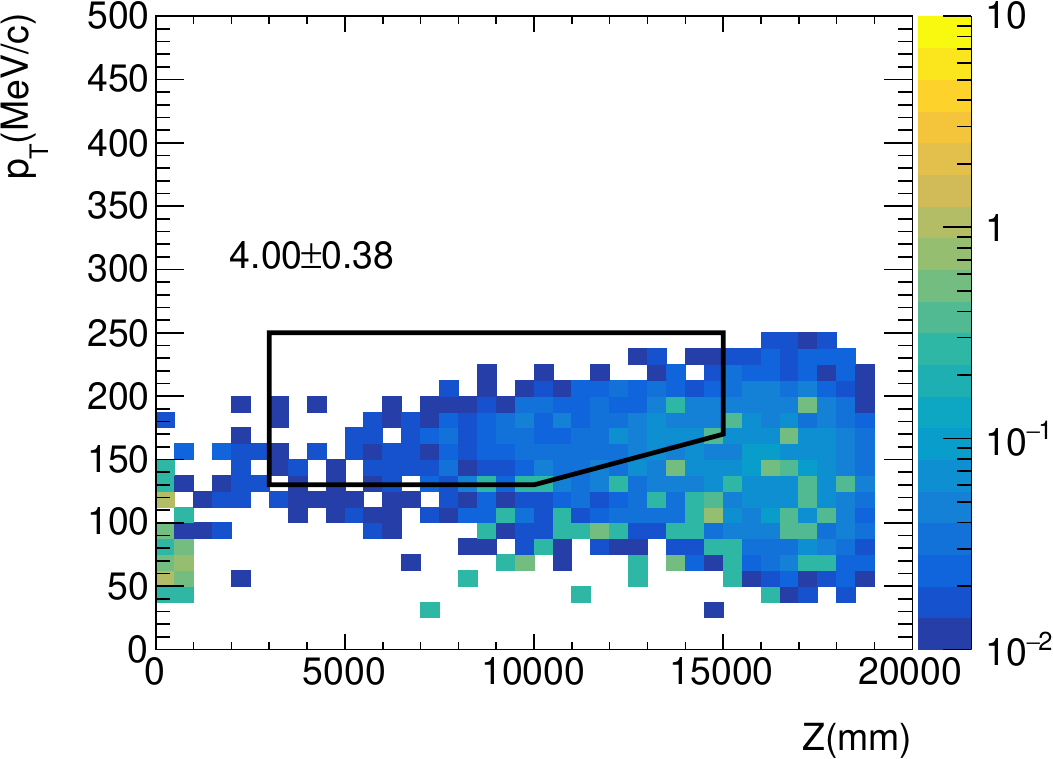}
 \caption{Distribution of $K^\pm\to\pi^0 e^\pm\nu$ events in the $\zvtx$-$\pt$ plane for a running time of $3\times 10^7$~s.
 All cuts except for the $\pt$ and $\zvtx$ cuts are applied.}
 \label{fig:ptz_Kplus}
\end{figure}

We have evaluated the number of the background events to be 4.0, as shown
in Fig.~\ref{fig:ptz_Kplus}.

\begin{figure}[h]
 \centering
 \includegraphics[page=3,width=0.45\textwidth]{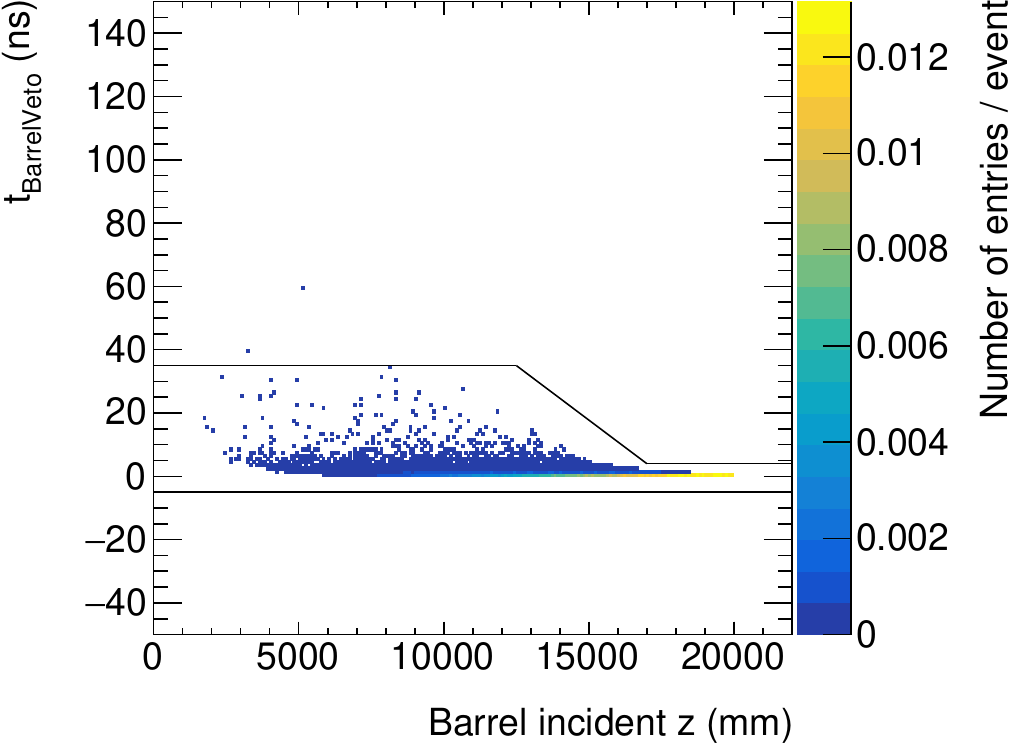}
 \caption{
 Correlation between
 $t_{\mathrm{BarrelVeto}}$ and the position of incidence in $z$ on the barrel
 for $e^\pm$ from $K^\pm\to\pi^0 e^\pm\nu$ background events
 after imposing all the cuts.
 Time smearing is not applied.
 }
 \label{fig:timing_Kplus}
\end{figure}

The veto timing of the barrel detector is essential also for reduction of this background.
Figure~\ref{fig:timing_Kplus} shows
the correlation between the barrel hit-$z$-position and
$t_{\mathrm{BarrelVeto}}$ for $K^\pm$ events.
The lower momentum electrons or positrons contribute to the events with larger $t_{\mathrm{BarrelVeto}}$
due to the backward-going configuration similarly to $\klpiopio$.
Unlike for photons,
the detection efficiency for $e^\pm$ is high,because
a few-MeV electron is still a minimum-ionizing particle.
Therefore, the loss of the low-momentum particles outside the veto window can significantly increase the number of background events.
The 40-ns veto window from $-5$~ns to $35$~ns is adopted
because the number of events with 
$t_{\mathrm{BarrelVeto}}>35~\mathrm{ns}$ is small (0.5 events).
The number of background events increases to 322, for example,
if we set a 20-ns veto window 
($-5~\mathrm{ns}<t_{\mathrm{BarrelVeto}}<15~\mathrm{ns}$)
instead of the 40-ns veto window.

For this report, we use the same veto timing for the charged veto as for the photon veto.
There is room to optimize the veto timings independently.

\subsubsection{Hadron cluster}\label{sec:hadronCluster}
The hadron cluster background arises when a halo neutron hits the calorimeter to produce a first hadronic shower,
and another neutron from the shower travels inside the calorimeter, producing a second hadronic shower at some distance from the first one.
These two hadronic clusters mimic the signal.

We have evaluated this background using halo neutrons prepared
with the beam line simulation,
and with a calorimeter composed of 50-cm-long CsI crystals.
A full-shower simulation was performed with those neutrons.

\begin{figure}[h]
 \centering
 \includegraphics[page=2,width=0.5\textwidth]{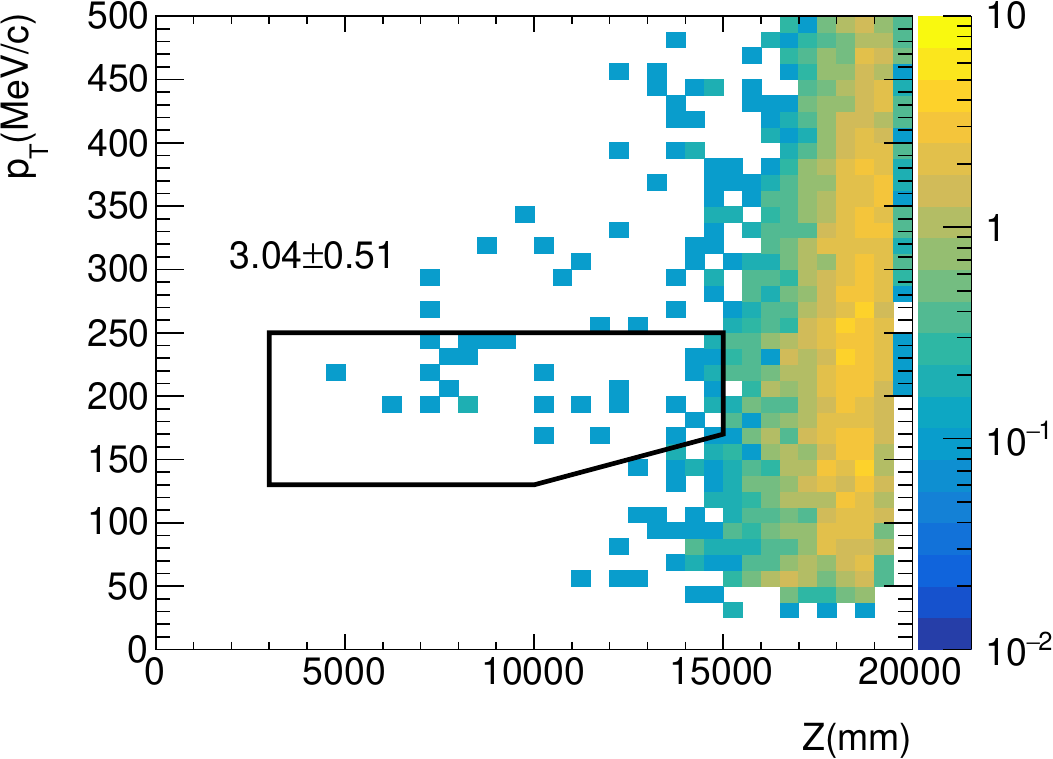}
 \caption{Distribution of hadron cluster background
 events in the $\zvtx$-$\pt$ plane
 for a running time of $3\times 10^7$~s.
 All cuts other than the $\pt$ and $\zvtx$ cuts are applied.}
 \label{fig:ptz_NN}
\end{figure}

We evaluated the number of background events to be 3.0
as shown in Fig.~\ref{fig:ptz_NN}.
Here we assume that the hadron cluster background can be reduced by a factor of $10^{-7}$ with
the cluster-shape, pulse-shape,
and shower-depth information in the calorimeter.
In KOTO, we have achieved background reduction by a factor of
$\times ((2.5 \pm 0.01)\times 10^{-6})$
with 72\% signal efficiency
by using cluster and pulse shapes.
By using the shower depth information,
we achieved an additional reduction factor of  $\times (2.1\times 10^{-2})$
with 90\% signal efficiency and only a small correlation with
the cluster and pulse shape cuts.
In total, a reduction factor of $10^{-7}$ is feasible.
This reduction power is one of the requirements on the calorimeter design.

\subsubsection{$\pi^0$ production at the Upstream Collar Counter}
If a halo neutron hits the Upstream Collar Counter, and produces a $\pi^0$ that decays into two photons, it can mimic the signal.

\begin{figure}[h]
 \centering
 \includegraphics[page=2,width=0.5\textwidth]{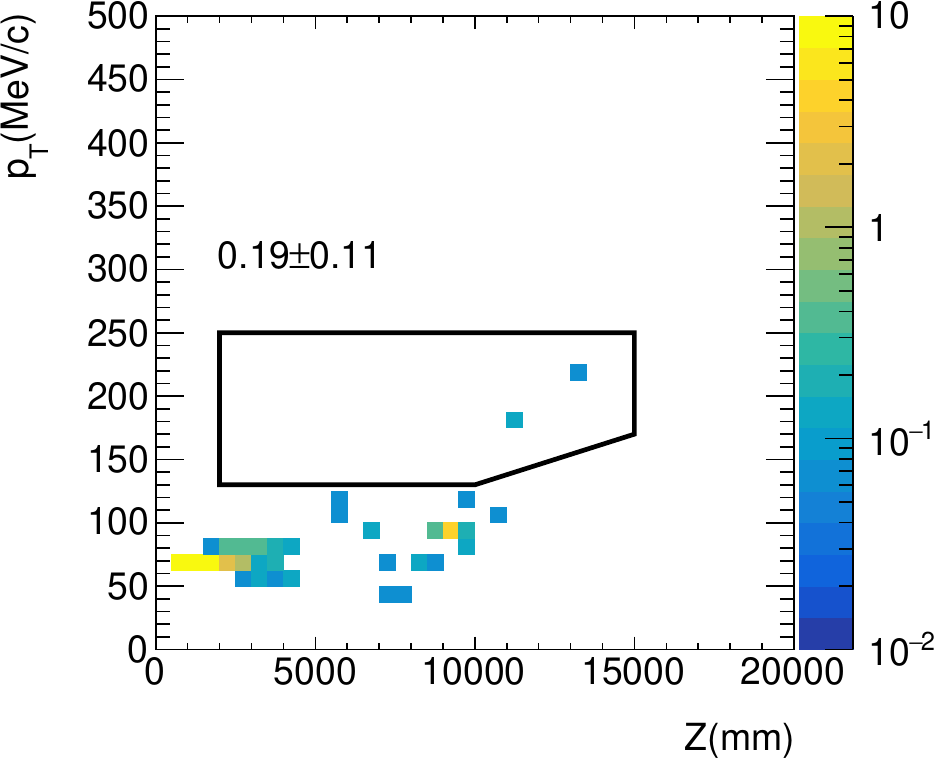}
 \caption{Distribution of background events from $\pi^0$ production at the
 Upstream Collar Counter
 in the $\zvtx$-$\pt$ plane
 for a running time of $3\times 10^7$~s.
 All cuts other than the $\pt$ and $\zvtx$ cuts are applied.}
 \label{fig:ptz_NCCpi0}
\end{figure}

Halo neutrons obtained from the beam line simulation are used
to simulate $\pi^0$ production in the Upstream Collar Counter.
We assume fully-active CsI crystals for the detector.
Other particles produced in the $\pi^0$ production can deposit energy
in the detector, and can veto the event.
In the simulation, such events were immediately discarded.
In the next step, only the $\pi^0$-decay was generated
in the Upstream Collar Counter. The two photons from the $\pi^0$ can also interact in 
the Upstream Collar Counter, and deposit energy in the counter.
Such events were also discarded.
Photons from $\pi^0$ production near the downstream surface of the
Upstream Collar Counter mainly survive.
Finally, when the two photons hit the calorimeter,
a full shower simulation was performed.
In this process, photon energy can be mismeasured due to
photonuclear interactions.
Accordingly, the distribution of the events in the $\zvtx$-$\pt$ plane
was obtained as shown in Fig.~\ref{fig:ptz_NCCpi0}.
We evaluated the number of expected background events to be 0.19.

\subsubsection{$\eta$ production at the Charged Veto Counter}
This background arises when a halo neutron hits the Charged Veto Counter and produces an $\eta$.
The $\eta$ decays into two photons with a branching fraction of 39.4\%,
which can mimic the signal.
The decay vertex will be reconstructed upstream of the
Charged Veto Counter because the $\eta$ mass is
four times larger than the $\pi^0$ mass.

\begin{figure}[h]
 \centering
 \includegraphics[page=2,width=0.5\textwidth]{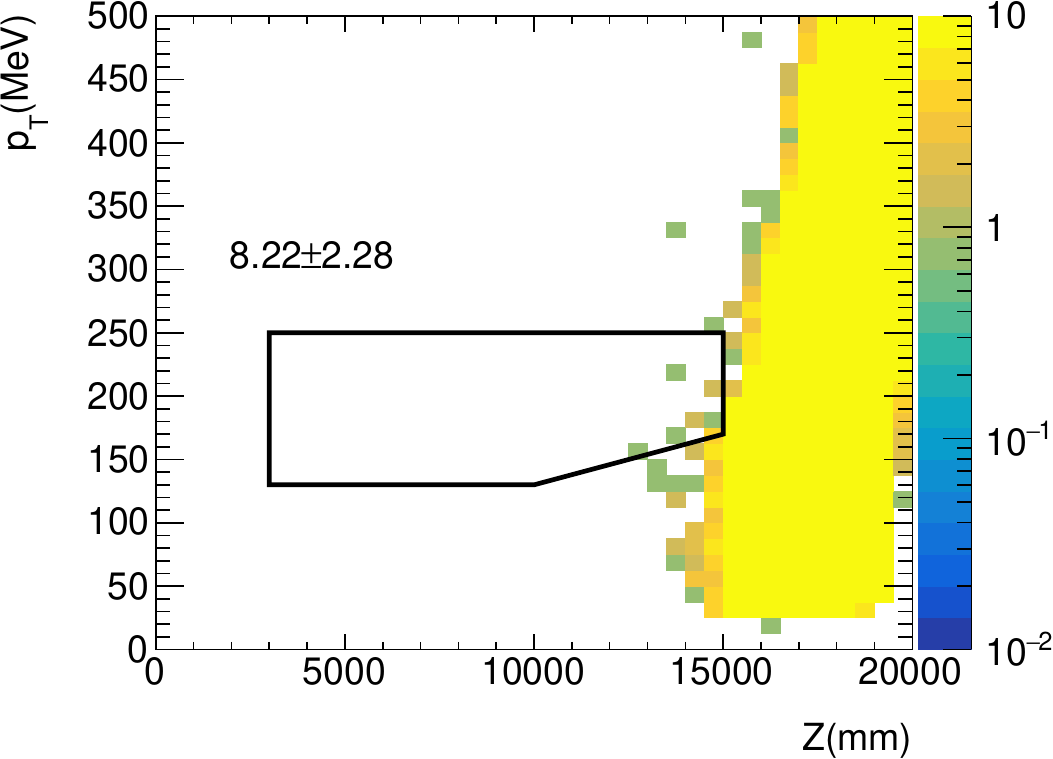}
 \caption{Distribution 
 of background events from $\eta$ production at the Charged Veto Counter
 in the $\zvtx$-$\pt$ plane
 for a running time of $3\times 10^7$~s.
 All cuts other than the $\pt$ and $\zvtx$ cuts are applied.}
 \label{fig:ptz_CVeta}
\end{figure}

Halo neutrons obtained from the beam line simulation are used
to simulate $\eta$ production in the Charged Veto Counter.
We assume the Charged Veto Counter to be a 3-mm-thick plastic scintillator at 30-cm upstream of
the calorimeter.
Other particles produced together with the $\eta$ can deposit energy in the
Charged Veto Counter and can veto the event.
In the simulation, such events were discarded immediately.
In the next step, only the $\eta$-decay was generated.
When the two photons from the $\eta$ hit the calorimeter,
a full shower simulation was performed.
Two clusters were reconstructed, and 
the distribution of the events in the $\zvtx$-$\pt$ plane
was obtained as shown in Fig.~\ref{fig:ptz_CVeta}.
We evaluated the number of background events to be 8.2.
This background can be suppressed more 
with cluster shape information, because 
the cluster shape should be different 
from the shape expected with the incident angle obtained from the reconstructed vertex. Further reduction is possible 
if the incident photon angle is reconstructed.

\subsubsection{Summary of the background estimation}
The estimated numbers of background events from each source are collected in Table~\ref{tab:bg}.
The total number of background events is
$40.0\pm 2.7$, with the largest contribution of $16.9\pm 1.1$ coming from
the $\klpiopio$ decay.
\begin{table}
 \centering
 \caption{Summary of background estimates.}\label{tab:bg}
 \begin{tabular}{lrr}\hline
  Background&Number&\\\hline
  $\klpiopio$&16.9 &$\pm$1.1 \\
  $\klppm$&2.5 &$\pm$0.4 \\
  $\kpien$& 0.08 & $\pm$0.0006\\
  halo $K_L\to 2\gamma$&4.8 &$\pm$0.2 \\
  $K^\pm\to\pi^0 e^\pm\nu$&4.0 &$\pm$0.4 \\
  hadron cluster&3.0 &$\pm$0.5 \\
  $\pi^0$ at upstream&0.2 &$\pm$0.1 \\
  $\eta$ at downstream&8.2 &$\pm$2.3 \\ \hline
  Total& 40 & $\pm$2.7 \\\hline
 \end{tabular}
\end{table}

\subsection{Sensitivity and impact}
We assume a running time of $3\times 10^7$ s with a 100 kW proton beam on a 1-interaction-length T2 target,
where the $K_L$ flux is
$1.1\times 10^7$ per $2\times 10^{13}$ protons on target.

The sensitivity and the impacts are calculated and 
summarized in Table~\ref{tab:sensitivity}.
Here we assume
that 
the statistical uncertainties in the numbers of events
are dominant.

The single event sensitivity is evaluated to be 
$8.5\times 10^{-13}$.
The expected number of background events is 40.
With the SM branching fraction of $3\times 10^{-11}$,
35 signal events are expected with a signal-to-background ratio (S/B) of 0.89.
 \begin{itemize}
  \item Observation at the level of $5.6\sigma$ is expected
	for the signal branching fraction
	$3\times 10^{-11}$.
  \item The measurement would indicate new physics at the 90\% confidence level (C.L.) if
	the new physics gives a
	$40\%$ deviation on the BR from the SM prediction.
  \item The measurement corresponds to $12\%$ measurement of the CP-violating CKM parameter $\eta$ in the SM (The branching fraction is proportional to $\eta^2$). 
 \end{itemize}

\begin{table}[h]
 \centering
 \caption{Summary of sensitivity and impact.}
 \label{tab:sensitivity}
 \begin{threeparttable}
 \begin{tabular}{lcc}\hline
 & Formula & Value \\
 \hline
  Signal (branching fraction : $3\times 10^{-11}$) & $S$ & $35.3 \pm 0.4$\\
  Background & $B$ & $40 \pm 2.7$ \\
  \hline
  Single event sensitivity & $(3\times 10^{-11})/S$ & $8.5\times 10^{-13}$\\
  Signal-to-background ratio & $S/B$ & 0.89 \\
  Significance of the observation&  $S/\sqrt{B}$ & $5.6 \sigma$\\
  90\%-C.L. excess / deficit & $1.64\times \sqrt{S+B}$ & 14 events \\
  & $1.64\times \sqrt{S+B}/S $ & 40\% of SM\\
  Precision on branching fraction & $\sqrt{S+B}/S$ & 25\%\\
  Precision on CKM parameter $\eta$ &$0.5\times \sqrt{S+B}/S$ &12\%\\
  \hline
 \end{tabular}
 \begin{tablenotes}\footnotesize
 \item[*] A running time of $3\times 10^7$~s is assumed in the calculation.
 \end{tablenotes}
 \end{threeparttable}
\end{table}

\clearpage


\section{Detector Feasibility Study\label{sec:dfeasibility}}
\subsection{Angle measurement of photons}
The $K_L \rightarrow \pi^0 \nu \bar{\nu}$ decay is defined with a $\pi^0$ reconstructed from the two photons in the calorimeter with an assumption that those are generated from a single $\pi^0$ decay on the beam axis. This method is not working correctly in the off-axis events. For example, when a $K_L$ enters the detector far from the beam axis due to interaction with beamline materials and decays into two photons (halo $\klgg$ decay), the reconstruction gives a wrong answer, and it could be a background event.

\begin{figure}[h]
\includegraphics[width=1.0\textwidth]{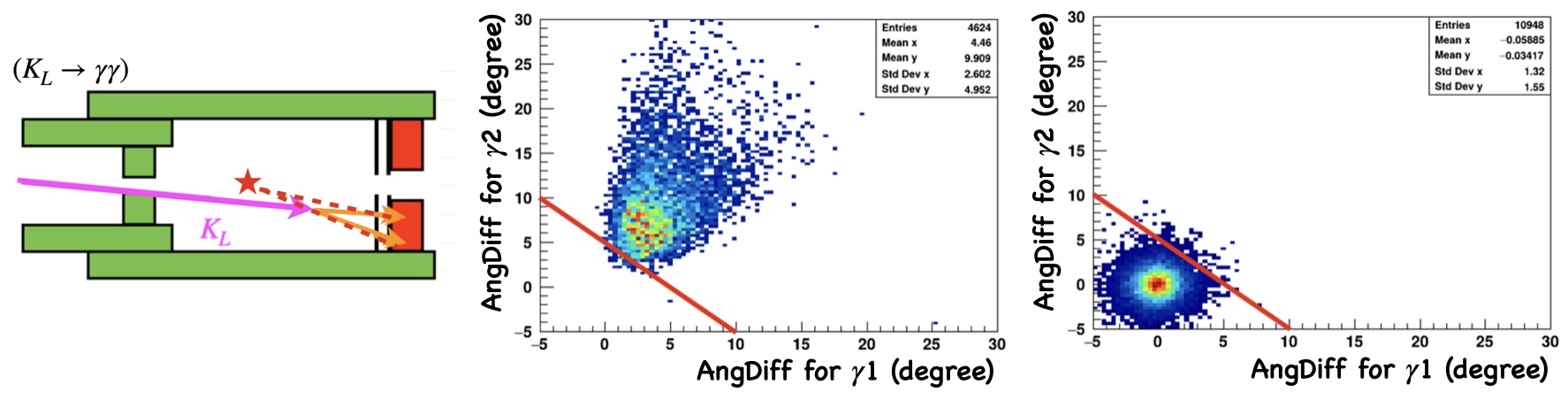}
\caption{The  {\it halo $K_L \rightarrow \gamma \gamma$} decay is reconstructed incorrectly due to the off-axis decay position and heavier mass (left). The difference between angles directly measured in the calorimeter and calculated from the $\pi^0$ reconstruction is significant in the {\it halo $K_L\rightarrow \gamma \gamma$ decay} (middle), while the $K_L \rightarrow \pi^0 \nu \bar{\nu}$ decay shows good consistency (right). }
\label{fig:Ang_diff}
\end{figure}

\begin{figure}[ht]
\centering\includegraphics[height=0.3\textwidth]{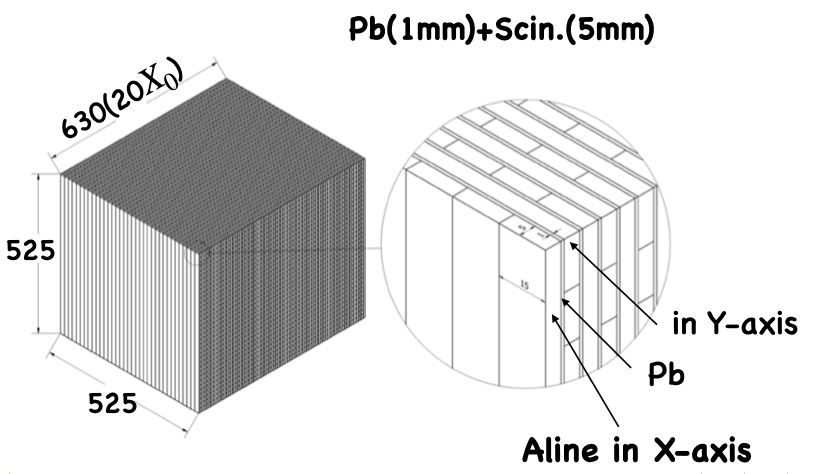}
\centering\includegraphics[height=0.3\textwidth]{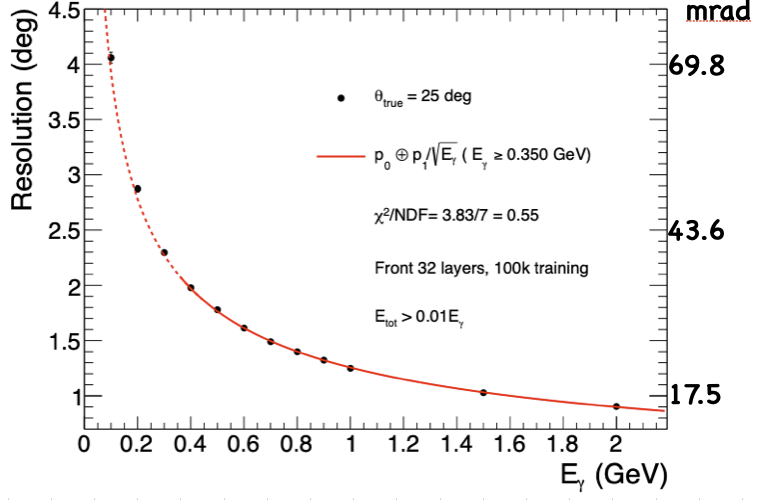}
\caption{A toy model to study the feasibility of measuring photon-angle with finely segmented sampling calorimeter (left) and the expected angular resolution (right)\cite{ref:PAScal}.}
\label{fig:MCstudy} 
\end{figure}

Figure~\ref{fig:Ang_diff} shows how to separate the halo $\klgg$ decay from the $K_L \rightarrow \pi^0 \nu \bar{\nu}$ decay if we can measure the incident angle of photons in the calorimeter. The vertex of 
the halo $K_L$ decay is obtained incorrectly, resulting in wrong incident angles of the photons (left plot). When we directly measure the incident angle of photons in the calorimeter, the decay will be rejected by comparing the calculated and measured angles (middle) without losing the signal acceptance (right). The simulation study implies the achievable angular resolution of 1.3 degrees for 1 GeV photons as shown in Fig.~\ref{fig:MCstudy}
(from the reference \cite{ref:PAScal}), 
which enables us to reduce the background events by two orders of magnitude with a 2\% of the signal loss.

The photon-angle measurement is performed by analyzing the shape of the electromagnetic shower produced in the calorimeter. The shower shape can be obtained by a calorimeter of thin and narrow modules alternatively aligned along the photon direction in horizontal and vertical directions. The distribution of energy deposits in each module represents the shower shape, and the incident angle is deduced by analyzing it with a machine learning package (XGBoost).  Figure ~\ref{fig:MCstudy} shows a schematic drawing of the toy calorimeter used in the simulation studies, consisting of lead (1 mm) and plastic scintillator (5 mm) modules in a dimension of $500~\rm{mm} \times 15~\rm{mm} \times 6~\rm{mm(t)}$ (right) and the expected angular resolution (left).  When we focus only on the angle measurement, the detector does not need to be thick to contain whole shower particles for the energy resolution. Therefore, we place a finely segmented calorimeter of the five radiation lengths (5$X_0$), pre-shower detector, in front of the traditional calorimeter to simultaneously measure the angle and energy. This configuration provides an angular resolution similar to that of the entire simulation model setup, minimizing energy resolution degradation due to the sampling fluctuation.

\begin{figure}[ht]
\centering\includegraphics[height=0.36\textwidth]{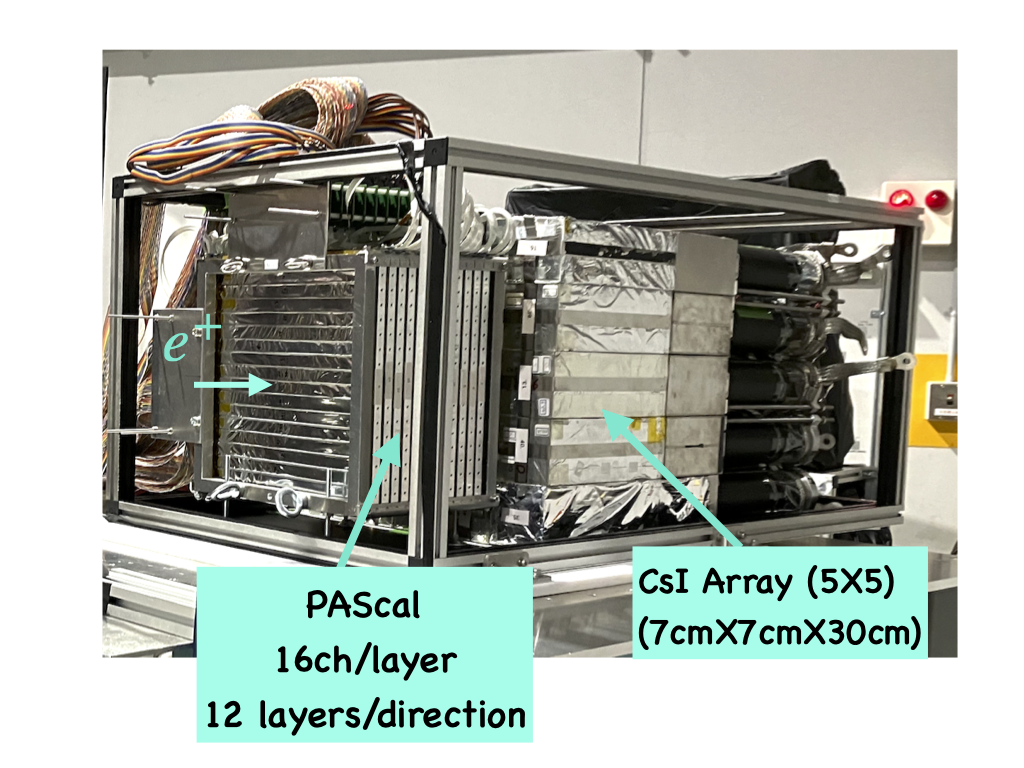}
\centering\includegraphics[height=0.36\textwidth]{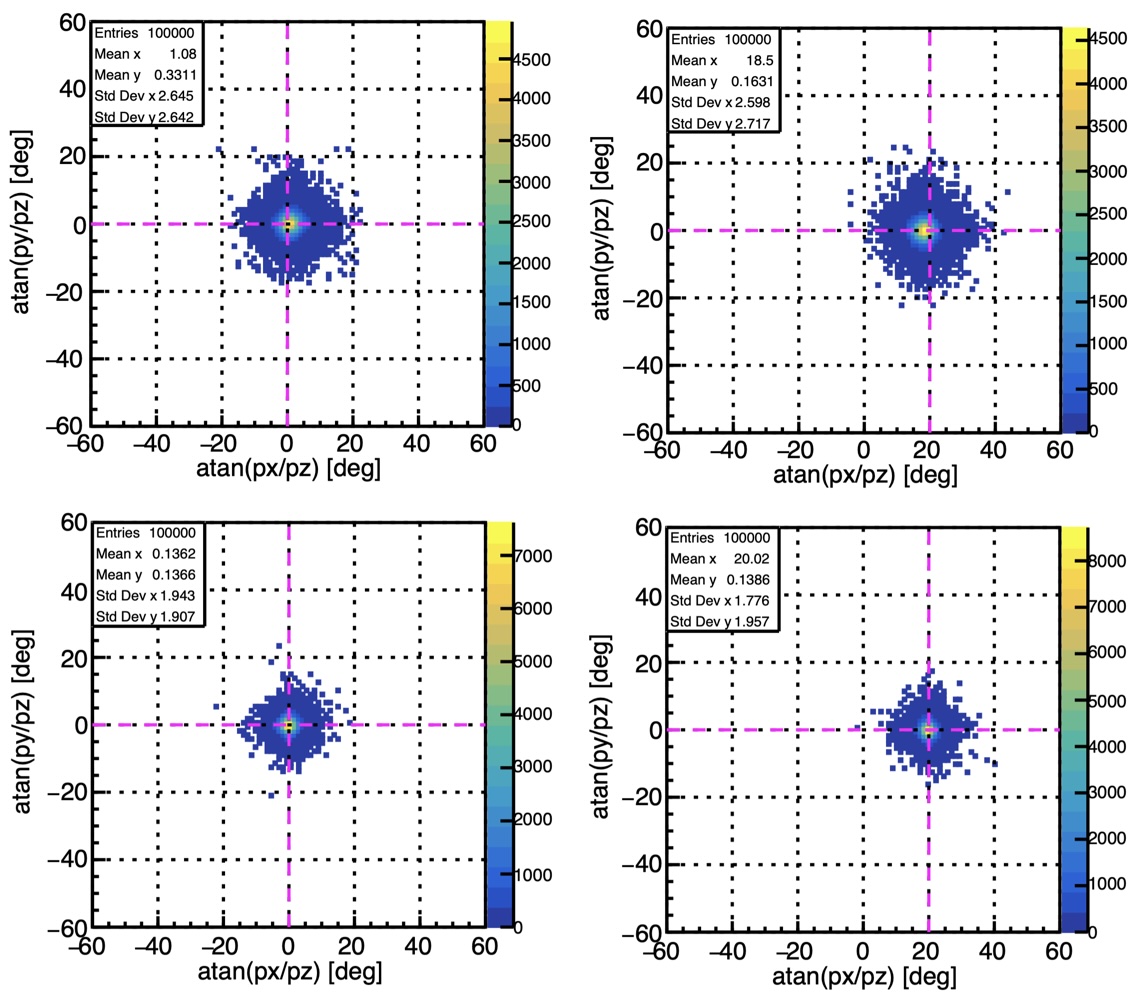}
\caption{Left: A photo of the pre-shower detector(indicated as PAScal) installed in the positron beamline for its performance test. Behind the detector, a homogeneous calorimeter made of the CsI crystals is prepared to study the effect of the pre-shower on the energy resolution. Right: Reconstructed incident angle for 0 degrees (left) and 20 degrees (right). The angular resolution of the data (up) is worse than that of the simulation (down) as 30-40\%.}
\label{fig:ELPH} 
\end{figure}

To validate the simulation results, we fabricated a pre-shower detector with 24 layers, each with 16 modules alternatively aligned horizontally and vertically. The module consists of five layers of alternating 1 mm-square scintillating fibers and 0.15 mm-thick tungsten strips. Compared to the simulation model, its layers are thinner to reduce the sampling fluctuation, which is the main factor determining the energy resolution of the sampling calorimeter.
The total number of modules is 384, and the detector has an effective cross-section of $\rm 239~mm\times239~mm$ and a thickness of 167 mm, corresponding to 5.36 radiation lengths.  
The detector performance was tested with a positron beam produced by the 1.3 GeV electron synchrotron at the Research Center for Electron Photon Science  (ELPH), Tohoku University. 
Figure~\ref{fig:ELPH} shows a photo of the detector installed in the positron beamline (left) and a snapshot of its analysis results. The incident angles are correctly reproduced for 0 and 20 degrees. Still, the data show a worse angular resolution of 30-40\% compared to the expectation from the simulation, and a detailed analysis is underway to understand the reason.

The feasibility study above clarifies that a finely segmented sampling calorimeter can measure a photon's angle. This enables us to provide a tool for further suppressing the background events caused by the off-axis decays. The detector needs many channels to analyze the shape of the shower, which requires large-scale fabrication costs. The detector configuration and fabrication method will be carefully optimized to ensure timely utilization in KOTO~II.  

\subsection{Discussion on the calorimeter performance}
In the background estimation, we assumed that the new calorimeter for the KOTO~II experiment
consisted of 50-cm-long CsI crystals the same as the KOTO experiment.
However, the cost to make the whole new calorimeter of CsI crystals is high because
the diameter of the new calorimeter is 3 m and the area is 2.2 times larger than that of the KOTO experiment. 
It is thus important to consider an alternative option which reduce the cost while keeping the signal and noise ratio.
In this study, we evaluated the numbers of the $\klpiopio$ and $\klppm$ background events
by changing the energy resolution used in the MC simulation. 
We considered sampling calorimeters such as the calorimeters used in the KLOE experiment at DA$\Phi$NE,
and the calorimeter designed for the KOPIO experiment at BNL. The basic parameters on those calorimeters are listed in Table~\ref{table:EnergyTimeReso}.
The results are shown in Table~\ref{table:RelativeBG}.  
We can keep the current background level by using one of the above-mentioned calorimeters
in terms of energy and position resolutions. However, the photon detection inefficiency 
is worse in the sampling calorimeter, and the effect should be studied.
The effect on other background events will be investigated in the future.

The new calorimeter is also required to distinguish photon clusters from hadron clusters to reduce the hadron cluster background
events. In the KOTO experiment, the discrimination is achieved by using the information of cluster shapes, pulse shapes and
the depth of the interaction. The depth of the interaction is estimated by using the timing difference between the photo sensors attached 
at the upstream and downstream surfaces of the crystals. The timing resolution is a key to get a better discrimination from the depth 
information. As shown in Table~\ref{table:EnergyTimeReso}, sampling calorimeters can achieve the similar timing resolution to that of the calorimeter 
used in the KOTO experiment and thus same discrimination power.
The discrimination power of a sampling calorimeter by using the information of cluster shapes and pulse shapes
still need to be studied.  We plan to make a prototype detector and
evaluate the performance of the discrimination between photon clusters and neutron clusters with electron and neutron beams.

\begin{table}[htbp]
\centering
  \caption{Energy and timing resolutions of sampling calorimeters together with the technologies and the depth of the calorimeters.
  The performance of the CsI calorimeter used in KOTO is also shown as a reference. $E$ is in GeV. The energy and timing resolutions of the 
  calorimeters are referred from \cite{ref:PDG,ref:KOPIO,ref:Calorimeter,ref:CsITime}. }
  \label{table:EnergyTimeReso}
\begin{tabular}{llll}
\hline \hline
  Technology (Experiment) & Depth & Energy resolution & Timing resolution\\
 \hline 
   CsI (KOTO) & 27$X_{0}$  &   $2\%/\sqrt{E}\oplus1\%$  & $115~\mathrm{ps}/\sqrt{E}\oplus5~\mathrm{ps}/E$ \\
   & & & $\oplus130~\mathrm{ps}$ \\
   \hline
   Scintillator/Pb (KOPIO) & 16$X_{0}$  &   $3\%/\sqrt{E}$ & $90~\mathrm{ps}/\sqrt{E}$ \\
    \hline
   Scintillator fiber/Pb & 15$X_{0}$  &   $5.7\%/\sqrt{E}\oplus0.6\%$ & $54~\mathrm{ps}/\sqrt{E}\oplus140~\mathrm{ps}$  \\
   spaghetti (KLOE)    &                   &                         &                \\    
   \hline
  \end{tabular} 
\end{table}

\begin{table}[htbp]
\centering
  \caption{ The relative numbers of background events by changing the energy resolution assumed in the simulation.
  The numbers in the table are normalized by the numbers of expected background events  with the default calorimeter consisting 
  of CsI crystals.
}
\label{table:RelativeBG}
\begin{tabular}{lll}
\hline \hline
  Calorimeter type & \# of $\klpiopio$ BG &  \# of $\klppm$ BG  \\
 \hline 
  KOPIO & 1.01  &   0.96 \\
  KLOE   & 0.99  &   1.02 \\
   \hline
  \end{tabular}  
\end{table}


\subsection{Thermal neutron blind calorimeter}
\begin{figure}[tb]
\centering
\includegraphics[width=\linewidth, trim=0cm 4.0cm 0cm 4.0cm]{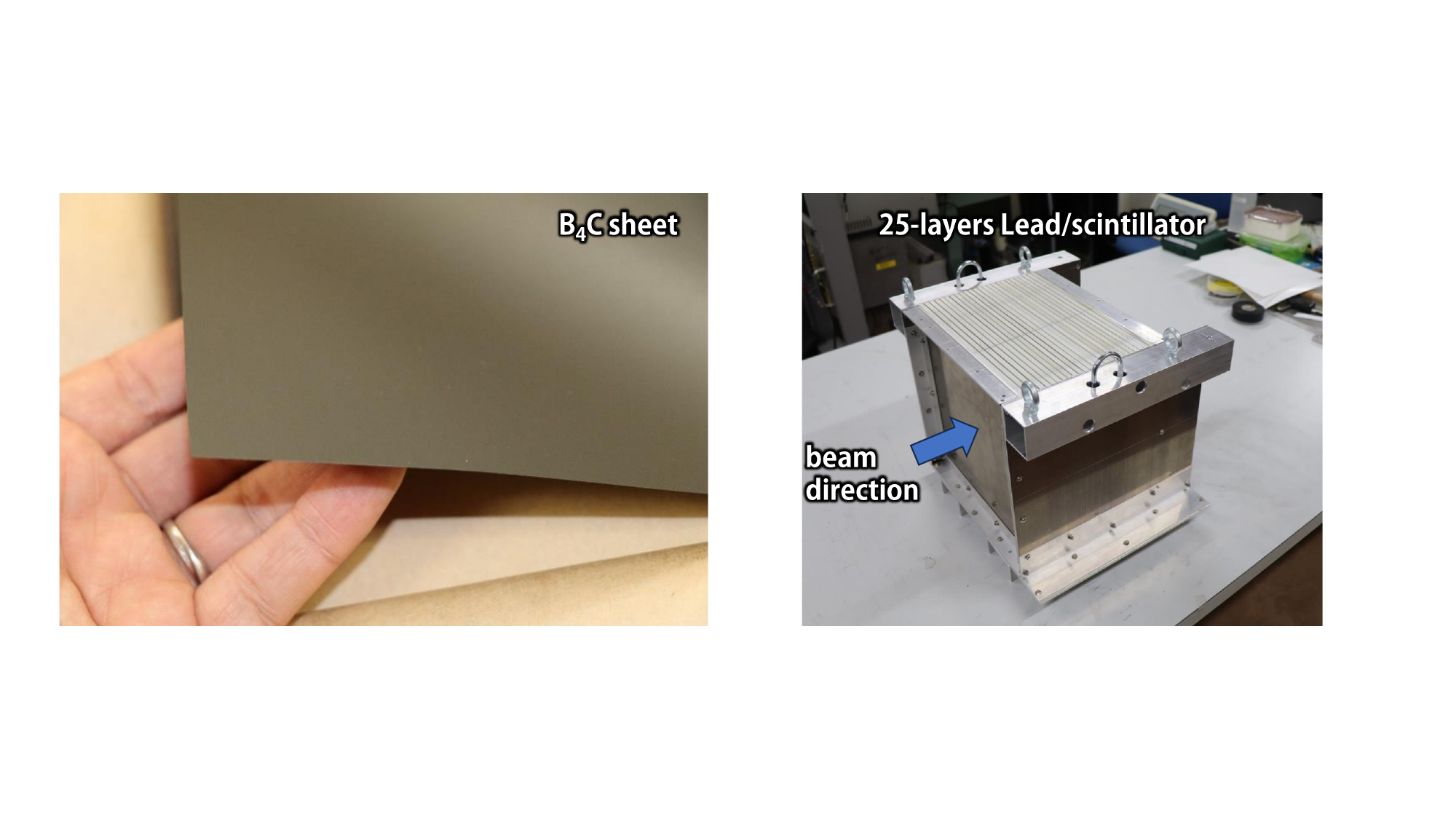}
\caption{(left) A commercial B$_4$C sheet with 0.1~mm thickness. (right) A prototype detector for validation of the effect of B$_4$C Mylar sheet. A acrylic light guide and a 5~inch PMT is optically attached to the top of scintillator surface (not shown in the figure). By injecting a low-intensity $\pi^-$ beam, we can observe delayed coincidence signals due to neutron capture reactions with protons in the scintillator. With B$_4$C Mylar sheets, these signals will disappear.} \label{fig:nblind-1}
\end{figure}

In the KOTO experiment, we found that the pion-induced neutrons generated in the Barrel Veto counters could enhance the accidental rate due to low-energy $\gamma$'s originating from neutron capture reactions. 
A simulation study showed that approximately 50\% of the counting rate of 5~MHz in the Front-Barrel Veto counter was caused by the thermal-neutron capture reactions~\cite{Matsumura_2023}. 
In the KOTO II experiment, we plan to insert thin B$_4$C films into the sandwich layers of the Barrel Veto counter to suppress the accidental rate. Boron-10, which is contained in 20\% of natural boron, has a thermal neutron capture cross section of 3840~b due to the $^{10}$B(n,$\alpha$)$^7$Li$^*$ reaction, so it can absorb thermal neutrons before they are captured by protons thorough the $^1$H(n,$\gamma$)$^2$H reaction, which emits a 2.2~MeV of $\gamma$. 
Figure~\ref{fig:nblind-1} shows a photo of a commercially available 0.1~mm thick B$_4$C Mylar sheet. By putting the sheet between the lead and reflector of a Barrel Veto detector, accidental counts caused by thermal neutrons can be reduced.

To validate the effect of the B$_4$C Mylar insertion, we fabricated a prototype detector consisting of 25 layers of 1-mm-thick lead converter and 5-mm-thick scintillator plates (Fig.~\ref{fig:nblind-1}). 
The detector has a size of $25\times 25$~cm$^2$ in the area and 16.7~cm in the length.    
We plan to irradiate a 1-GeV $\pi^-$ beam generated from the K1.8 beamline at J-PARC. 
We measure delayed coincidence signals coming from thermal neutrons with and without B$_4$C Mylar. 
With this experiment we can confirm the existence of captured $\gamma$-rays caused by pion-induced thermal neutrons and the elimination of them with B$_4$C Mylar. 

\subsection{Beam hole photon veto counter}
\label{sec:beamHolePhotonVetoCounter}
A beam-hole photon-veto (BHPV) counter detects a photon from the $\klpiopio$ decay
in the neutral beam, and reduces the background.
To operate it in the neutral beam,
it should be less sensitive to neutrons, and photons with energies
lower than those from the $\klpiopio$ decay.
In this section, we discuss the design and rate of the beam-hole photon-veto counter.

\subsubsection{The design of beam hole photon veto counter}
\begin{figure}[h]
 \includegraphics[width=0.9\textwidth]
 {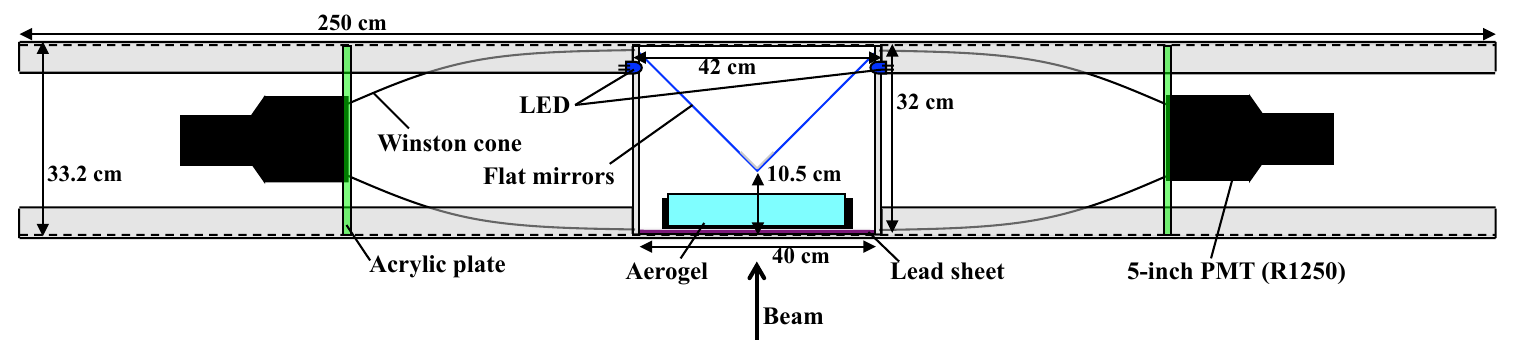}
 \caption{Structure of a module for the beam-hole photon-veto counter in KOTO step-1
 ~\cite{Maeda:2014pga}.}\label{fig:BHPVmodule}
\end{figure}

The beam-hole photon-veto counter consists of an array of a module
with a lead photon-converter and aerogel Cherenkov radiators.
The module structure which is the same as that in KOTO
is shown in Fig.~\ref{fig:BHPVmodule}.
A high-energy photon is converted to $e^+$ and $e^-$,
and these generate Cherenkov light in the aerogel radiator.
The Cherenkov light is guided by mirrors to a PMT 5-inch in diameter.
The Cherenkov light is generated less for slow particles
such as protons or charged pions from the hadronic interaction
of beam neutrons.  Several modules are lined up along the beam
and three-consecutive coincident hits in the modules are required,
which keep high efficiency against high energy photons while causing 
low energy photons and neutrons to be less sensitive.  
This is because electro-magnetic showers of high energy photons
develop in the forward direction while hadronic showers  develop 
isotropically and electro-magnetic showers of low energy photons cannot
develop in three or more modules.

\begin{figure}[h]
 \includegraphics[width=0.9\textwidth]
 {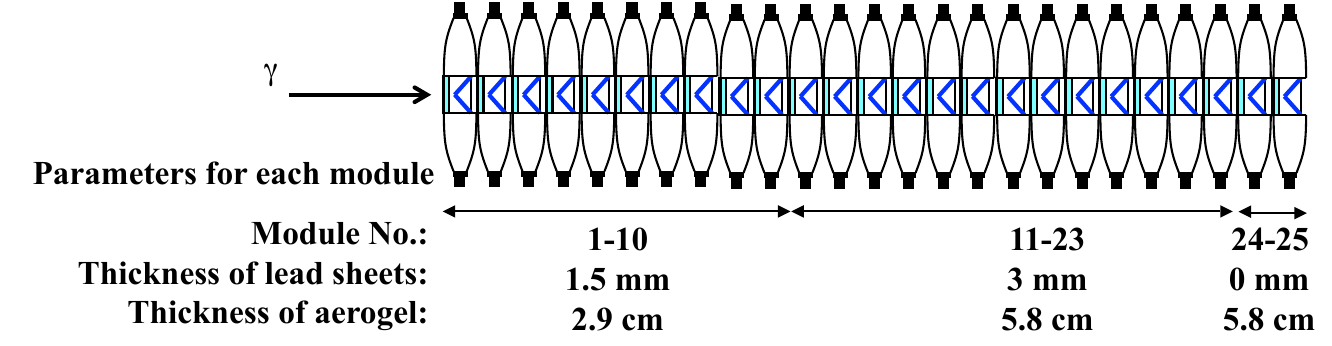}
 \caption{Configurations of the modules of the beam-hole photon-veto counter.}
 \label{fig:BHPVconfig}
\end{figure}

The configuration of the beam-hole photon-veto counter in KOTO~II is shown in  Fig~\ref{fig:BHPVconfig}.
The thickness of lead and aerogel sheets  is changed according to the location of the module 
to balance the hit rates among the modules. The total thickness of the lead sheets
is 54~mm, which corresponds to 9.6 $X_{0}$, to keep the photon punch-through inefficiency to be less than $10^{-3}$.

We evaluated the inefficiency of the beam-hole photon-veto counter against photons as a function of the incident energy by using a full-shower simulation
as already shown in Fig.~\ref{fig:bhpvinef}.
We can achieve $10^{-3}$ inefficiency for 2000~MeV photons with a threshold of 5.5 p.e.
which is considered as the default threshold.


\subsubsection{Expected counting rate and accidental loss in KOTO step-2}
A single module is sensitive to 1-MeV photons.
The incident rate of photons is 0.75~GHz for the energy larger than 1~MeV.
If the detection efficiency is $\sim 5$\%, the hit rate is $O(10)$~MHz.

\begin{figure}[h]
 \centering
 \includegraphics[page=26,width=0.45\textwidth]
 {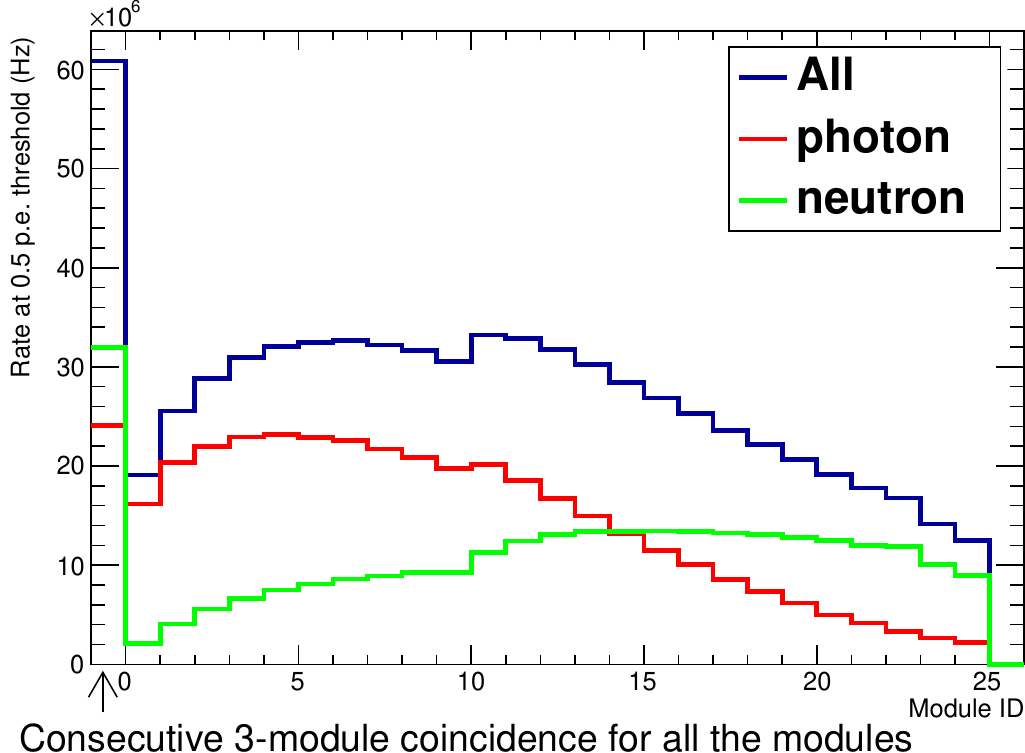}
 \caption{Hit rate of each module and counter (consecutive three-module coincidence)
 are shown with 0.5-p.e.-threshold. 
 The first bin of the histogram corresponds to the counter rate.}
 \label{fig:BHPVrate}
\end{figure}

We injected all the particles collected in the beam-line simulation
to the beam-hole photon-veto counter with 25 modules.
A full shower simulation and optical-photon tracking to the PMTs were performed,
and the observed number of photoelectrons was recorded.
For the module hit, ``OR'' of the PMTs on both sides is used
for a certain photoelectron (p.e.) threshold.
Consecutive three-module coincidence with a certain p.e. threshold
(counter rate) was also calculated.
Those module hit-rate and the consecutive three-module-coincidence rate (counter rate)
are already shown in Fig.~\ref{fig:bhpvrate} for the 5.5-p.e.\ threshold as a base design.
Those for the 0.5-p.e.\ threshold 
are shown in Fig.~\ref{fig:BHPVrate}.
Average module hit-rate in 2-sec spill is more than 30~MHz.
The counter rate at the default threshold is $35.2~\mathrm{MHz}$.
The resultant accidental loss is 19\% with the 6-ns veto window, which correspond to 
$\pm 5 \sigma$ of the timing resolution of 0.59~ns.
It was achieved with the beam-hole photon-veto
counter at KOTO by requiring the three-module coincidence with the 2.5-p.e.\ threshold~\cite{Maeda:2014pga}.

\subsubsection{PMT operation and expected waveforms}
\begin{figure}[h]
 \centering
 \subfloat[]{
 \includegraphics[page=36,width=0.45\textwidth]
 {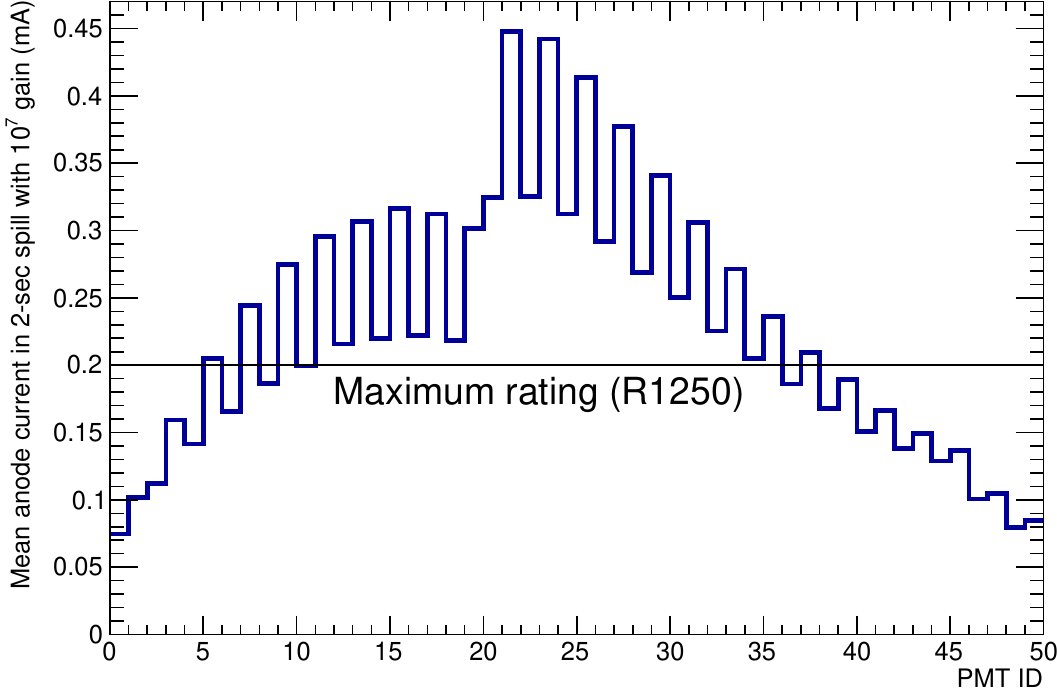}
 }
 \subfloat[]{
 \includegraphics[page=11,width=0.45\textwidth]
 {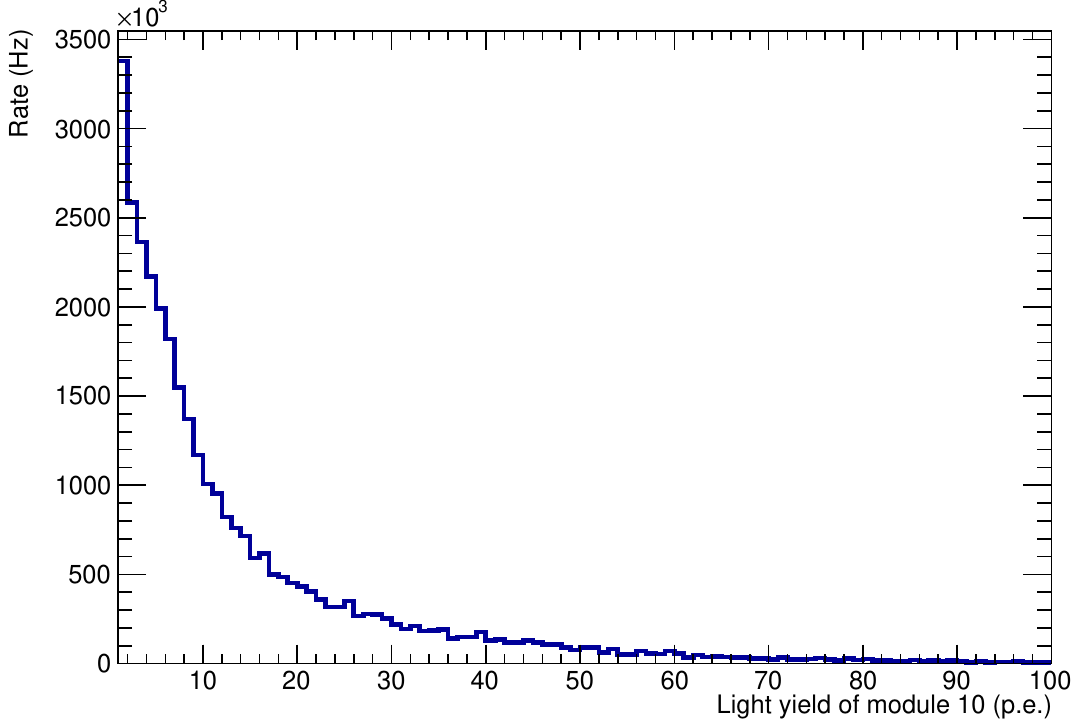}
 }
 \caption{(a)Expected average anode-current for each PMT
 with the gain of $10^7$. (b)Distribution of number of p.e. for a module (ID:10).}
 \label{fig:AnodeCurrent} 
\end{figure}

We evaluated the average PMT-anode current
in Fig.~\ref{fig:AnodeCurrent}(a) based on the p.e. yields
from the full simulation with the particles in the beam
(Module ID 10 for example in Fig.~\ref{fig:AnodeCurrent}(b)).
A $10^7$ gain of the PMT is assumed.
In more than half of PMTs, the anode currents exceed
the maximum rating (0.2~mA) of the current PMT
used in KOTO.
We may use a pre-amplifier to reduce the gain of the PMT
and keep the single-p.e.sensitivity.

\begin{figure}[h]
 \centering
 \subfloat[]{
 \includegraphics[page=101,width=0.32\textwidth]
 {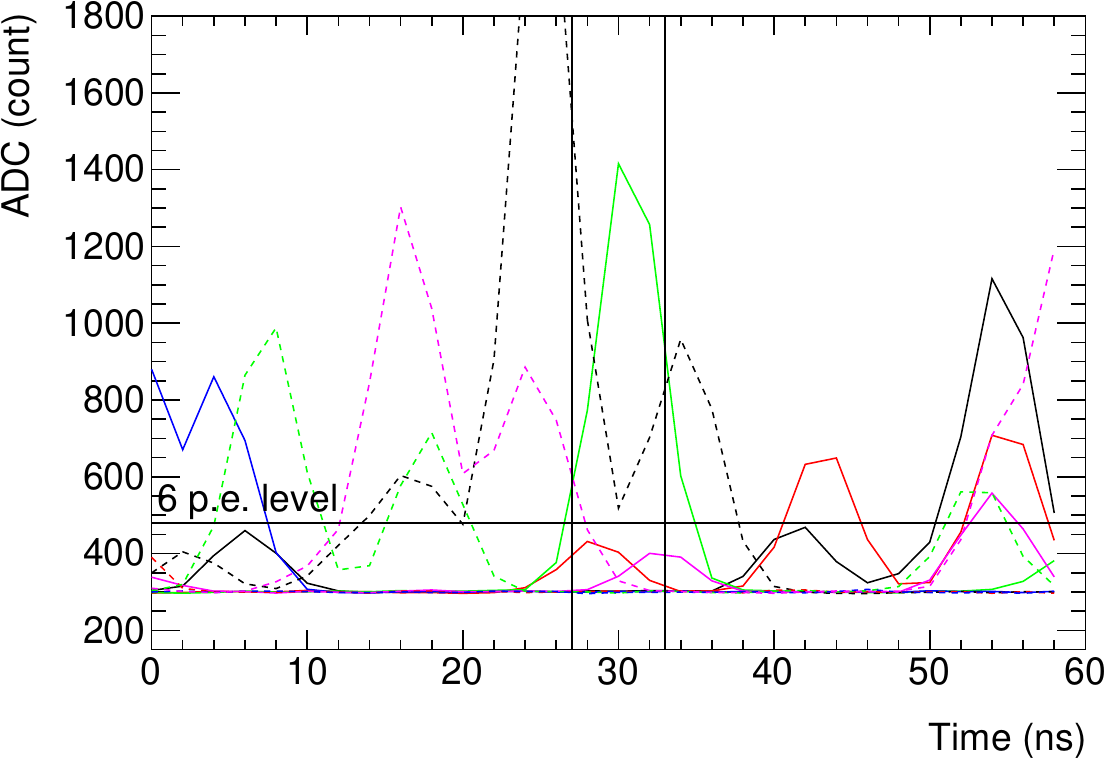}
 }
 \subfloat[]{
 \includegraphics[page=102,width=0.32\textwidth]
 {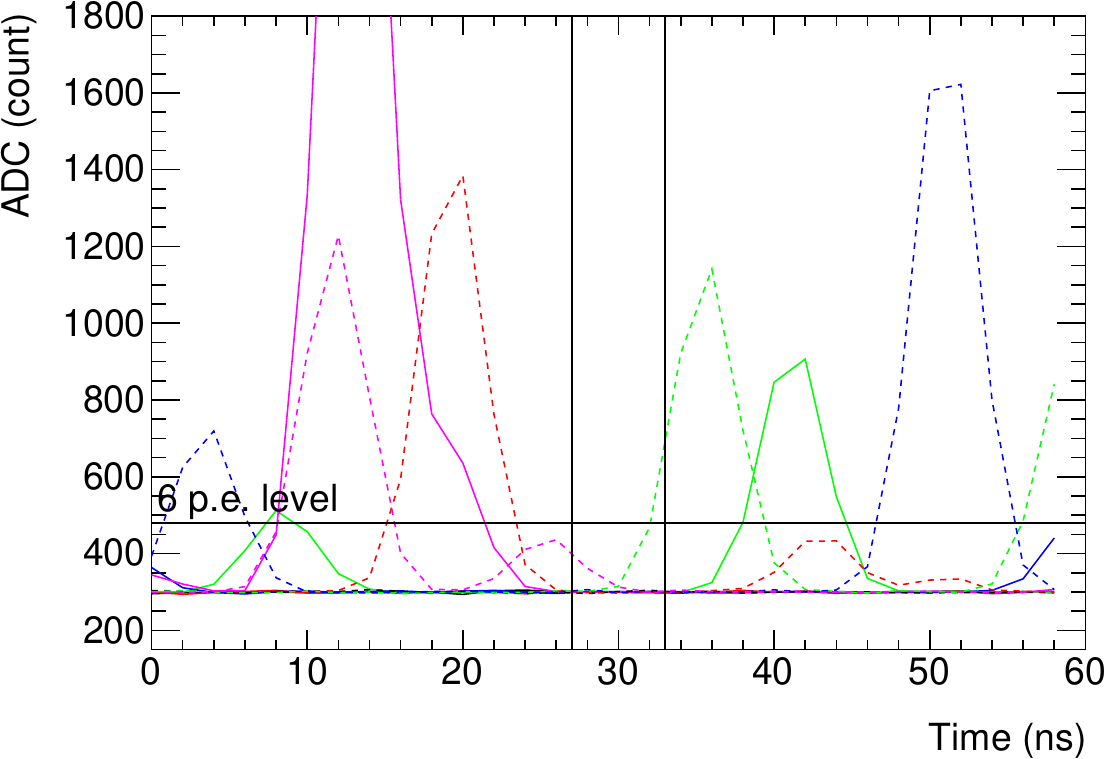}
 }
 \subfloat[]{
 \includegraphics[page=103,width=0.32\textwidth]
 {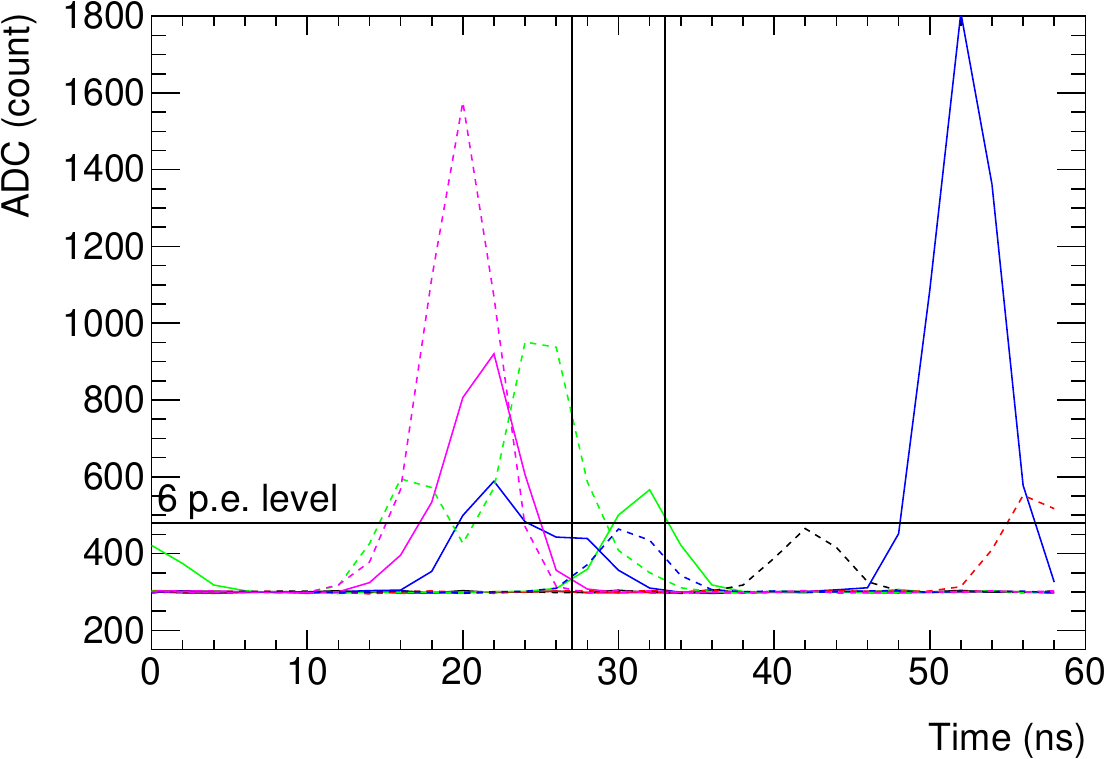}
 }

 \caption{Ten expected waveforms are overlaid in each panel. 
 Two vertical lines indicate the 6-ns veto window.
 A horizontal line shows 6-p.e.-level pulse height.}
 \label{fig:waveform}
\end{figure}

Expected waveforms after summing up the PMTs on both sides
are shown in Fig.~\ref{fig:waveform}.
In each panel, 10 waveforms are overlaid.
The waveforms do overlap, but peaks
can still be separated thus photons can be vetoed.
Thus, we can still operate the beam-hole photon-veto counter  in KOTO~II despite the harsh environment.


\subsubsection{Alternative design based on fast crystals}

The lead-aerogel Cherenkov design for the BHPV has proven to be successful in KOTO. However, for high-intensity operation at KOTO~II, it would be desirable to reduce the losses because of accidental vetoes from beam photons and neutrons. The principal limitation of the existing BHPV design is the time resolution of $\sim$600~ps, which necessitates a veto window of 6~ns. This in turn induces a 19\% loss of effective $K_L$ flux (Table~\ref{tab:rateWidth}). 

An alternative design that could be considered for use at KOTO~II is a compact crystal calorimeter with high-granularity transverse and longitudinal segmentation, along the lines of the small-angle calorimeter (SAC) proposed for the HIKE experiment~\cite{HIKE:2023ext}.
The conceptual design of the HIKE SAC is closely related to that for CRILIN, an independently proposed concept for calorimetry at a future muon collider~\cite{Ceravolo:2022rag}. Much R\&D work for the HIKE SAC and CRILIN projects has been carried out in common~\cite{Cantone:2023fac,Cantone:2024aix,Cantone:2024qbx}.

The proposed design for the HIKE SAC featured 4 layers of $10\times10$ ultrafast lead tungstate (PWO-UF) crystals, each $20\times20\times40$ mm$^3$ in dimension. PWO-UF is a recently developed formulation of lead tungstate with a decay time constant of 640 ps, good light yield, and high radiation tolerance~\cite{Korzhik:2022xln}. Because of the high density of lead tungstate, 3 layers of crystals provide a depth of 13.5$X_0$. This corresponds to only 0.6 nuclear interaction lengths, so about 55\% of beam neutrons will not interact in the SAC at all, and, based on extrapolations from HIKE, no more than 10-20\% of beam neutrons will leave signals in excess of the 100~MeV threshold for photon showers. The high granularity of the SAC (both transverse and longitudinal) provides abundant information for $n/\gamma$ discrimination offline, so that neutron interactions would not be expected to contribute significantly to the accidental veto rate.

Single PWO-UF $10\times10\times 40$~mm$^2$ crystals for this prototype coupled to silicon photomultipliers were tested with high-energy electrons at the CERN SPS in 2022 and 2023 and compared with lead fluoride (PbF$_2$) crystals~\cite{Cantone:2023fac}. A two-layer, $3\times3$ crystal prototype with PbF$_2$ crystals and silicon photomultiplier readout was tested with 450~MeV electrons at the Frascati BTF in 2023~\cite{Cantone:2024aix}. A photoelectron yield of $\sim$0.6/MeV was obtained with PWO-UF; extrapolating the results to the case of 100 MeV electrons, a time resolution of about 150~ps is expected. This is of the same order as the resolution for the determination of the event time from the KOTO II calorimeter, which opens the possibility of reducing the BHPV veto window to as low as 2.5~ns and cutting expected accidental veto losses from 19\% to 8\%. 

PWO-UF crystals were exposed to $\gamma$-rays from $^{60}$Co at the ENEA Calliope facility in 2023~\cite{Cantone:2024aix} and 2024~\cite{Cemmi:2024ojv}: no significant degradation in transparency from ionizing doses up to 2100~kGy was observed.

Simulation studies to adapt the design to KOTO II are underway, and a new prototype is under construction with full size crystals and readout with Hamamatsu R9880 photomultiplier tubes, which are expected to confer better radiation robustness and faster time resolution than silicon photomultipliers.



\subsection{Beam Hole Charge Veto}
The forward charged-particle veto in KOTO II sits in the beam, in the region downstream of the main calorimeter, and should veto charged particles that would otherwise escape and create a potential background to $K_L \rightarrow \pi^0 \nu\nu$ analysis. This veto is followed by a neutral-particle veto (see previous section) with an analogous role. However, the material budget should be kept low to reduce hits from photons and neutrons in the neutral veto. 

In KOTO, the Beam Hole Charge Veto is a Thin-Gap Chamber of dimensions $30 \times 30$ cm$^2$ operating in air, with 3 layers used with a ``2-out-of-3'' logic giving 99.5\% efficiency for the veto. A layer has a 2.8 mm thick gas volume and two 50-$\mu$m thick Kapton cathode foils coated with graphite, equivalent to about 0.03 \% $X_{0}$. 
The granularity is 2 $\times$ 30 cm. The veto has a time jitter of about 20 ns and is used with a time window of 30 ns, which results in an 8.3\% random veto.

In KOTO II, the default design could be replaced by a silicon pixel detector comprising 2 planes of 200 $\mu$m thickness each.
The level of radiation in that location is $10^{13}$ 1 MeV neutron-equivalent per cm$^2$, which is a modest exposure for modern silicon detectors. In the KOTO~II charged-particle veto, a charged particle hit rate of 0.2 MHz is expected (and a hit rate of 40 MHz in total including photons and neutrons): such a rate is compatible with various existing technological solutions. The geometrical configuration can be made to ensure a high detection efficiency ($>99.5\%$). An improved time resolution of 1 ns or better is also achievable, allowing the time window to be narrowed to at most 5 ns and therefore reducing the random veto.
Preliminary simulation studies validate this solution showing that the random veto (signal loss) will be at most less than 4\%.

Other technological silicon options can be considered.
Thinner silicon sensors 
(\SI{50}{\micro m} thickness corresponding to 0.054\% $X_0$) 
such as Monolithic Active Pixel Sensors (MAPS)~\cite{ NA60:2022sze,The:2890181} could be used to reduce the amount of material.
For faster timing, 
Low-Gain Avalanche Diode (LGAD) sensors could be considered
~\cite{Butler:2019rpu,CERN-LHCC-2020-007}.
Both types of sensors are already in use elsewhere or under a mature stage of R\&D.

\subsection{Additional tracking in the forward region}
\label{feas:additionaltracking}

While the default KOTO~II setup, as described in this document, is largely optimized for detection of neutral particles, it is possible to envisage additional detectors, particularly with the aim of detecting charged particles.

\begin{figure}
\centering    \includegraphics[width=0.5\linewidth]{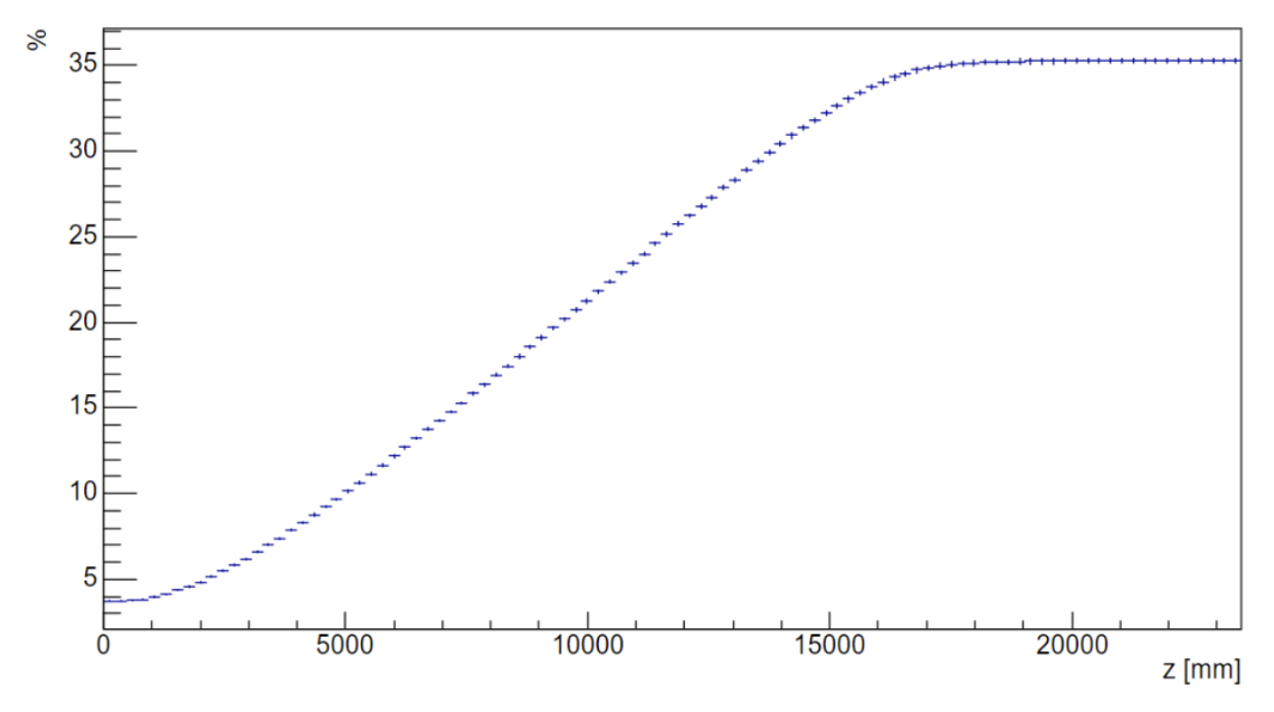}
\caption{Estimated acceptance for $K_L\to\pi^0e^+e^-$ decays in percent versus barrel instrumentation ($z=0$ is at the surface of the main calorimeter).}
\label{fig:barrel}
\end{figure}

These additional detectors would help to increase the sensitivity for $K_L\to \pi^0\ell^+\ell^-$ decays, and more generally for $K_L$ decays with charged particle tracks.
Preliminary studies show that the acceptance for electron tracks from $K_L\to \pi^0e^+e^-$ decays would be significantly enhanced, from about 5\% to a maximum of about 35\% in presence of tracking that covers the forward region in front of the calorimeter and the last 16~m of the barrel region (Figure~\ref{fig:barrel}). The signal acceptance already includes kinematical cuts against the Greenlee background and the condition $m_{ee} > 140$~MeV/$c^2$, which together account for a reduction from 100\% to 55\%; a further 10\% is lost in the forward beam hole region. 

Two possible approaches for the detector setup in the forward region can be considered.

The first approach implies that additional detectors can satisfy the stringent constraints on extra material in front of the calorimetry system dictated by the $K_L\to\pi^0\nu\bar\nu$ analysis, while maintaining their advantage for charged particles; in this approach, data can be taken simultaneously for all the $K_L$ decays of interest. In this case, the functionality of the Charged Veto Counter and of the forward-region tracking could be combined, altogether consisting of two planes of 3~m diameter located just upstream of the main calorimeter. The material thickness must be kept very low in this case (in KOTO, the thickness of the Charged Veto Counter is 3~mm or 1\%$X_0$, with a granularity of 7~cm and a time resolution of 2~ns).

The second approach involves dividing the data taking into two separate stages, dedicated to $\pi^0\nu\bar\nu$ and $\pi^0\ell^+\ell^-$ respectively. In this case, the requirements for the sub-detectors that serve only one stage can be neatly separated and optimized differently, while the sub-detectors serving both stages would still be optimized simultaneously and remain the same.

Studies are on-going to establish the optimal approach that maximize the physics return and minimizes cost. Possible technical solutions able to provide sufficient time resolution, efficiency and granularity include: a light-material tracking device in vacuum, based on an evolution from the Straw Tracker in the NA62 experiment at CERN~\cite{NA62:Det, HIKE:2023ext}; and a hybrid tracking system like the Mighty Tracker from the LHCb Upgrade2 experiment at CERN, comprising silicon pixels in the inner region and scintillating fibers in the outer region~\cite{LHCb:Mighty}.
The former would have the advantage of reducing the material budget to $0.5 X_0$ or less, while the latter would allow for a time resolution of the order of hundred(s) ps.

\subsection{Central Barrel Counter with Tracking}
\label{feas:centralbarreltracking}

The design of the Central Barrel Counter is based on the experience with the KOTO experiment and is planned to be a lead-scintillator sandwich calorimeter with 1~mm thick lead plates interleaved with 5~mm thick scintillator plates. As shown above (Sec.~\ref{det:inefficiency}), such a design is sufficient for vetoing photons down to energies of 1~MeV, which is necessary for the $\klpionn$ measurement. 

However, with little effort, parts of the Central Barrel Counter could be used as a simple tracking device to be able to extend the acceptance for $\klpioll$ and in particular $K_L\to\pi^0e^+e^-$. Mostly, the $K_L\to\pi^0 e^+ e^-$ decay products are strongly pointing into the forward direction, but, since this is a four-body decay, the acceptance would considerably increase when using the final part of the Barrel Counter in addition to the main calorimeter in the end cap (see Sec.~\ref{feas:additionaltracking}).
In considering the Barrel Counter for use in electron and photon detection and tracking, it should first be noted that in its planned design the Barrel Counter will have an excellent energy resolution with a stochastic term of $< 10\,\%/\sqrt{E\, \text{[GeV]}}$. Second, for use as a tracking device, the end part of the Barrel Counter can be finely segmented in scintillating strips of 1~cm width, both in the $z$ and $\varphi$ directions, to obtain an angular resolution of 20~mrad or better on electron and photon showers~\cite{ref:PointingCalorimeter}. Alternatively, layers of scintillating fibers (SciFi) could be used to measure shower directions with high precision. Finally, a rough distinction between electron and photon showers is achieved using the first scintillating layers, where electrons leave MIP signals but photons (with energies of {\cal O}(GeV) only convert with a $10\,\%$ probability.
\subsection{In-beam Upstream Charged Veto}
The charged kaon background
typically
occurs when a charged kaon from the beam line undergoes $K^\pm\to \pi^0 e^\pm \nu$ decay, and the $e^\pm$ goes undetected.
The number of such background events
can be reduced to be 4.8
with two magnets in the beam line
and the efficient detection of the $e^\pm$.
Therefore, charged particle detectors line the inside of the Barrel Counters in the baseline design.

\begin{figure}[h]
\centering
\subfloat[]{
\includegraphics[width=0.54\textwidth]{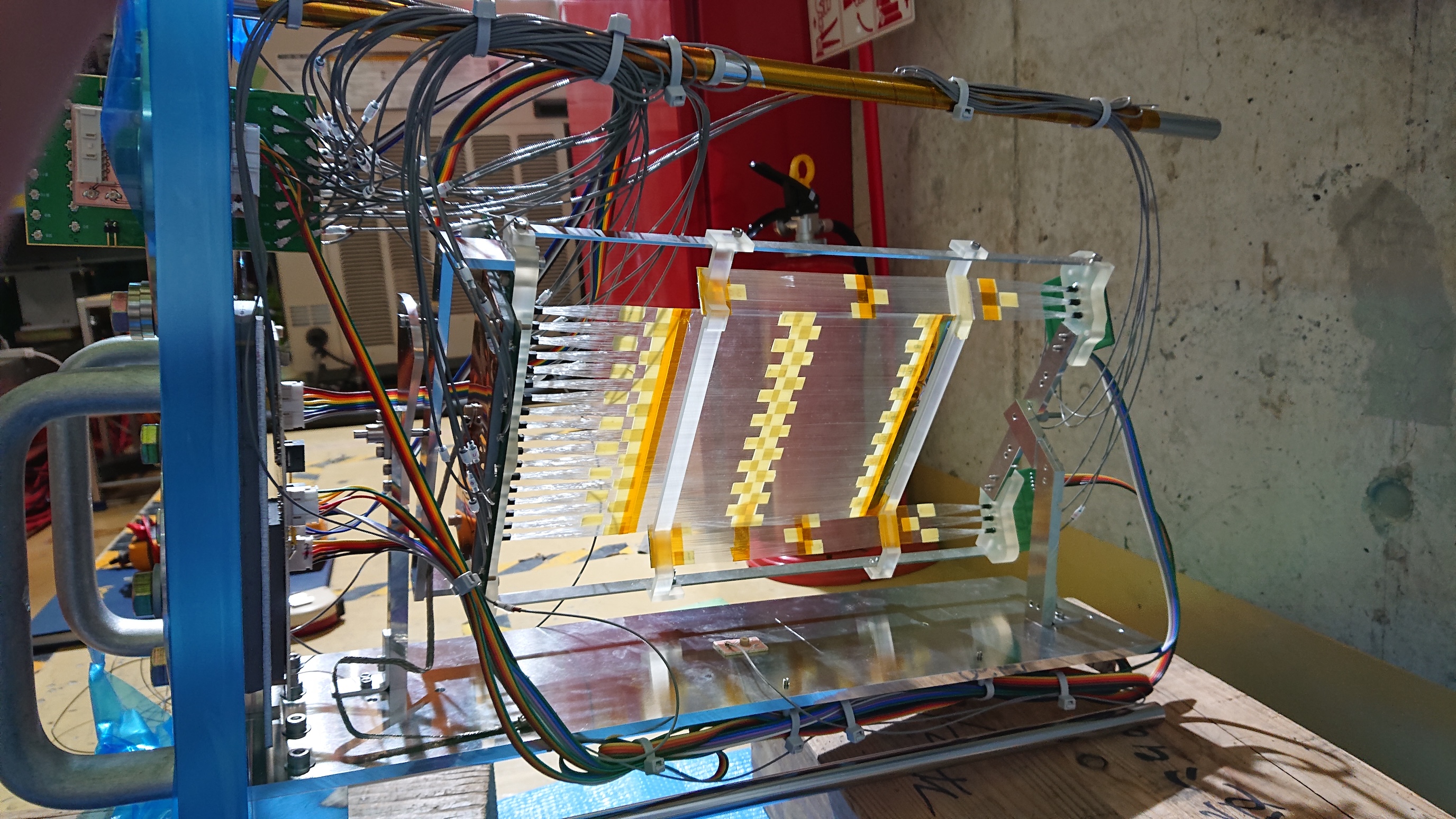}
}
\subfloat[]{
\includegraphics[width=0.42\textwidth]{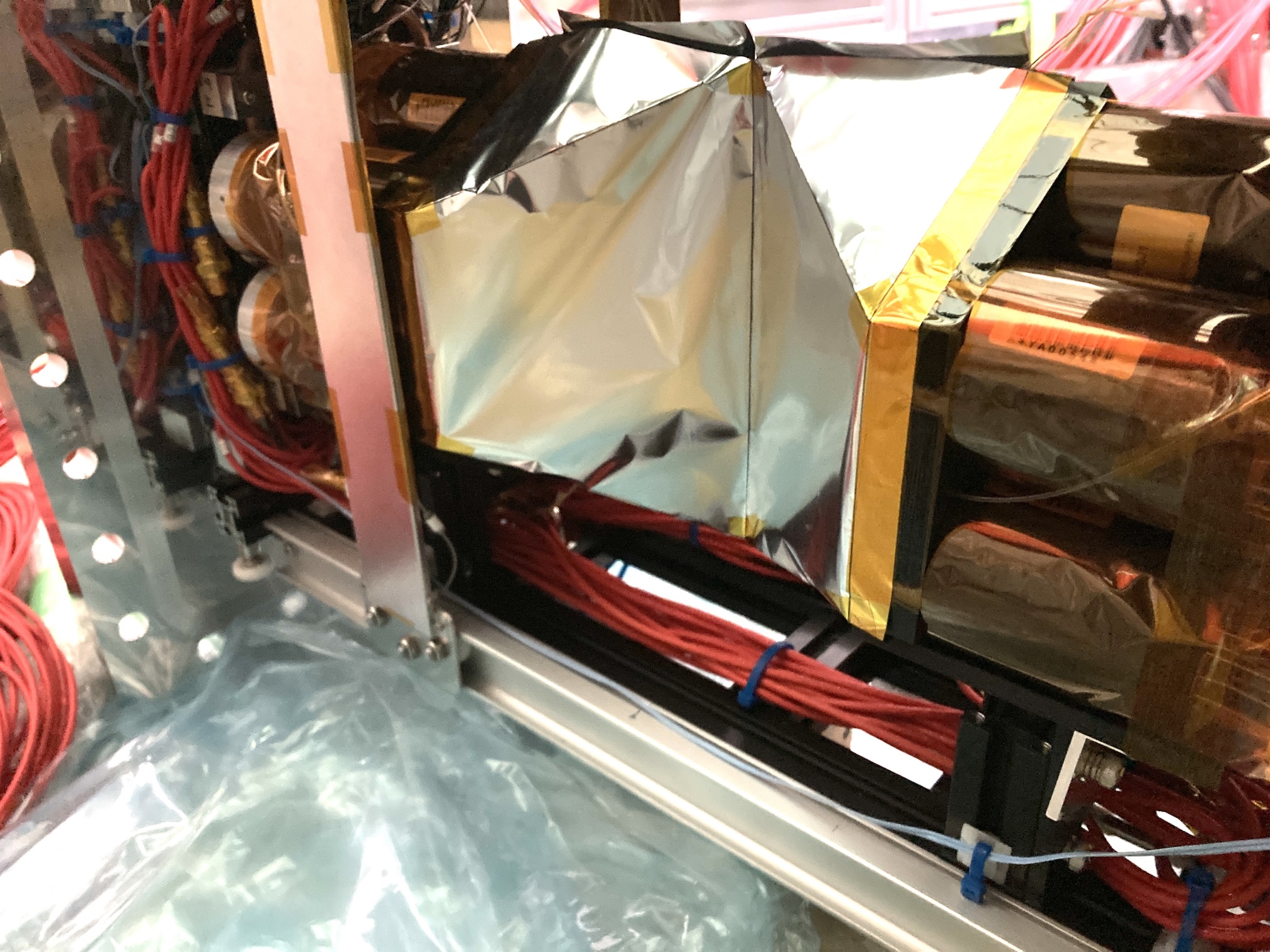}
}
\caption{Upstream Charged Veto (UCV) in KOTO.
(a) First version with 0.5-mm-square scintillating fibers.
(b) Second version with 0.2-mm-thick scintillator film surrounded by a \SI{12}{\micro m}-thick aluminized mylar foil reflector.
\label{fig:ucv}}
\end{figure}

This background can be further suppressed if charged kaons at the entrance of the KOTO~II detector can be detected. This requires a charged-veto counter in the beam that is thin enough to minimize interactions with neutral particles (photons, neutrons, and $K_L$s).
An upstream charged veto counter (UCV)
shown in Fig.~\ref{fig:ucv} was
installed in the KOTO detector
for this purpose.

The first version (Fig.~\ref{fig:ucv} (a) ) 
consists of 0.5-mm square scintillating fibers.
The fiber plane is tilted to reduce the gaps in the sensitive area
from the fiber cladding.

The second version (Fig.~\ref{fig:ucv} (b) )
is thinner and 
consists of a 0.2-mm thick scintillator film surrounded by a reflector of \SI{12}{\micro m} thick aluminized mylar.
The scintillation light escaping out from the surface of the scintillation film
is collected with the aluminized reflector and directed
to PMTs.
The obtained light yield is 18.5 p.e.\ for minimum ionizing particles (MIPs)
and the detection efficiency is 99.9\% with a threshold corresponding to 0.4 MIP.

\subsection{Design of vacuum tank}
\begin{figure}[h]
\includegraphics[width=0.48\textwidth]{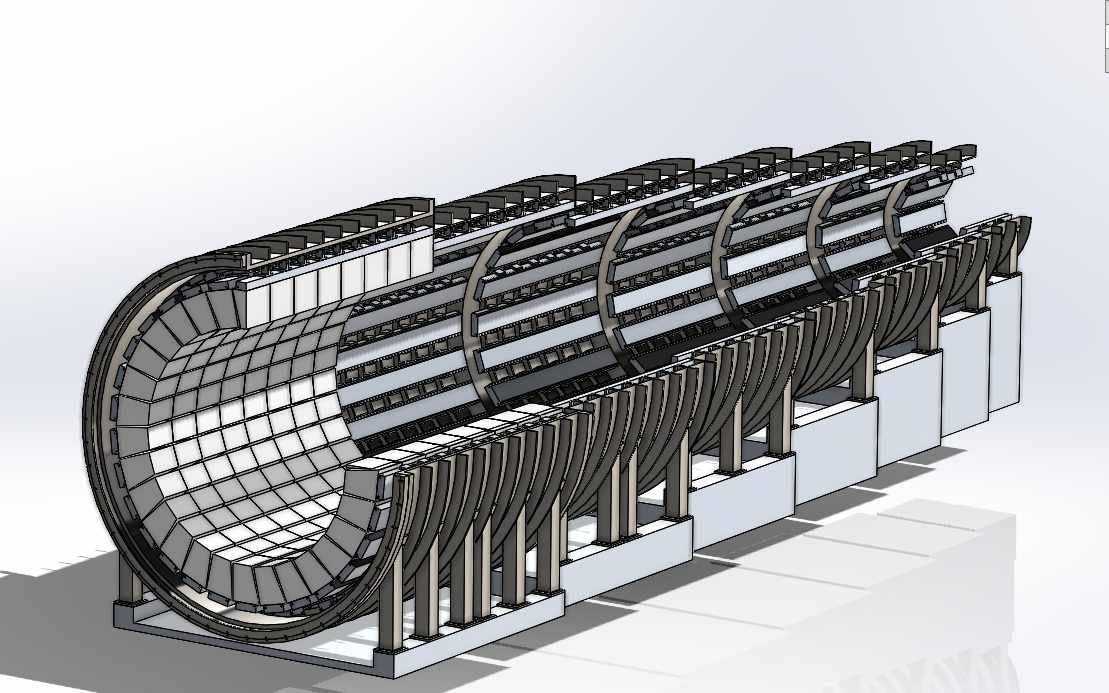}
\includegraphics[width=0.48\textwidth]{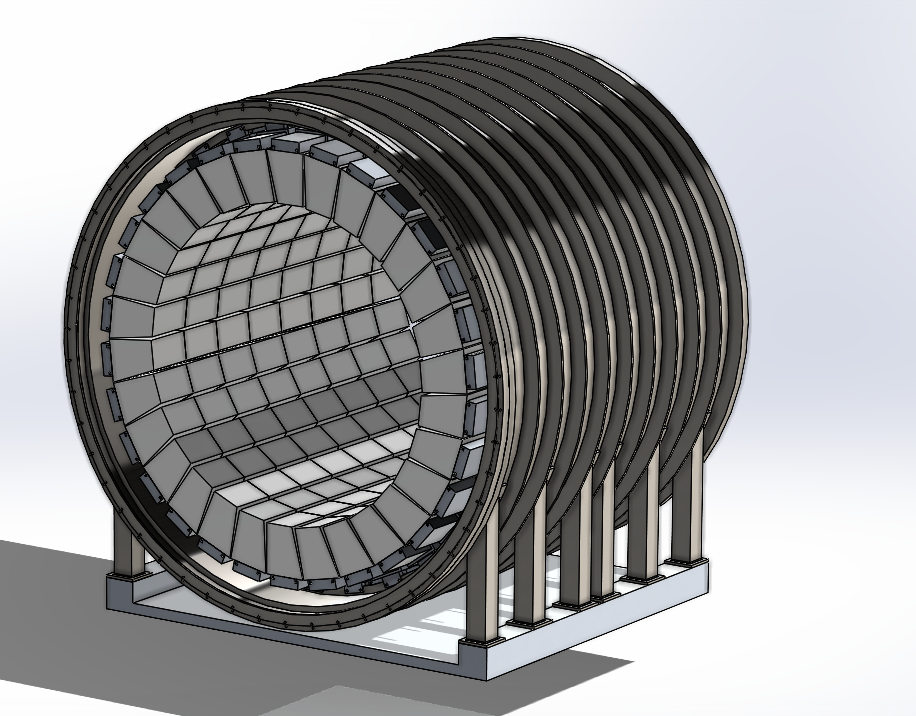}
\caption{Vacuum tank under design.\label{fig:vacuumTank}}
\end{figure}

The vacuum tank with a structure to hold the Barrel Counter
is under design with an engineer team.
The current version is shown in Fig.~\ref{fig:vacuumTank}.
The thickness of the wall, reinforcement on the wall,  
and a system to hold the modules are studied by considering the mechanical strength and stability.
The vacuum tank should be moved in z direction 
to open the tank in the installation and maintenance.
The shape and the arrangement of the Barrel Counter modules
should be designed to have no gap for the detection of photons.  
The module shape and its support system are also related to the installation scenario.
Such a module shape, an installation scenario, and a maintenance scenario  will be studied more.

\clearpage
\section{Trigger and DAQ}
The KOTO~II DAQ system design is based on the KOTO experience: The analog inputs are continuously digitized and pipelined in the on-chip memory, regardless of the triggers. The size of the memory governs the depth of the pipeline and, in turn, determines the time allowed for the trigger evaluation. The major advantages of such a design are the following: The trigger evaluation is more accurate because the calculation is based on the digitized data. The baseline fluctuation is smaller due to the elimination of long delay cables.  

\subsection{Digitizer}
\begin{figure}[h]
    \centering
    \includegraphics[width=0.5\linewidth]{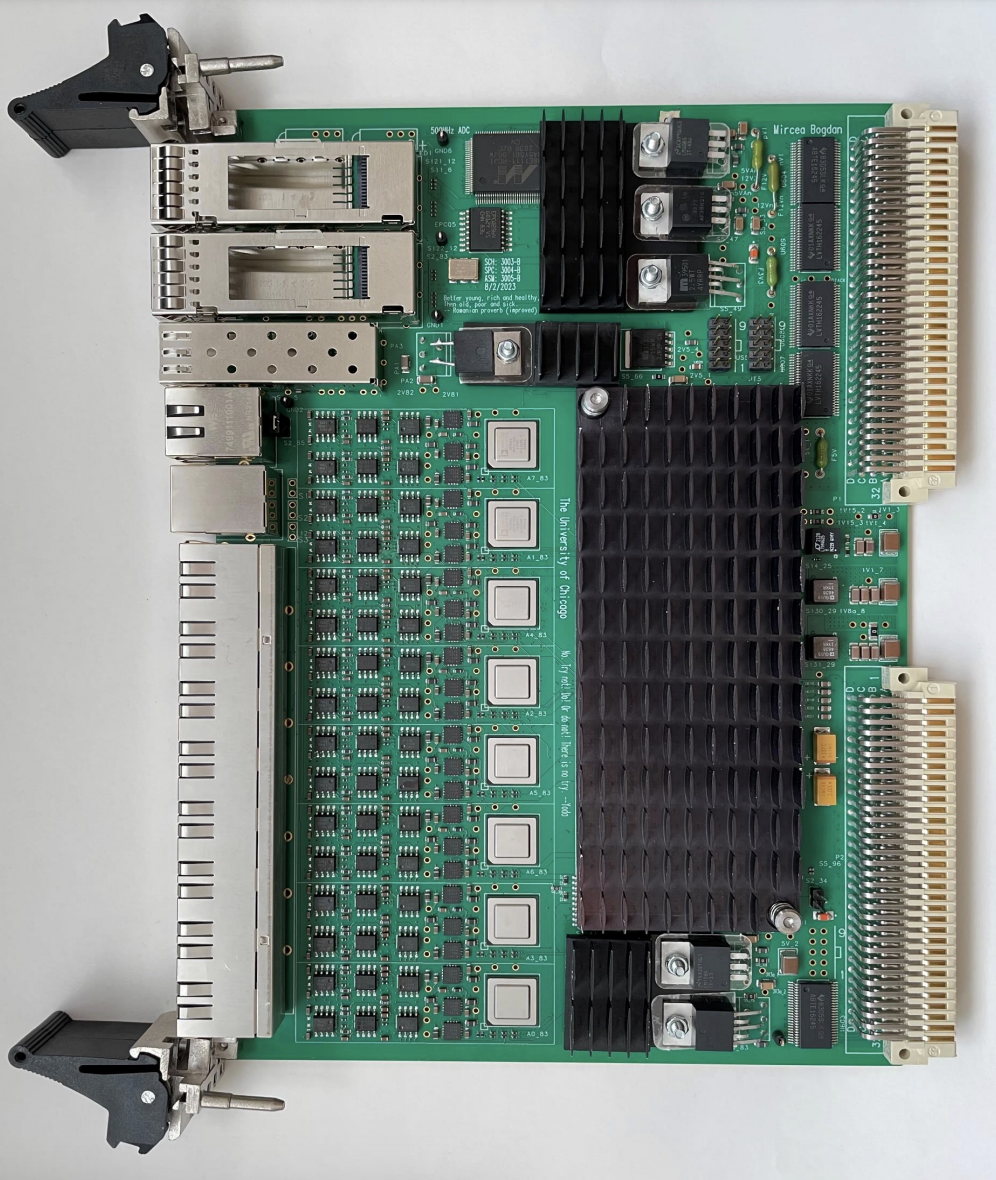}
    \caption{Prototype of a 14-bit 500-MHz digitizer for the KOTO~II experiment.}
    \label{fig:koto2_adc}
\end{figure}

\begin{figure}[h]
    \centering
    \includegraphics[width=0.95\linewidth]{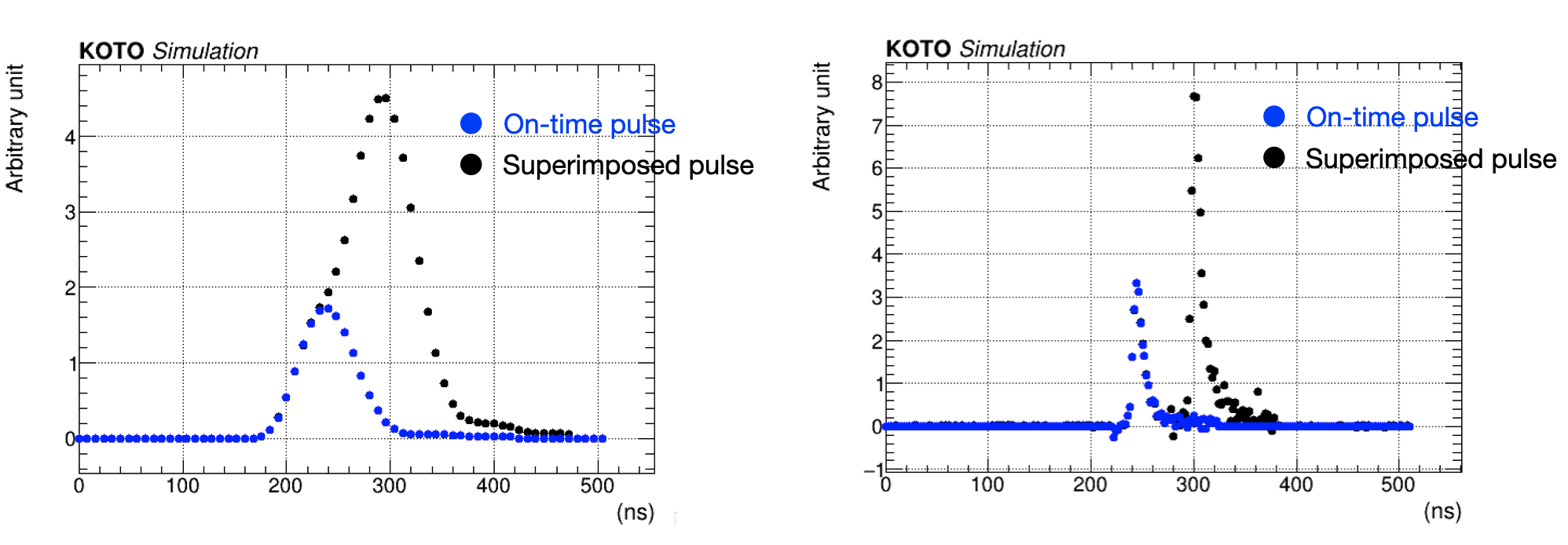}
    \caption{Pulse shape demonstration of a 125-MHz digitizer for KOTO (left plot) and a 500-MHz digitizer for KOTO~II (right plot).}
    \label{fig:pulse_shape_compare}
\end{figure}

\begin{figure}[h]
    \centering
    \includegraphics[width=0.6\linewidth]{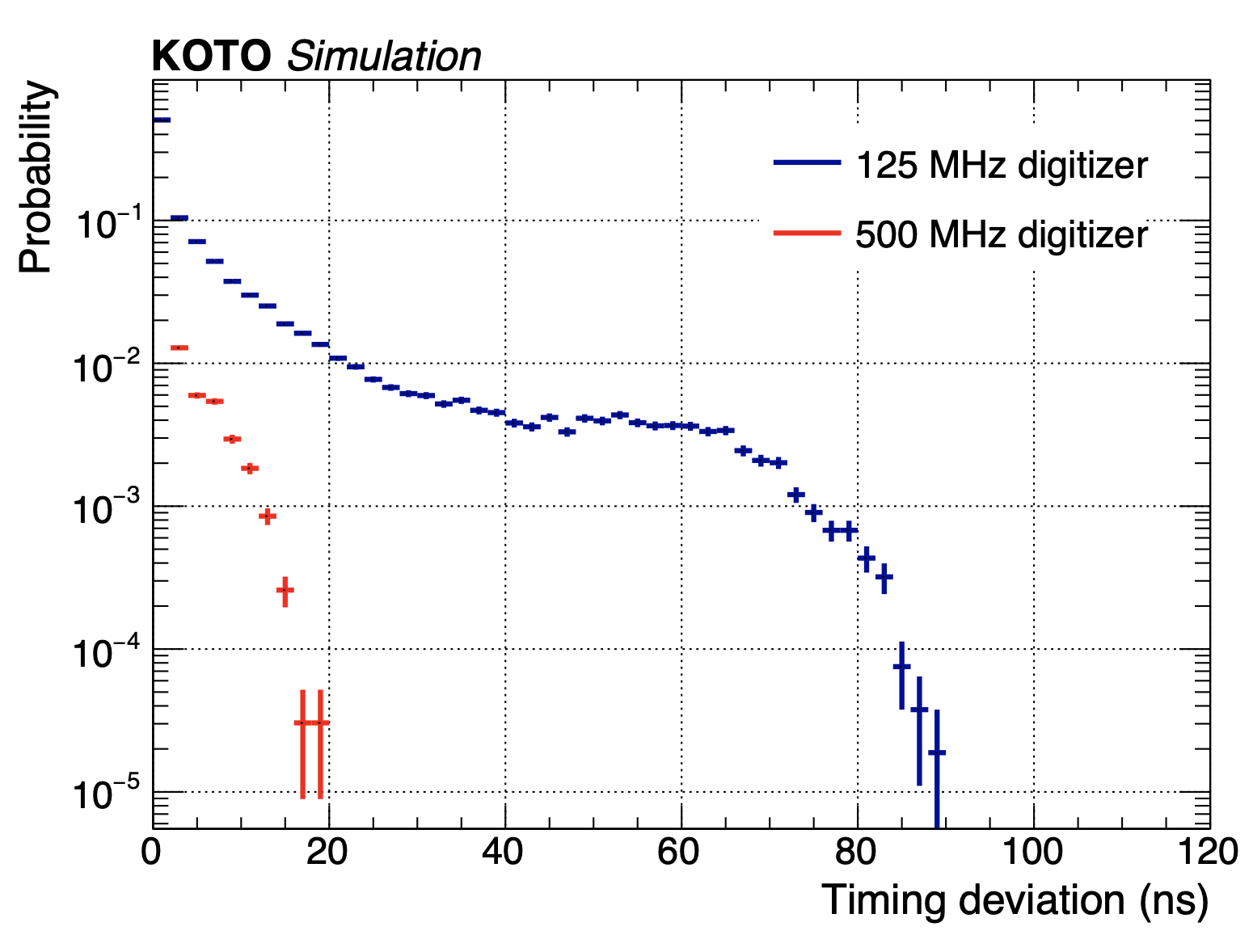}
    \caption{Timing deviation due to an overlapped pulse versus the probability. This is modeled based on the energy and timing distributions of both on-time veto hits and accidental hits from data.}
    \label{fig:deltaT_compare}
\end{figure}

The 14-bit 500-MHz homemade waveform digitizers, as shown in Figure~\ref{fig:koto2_adc}, will be utilized to collect the signals from all detector components. In KOTO, in order to measure the timing with a 125-MHz digitizer, a pulse shaper was introduced to slow down the rising edge. The 500-MHz digitizers do not include the shaper and the waveforms are therefore narrower. Figure~\ref{fig:pulse_shape_compare} demonstrates the pulse shape comparison, and Figure~\ref{fig:deltaT_compare} shows the quantitative improvement from a 125-MHz digitizer to a 500-MHz one. No large timing deviation due to an overlapped pulse is expected.

This digitizer is also equipped with high-speed optical links, which are primarily used for trigger processing. Compared with KOTO digitizers, the speed is improved by a factor of 8. This feature is expected to completely release the \SI{0.16}{\micro s} system dead time for online clustering, and capable of collecting more data at the front-end. 

The file size of a KOTO~II event is estimated to be 0.2~MB, inferred from the following assumptions: The number of readout channels is estimated to be 8000, which is doubled of the KOTO. The number of samples recorded toward a trigger is 128 (256-ns window), which is half of the KOTO. The file size can be compressed by a factor of 10 if an aggressive noise suppression algorithm is applied to the KOTO~II data. 

If a normal KOTO~II operation generates the data rate of 0.2~PB/(month = 25 days), the final trigger rate of 2 k/(spill = 4 sec) has to be managed. A multi-stage trigger is considered to reduce the rate. 

\subsection{Trigger and Data Transfer}
\begin{figure}[h]
    \centering
    \includegraphics[width=0.8\linewidth]{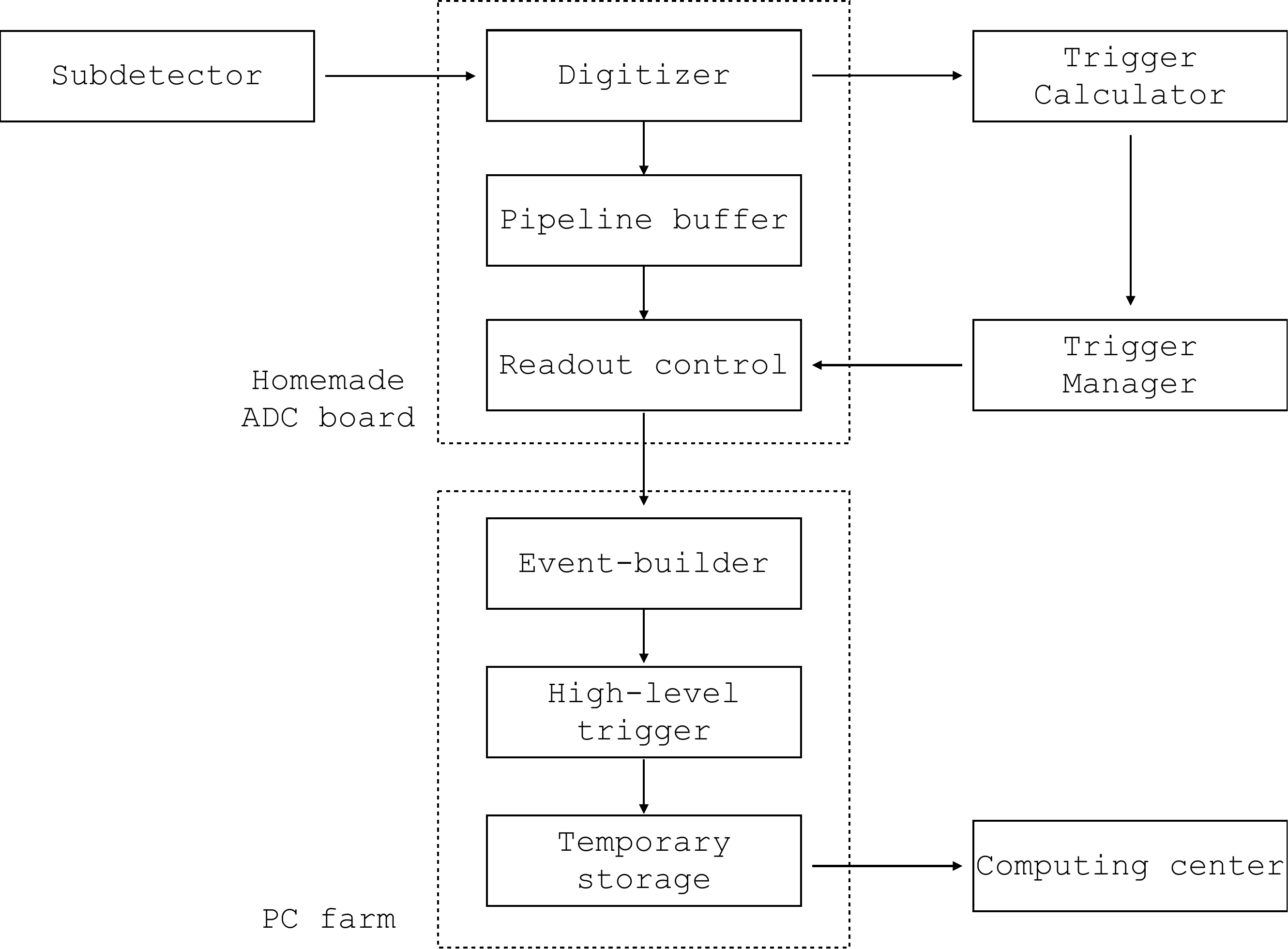}
    \caption{Schematics of the KOTO~II DAQ system.}
    \label{fig:koto2_daq_architecture}
\end{figure}

Figure~\ref{fig:koto2_daq_architecture} shows the architecture of the KOTO~II DAQ system. All the signals from the KOTO~II detector are connected to the homemade ADC boards. After the digitization, the energy and timing of each readout channel are calculated at the on-board FPGA and sent to the trigger calculator via optical links. The trigger manager collects all the trigger calculation results and perform the trigger decisions. 

The PC farm receives the packages from ADC boards and builds events. A high-level trigger can be subsequently evaluated with complete events. Events are then compressed, buffered in the temporary storage, and sent to the computing center for a long-term storage.  

The trigger system is categorized into two parts: electronics-based triggers and PC-based triggers. 

\begin{itemize}
    \item \textbf{Electronics-based Trigger}. The level-1 trigger is determined based on the energy sum of the calorimeter (total energy, ET), the number of hits at each veto counter, the number of clusters in the calorimeter, and the distance between the center of energy and the beam axis (RCOE). These variables are evaluated in parallel every system clock of 8~ns. A $K_L^0 \to \pi^0 \nu \overline{\nu}$ candidate event should have ET larger than a certain threshold, absence of hits in veto counters, two clusters in the calorimeter, and large RCOE. The target is to reduce the rate at few tens kHz at this stage. 

    The level-2 trigger is processed after the level-1 trigger is determined. The cluster position, cluster distance, or cluster energy are the candidates to be calculated at this stage. Alternatively, the FPGA-based AI triggers may be performed. The target is to further reduce the rate at 10 kHz at this stage.

    \item \textbf{PC-based Trigger}. The high-level trigger is processed after the data received by the PC farm. Besides the event-building and the data compression, a more sophisticated trigger can be processed, including the selection based on the reconstructed kinematic variables, the cluster shapes in the calorimeter, etc. The target of the final trigger rate is few kHz. Eventually, the noises will be highly suppressed in the data to reduce the file size and reduce the burden of the data transfer between the PC farm and the computing center.

\end{itemize}


\section{Cost}
\subsection{Detector cost}
The 3-m diameter KOTO~II calorimeter
requires new undoped CsI crystals in addition to the existing ones in KOTO.
The necessary number of new modules is 1456
if the cell cross-section is $\SI{5}{cm}\times \SI{5}{cm}$.
A new detector and a new vacuum tank
are needed for the longer Barrel Counter.
For the Collar Counters and Beam Hole Photon Veto,
the existing counters 
in KOTO may be used
with some modifications 
or with the addition of more modules.
For the baseline design, with undoped CsI crystals for the whole calorimeter,
the cost of the KOTO~II detector is evaluated
to be 23.4 M\$,
as shown in Table.~\ref{tab:costbase}
\footnote{The US Dollar-Yen rate used here is $\mathrm{\$}1=140~\mathrm{Yen}$.}.
\begin{table}[h]
\centering
\caption{Cost table for the detector base design.\label{tab:costbase}}
\begin{tabular}{lllll} \hline
Detector & Sum &Breakdown&Numbers&Cost \\
     & (M\$)    &         &        &(M\$)\\\hline
Endcap &&  Crystals &1456 &8.65\\
       &&  Outer veto &92&0.92\\
&&Additional PMTs & 244+1456+92 &0.90\\
&& HV supply &244+1456+92  &0.34\\
&&MPPC & 10944&0.43\\
&10.1&PreAmp& 4416&0.03\\ \hline
Charged Veto &&Scintillator  & &0.023 \\
&&MPPC & &0.031 \\
&0.07&Support & &0.014\\ \hline 
Barrel Veto& &Lead & 135 layers $\times$ 688 modules &0.26\\
&&Scintillator&135 layers $\times$ 688 modules &1.94\\
&&WLS fiber& 1700-4000 fibers/module&3.77\\
&&Fabrication &&1.0\\
&&Module assembly & & 0.05\\
&&PMT &$688 \times 2$ & 0.49\\
&7.7&Labor &  &0.21        \\\hline
Beam Hole& &Lead &22 &  \\
Photon Veto& &Aerogel &$15~\mathrm{cm}\times 15~\mathrm{cm} \times 3~\mathrm{cm}$: 72 tiles & 0.027\\
& &Frame&9 modules &0.002\\ 
&0.029 &PMT etc & existing& 0 \\ \hline 
Beam Hole & & & &  \\
Charged Veto &0.011 & &3 modules &0.011  \\ \hline 
Others &0.1 &Reuse / Fabrication & &0.1 \\ \hline
Vacuum tank&3&& &3 \\ \hline
Readout &&ADC &9200 channels &2.3\\ 
&2.4&Trigger &40 modules & 0.13\\\hline
Sum&23.4&&&\\
\hline
\end{tabular}
\end{table}

We are considering several alternative options, upgrades
or additional functionalities, as discussed in Section~\ref{sec:dfeasibility}.
The cost of shashlyk counters $\SI{5}{cm}\times \SI{5}{cm}$
in cross section
with 500 layers of 0.275-mm thick lead and 
1.5-mm thick plastic-scintillator plates ($26.4~X_0$)
is considered instead of the additional crystals in the calorimeter.
The cost of the additional $5X_0$ layers of pre-shower counters
for the photon angle measurement is also considered.
The cost of the Beam Hole Photon Veto (BHPV) design with 450 ultrafast lead tungstate (PWO-UF) crystals of dimension $20\times 20\times 40~\mathrm{mm}^3$
is considered.
The cost of the Beam Hole Charged Veto (BHCV) with two layers of 
\SI{200}{\micro m} thick silicon of area
$25~\mathrm{cm}\times 25~\mathrm{cm}$
is considered.
These cost scenarios are shown in Table.~\ref{tab:costalt}.
The option with shashlyk counters 
for the additional calorimeter region
reduces the cost to  $14.2$ M\$.

The cost for the additional tracking and other possible modifications to the barrel region for $K_L\to \pi^0\ell^+\ell^-$ will be considered later, 
and detectors can be prepared in the earlier phase of KOTO~II.

\begin{table}[h]
\centering
\caption{Cost for alternative design or additional function.
\label{tab:costalt}}
\begin{tabular}{l|ll|ll}
Detector &Baseline design&Cost&Alternative design/ & Cost\\
& & (M\$)&Additional function &(M\$)\\ \hline
Endcap &undoped CsI crystals&10.1& Shashlyk counters&0.92 \\
       &                    &    &Additional photon-angle &4.1\\
       &                    &    & \,\,pre-shower counter  & \\
BHPV &Lead/Aerogel&0.029& Lead tungstate crystal & 0.68\\
BHCV&Gas wire chamber&0.011 & Semiconductor & 0.30
\end{tabular}
\end{table}

\subsection{Beam line}
The cost for the beam line is shown in Table.~\ref{tab:costbeamline}.
The existing one magnet, one power supply, and one beam plug are reused.
\begin{table}[h]
\centering
\caption{Cost table for the beam line\label{tab:costbeamline}}
\begin{tabular}{lllll} \hline
Component & Cost (M\$) &&& \\
Magnet &  0.7   &&&  \\
Power supply& 0.2    &&&  \\      
Collimator& 0.79 &&&  \\      
Absorber  & 0.05 &&&  \\      
Beam plug  & 0.11&&&  \\
Vacuum & 0.5  &&&  \\      
Shield &    2 &&&  \\
Others &  0.58&&&\\ \hline
Sum    &   4.93   &&&
\end{tabular}
\end{table}

\section{Schedule}
Given that it may take time to prepare the KOTO II beam line
in the Hadron Experimental Facility at J-PARC
even after the start of beam line construction, 
we will continue the detector preparation until the entire beam-delivery system is complete. 

The schedule of the KOTO II experiment is 
considered
as shown in Table.~\ref{tab:schedule} 
after the KOTO II beam line is ready (defined as Year 1).
Physics data taking is assumed with two phases:
Phase I for $K_L\to\pi^0\nu\bar{\nu}$ and
Phase II for $K_L\to \pi^0\ell^+\ell^-$
with more a optimized setup for the latter decay.

We will perform a beam line survey in Year 1
when the beam line is ready.
A partial detector in terms of the physics run
and a dedicated setup for the measurement of the beam properties can be used.
In the next year, the KOTO II detector will be installed 
for the first physics run.
In the following years, the physics data taking of Phase I 
will take place
with the measurement of the $K_L\to\pi^0\nu\bar{\nu}$ branching ratio as the main target.
It will take 5 years to achieve
the goal of reaching a single event sensitivity of $8.5\times 10^{-13}$ for $K_L\to\pi^0\nu\bar{\nu}$
under the assumptions of $3\times10^7$~s of running time, i.e.,
4 months/year, 20 days/month, and using 90\% of the beam
for physics runs.

In Phase II, the branching ratio for the decay $K_L\to\pi^0\ell^+\ell^-$ will be measured. To improve the sensitivity in this channel, after Phase I, the detector will be modified to optimize its measurement. 


\begin{table}[h]
\centering
\caption{Schedule of the KOTO II experiment. Beam delivery starts from Year 1. A running time of 4 months/year and 20 days/month, with 90\% of the beam used for physics runs are assumed.\label{tab:schedule}}
\begin{tabular}{ll}
Year &Main object  \\ \hline
1 & Beam line survey \\
2 & Construction of the rest of the detector \\
3-6 & Phase I: Physics run for mainly $K_L\to\pi^0\nu\bar{\nu}$   \\
7 & Single event sensitivity 
will reach $8.5\times 10^{-13}$
for the $K_L\to\pi^0\nu\bar{\nu}$ search      \\
8 & Detector upgrade   \\ 
9-12 & Phase II: Physics run mainly for $K_L\to \pi^0\ell^+\ell^-$ with an optimized setup \\
13 &End of Phase II \\
\hline
\end{tabular}
\end{table}
\clearpage
\section{Conclusion}
Measurement of the CP-violating $\kl$ ultra-rare decay mode $\klpionn$ plays an important role in the study of flavor physics. 
It is one of the best probes for physics beyond the Standard Model. 
Following the ongoing search by the KOTO experiment,
which will reach a sensitivity of better than $10^{-10}$ in 3--4 years, the
KOTO~II experiment, designed to observe several tens of $\klpionn$ signal events, is proposed.
To achieve more than 100 times better sensitivity than the current KOTO experiment, a new beam line with a production angle of 5 degrees
is needed.

We have evaluated the experimental sensitivity and the background level with a modeled beam line and detector.
With a total of $6.3\times 10^{20}$ protons on target, which is equivalent to $3\times 10^7$ seconds of running with a proton beam power of 100~kW,
observation of 35 
events is expected
for the $K_L\to\pi^0\nu\bar{\nu}$ decay
with the SM value of the branching ratio ($3\times 10^{-11}$)
.
The corresponding single event sensitivity is $8.5\times 10^{-13}$.
The estimated number of background events is 40, which corresponds to a signal-to-background ratio of 0.89.
The significance of the observation for SM events is expected to be 5.6$\sigma$.
The measurement would indicate new physics at the 90\% confidence level
if the new physics gives a 40\% deviation of the BR from the SM prediction.

Relative to the existing KOTO detector, the KOTO~II detector will have significantly enhanced capabilities for charged-particle tracking, allowing the KOTO~II physics program to be optimized for the measurement of the $K_L\to\pi^0\ell^+\ell^-$ branching ratios in a second phase of running, after the measurement of $\klpionn$ is complete. This expansion of the physics program will allow KOTO II to obtain complete information on FCNC decays in the kaon sector, enhancing the potential for the discovery and understanding of any new physics present. All phases of KOTO~II running will provide opportunities for concomitant measurements of other rare kaon decays, including searches for new particles such as dark photons and axion-like particles.

Moving forward beyond the baseline design to obtain better sensitivity and signal acceptance, 
we are continuing to study and optimize many known effects and techniques, including R\&D for possible new detectors and analysis methods.

The importance of the physics program proposed here is recognized world-wide, and
deserves a world class experiment (KOTO II) at a world class accelerator facility (J-PARC and Hadron Experimental Facility).

\clearpage

\setcounter{secnumdepth}{\value{secnumdepthsave}}


\bibliographystyle{common/utphys}
\bibliography{common/koto2}

\providecommand{\href}[2]{#2}\begingroup\raggedright\begin{thebibliography}{10}

\bibitem{ref:Zepto}
A.~J. Buras, D.~Buttazzo, J.~Girrbach-Noe, and R.~Knegjens, ``{Can we reach the
  Zeptouniverse with rare $K$ and $B_{s,d}$ decays?},''
  \href{http://dx.doi.org/10.1007/JHEP11(2014)121}{{\em JHEP} {\bfseries 11}
  (2014) 121}, \href{http://arxiv.org/abs/1408.0728}{{\ttfamily arXiv:1408.0728
  [hep-ph]}}.

\bibitem{Brod:2021hsj}
J.~Brod, M.~Gorbahn, and E.~Stamou, ``{Updated Standard Model Prediction for $K
  \to \pi \nu \bar{\nu}$ and $\epsilon_K$},''
  \href{http://dx.doi.org/10.22323/1.391.0056}{{\em PoS} {\bfseries BEAUTY2020}
  (2021) 056}, \href{http://arxiv.org/abs/2105.02868}{{\ttfamily
  arXiv:2105.02868 [hep-ph]}}.

\bibitem{DAmbrosio:2022kvb}
G.~D'Ambrosio, A.~M. Iyer, F.~Mahmoudi, and S.~Neshatpour, ``{Anatomy of kaon
  decays and prospects for lepton flavour universality violation},''
  \href{http://dx.doi.org/10.1007/JHEP09(2022)148}{{\em JHEP} {\bfseries 09}
  (2022) 148}, \href{http://arxiv.org/abs/2206.14748}{{\ttfamily
  arXiv:2206.14748 [hep-ph]}}.

\bibitem{Buras:2022qip}
A.~J. Buras, ``{Standard Model predictions for rare K and B decays without new
  physics infection},''
  \href{http://dx.doi.org/10.1140/epjc/s10052-023-11222-6}{{\em Eur. Phys. J.
  C} {\bfseries 83} no.~1, (2023) 66},
  \href{http://arxiv.org/abs/2209.03968}{{\ttfamily arXiv:2209.03968
  [hep-ph]}}.

\bibitem{ref:Grossman}
Y.~Grossman and Y.~Nir, ``{$K_L \to \pi^0 \nu \overline{\nu}$ beyond the
  standard model},''
  \href{http://dx.doi.org/10.1016/S0370-2693(97)00210-4}{{\em Phys. Lett. B}
  {\bfseries 398} (1997) 163--168},
  \href{http://arxiv.org/abs/hep-ph/9701313}{{\ttfamily arXiv:hep-ph/9701313}}.

\bibitem{ref:newphysics}
A.~J. Buras, D.~Buttazzo, and R.~Knegjens, ``{$ K\to \pi \nu \overline{\nu} $
  and \ensuremath{\varepsilon}'/\ensuremath{\varepsilon} in simplified new
  physics models},'' \href{http://dx.doi.org/10.1007/JHEP11(2015)166}{{\em
  JHEP} {\bfseries 11} (2015) 166},
  \href{http://arxiv.org/abs/1507.08672}{{\ttfamily arXiv:1507.08672
  [hep-ph]}}.

\bibitem{ref:Banomalies}
D.~Marzocca, S.~Trifinopoulos, and E.~Venturini, ``{From B-meson anomalies to
  Kaon physics with scalar leptoquarks},''
  \href{http://dx.doi.org/10.1140/epjc/s10052-022-10271-7}{{\em Eur. Phys. J.
  C} {\bfseries 82} no.~4, (2022) 320},
  \href{http://arxiv.org/abs/2106.15630}{{\ttfamily arXiv:2106.15630
  [hep-ph]}}.

\bibitem{DAmbrosio:2024ewg}
G.~D'Ambrosio, A.~M. Iyer, F.~Mahmoudi, and S.~Neshatpour, ``{Theoretical
  implications for a new measurement of $K_L\to \pi^0 \ell\ell$},''
  \href{http://arxiv.org/abs/2409.06545}{{\ttfamily arXiv:2409.06545
  [hep-ph]}}.

\bibitem{Allwicher:2024ncl}
L.~Allwicher, M.~Bordone, G.~Isidori, G.~Piazza, and A.~Stanzione, ``{Probing
  third-generation New Physics with $K\to \pi \nu\bar\nu$ and $B\to K^{(*)}
  \nu\bar\nu$},'' \href{http://arxiv.org/abs/2410.21444}{{\ttfamily
  arXiv:2410.21444 [hep-ph]}}.

\bibitem{DAmbrosio:1998gur}
G.~D'Ambrosio, G.~Ecker, G.~Isidori, and J.~Portoles, ``{The decays
  $K\to\pi\ell^+\ell^-$ beyond leading order in the chiral expansion},''
  \href{http://dx.doi.org/10.1088/1126-6708/1998/08/004}{{\em JHEP} {\bfseries
  08} (1998) 004}, \href{http://arxiv.org/abs/hep-ph/9808289}{{\ttfamily
  arXiv:hep-ph/9808289}}.

\bibitem{Isidori:2004rb}
G.~Isidori, C.~Smith, and R.~Unterdorfer, ``{The Rare decay $K_L\to \pi^0 \mu^+
  \mu^-$ within the SM},''
  \href{http://dx.doi.org/10.1140/epjc/s2004-01879-0}{{\em Eur. Phys. J. C}
  {\bfseries 36} (2004) 57--66},
  \href{http://arxiv.org/abs/hep-ph/0404127}{{\ttfamily arXiv:hep-ph/0404127}}.

\bibitem{Mescia:2006jd}
F.~Mescia, C.~Smith, and S.~Trine, ``{$K_L\to\pi^0e^+e^-$ and
  $K_L\to\pi^0\mu^+\mu^-$: A binary star on the stage of flavor physics},''
  \href{http://dx.doi.org/10.1088/1126-6708/2006/08/088}{{\em JHEP} {\bfseries
  08} (2006) 088}, \href{http://arxiv.org/abs/hep-ph/0606081}{{\ttfamily
  arXiv:hep-ph/0606081}}.

\bibitem{NA481:2003cfm}
{\bfseries NA48/1} Collaboration, J.~R. Batley {\em et~al.}, ``{Observation of
  the rare decay $K_S\to\pi^0e^+e^-$},''
  \href{http://dx.doi.org/10.1016/j.physletb.2003.10.001}{{\em Phys. Lett. B}
  {\bfseries 576} (2003) 43--54},
  \href{http://arxiv.org/abs/hep-ex/0309075}{{\ttfamily arXiv:hep-ex/0309075}}.

\bibitem{NA481:2004nbc}
{\bfseries NA48/1} Collaboration, J.~R. Batley {\em et~al.}, ``{Observation of
  the rare decay $K_S\to\pi^0\mu^+\mu^-$},''
  \href{http://dx.doi.org/10.1016/j.physletb.2004.08.058}{{\em Phys. Lett. B}
  {\bfseries 599} (2004) 197--211},
  \href{http://arxiv.org/abs/hep-ex/0409011}{{\ttfamily arXiv:hep-ex/0409011}}.

\bibitem{AlvesJunior:2018ldo}
A.~A. Alves~Junior {\em et~al.}, ``{Prospects for Measurements with Strange
  Hadrons at LHCb},'' \href{http://dx.doi.org/10.1007/JHEP05(2019)048}{{\em
  JHEP} {\bfseries 05} (2019) 048},
  \href{http://arxiv.org/abs/1808.03477}{{\ttfamily arXiv:1808.03477
  [hep-ex]}}.

\bibitem{Aebischer:2022vky}
J.~Aebischer, A.~J. Buras, and J.~Kumar, ``{On the Importance of Rare Kaon
  Decays: A Snowmass 2021 White Paper},'' in {\em {Snowmass 2021}}.
\newblock 3, 2022.
\newblock \href{http://arxiv.org/abs/2203.09524}{{\ttfamily arXiv:2203.09524
  [hep-ph]}}.

\bibitem{isidori-seminar}
G.~Isidori, ``{Decoding flavour hierarchies: an essential key to physics beyond
  the SM}.'' CERN Theory Colloquia, 2013.

\bibitem{KTeV:2003sls}
{\bfseries KTeV} Collaboration, A.~Alavi-Harati {\em et~al.}, ``{Search for the
  rare decay $K_L\to\pi^0e^+e^-$},''
  \href{http://dx.doi.org/10.1103/PhysRevLett.93.021805}{{\em Phys. Rev. Lett.}
  {\bfseries 93} (2004) 021805},
  \href{http://arxiv.org/abs/hep-ex/0309072}{{\ttfamily arXiv:hep-ex/0309072}}.

\bibitem{KTEV:2000ngj}
{\bfseries KTEV} Collaboration, A.~Alavi-Harati {\em et~al.}, ``{Search for the
  Decay $K_L \to \pi^0 \mu^+ \mu^-$},''
  \href{http://dx.doi.org/10.1103/PhysRevLett.84.5279}{{\em Phys. Rev. Lett.}
  {\bfseries 84} (2000) 5279--5282},
  \href{http://arxiv.org/abs/hep-ex/0001006}{{\ttfamily arXiv:hep-ex/0001006}}.

\bibitem{Antel:2023hkf}
C.~Antel {\em et~al.}, ``{Feebly-interacting particles: FIPs 2022 Workshop
  Report},'' \href{http://dx.doi.org/10.1140/epjc/s10052-023-12168-5}{{\em Eur.
  Phys. J. C} {\bfseries 83} no.~12, (2023) 1122},
  \href{http://arxiv.org/abs/2305.01715}{{\ttfamily arXiv:2305.01715
  [hep-ph]}}.

\bibitem{Afik:2023mhj}
Y.~Afik, B.~D\"obrich, J.~Jerhot, Y.~Soreq, and K.~Tobioka, ``{Probing
  long-lived axions at the KOTO experiment},''
  \href{http://dx.doi.org/10.1103/PhysRevD.108.055007}{{\em Phys. Rev. D}
  {\bfseries 108} no.~5, (2023) 055007},
  \href{http://arxiv.org/abs/2303.01521}{{\ttfamily arXiv:2303.01521
  [hep-ph]}}.

\bibitem{Littenberg:1989ix}
L.~S. Littenberg, ``{CP-Violating decay $K^0_L \to \pi^0 \nu
  \overline{\nu}$},'' \href{http://dx.doi.org/10.1103/PhysRevD.39.3322}{{\em
  Phys. Rev. D} {\bfseries 39} (1989) 3322--3324}.

\bibitem{Graham:1992pk}
G.~E. Graham {\em et~al.}, ``{Search for the Decay $K_L \to \pi^0 \nu
  \overline{\nu}$},''
  \href{http://dx.doi.org/10.1016/0370-2693(92)90107-F}{{\em Phys. Lett. B}
  {\bfseries 295} (1992) 169--173}.

\bibitem{E799:1994amx}
{\bfseries E799} Collaboration, M.~Weaver {\em et~al.}, ``{Limit on the
  branching ratio of $K_L \to \pi^0 \nu \overline{\nu}$},''
  \href{http://dx.doi.org/10.1103/PhysRevLett.72.3758}{{\em Phys. Rev. Lett.}
  {\bfseries 72} (1994) 3758--3761}.

\bibitem{KTeV:1998taf}
{\bfseries KTeV} Collaboration, J.~Adams {\em et~al.}, ``{Search for the decay
  $K_L \to \pi^0 \nu \overline{\nu}$},''
  \href{http://dx.doi.org/10.1016/S0370-2693(98)01593-7}{{\em Phys. Lett. B}
  {\bfseries 447} (1999) 240--245},
  \href{http://arxiv.org/abs/hep-ex/9806007}{{\ttfamily arXiv:hep-ex/9806007}}.

\bibitem{E799-IIKTeV:1999iym}
{\bfseries E799-II/KTeV} Collaboration, A.~Alavi-Harati {\em et~al.}, ``{Search
  for the decay $K_L \to \pi^0 \nu \overline{\nu}$ using $\pi^0 \to e^+ e^-
  \gamma$},'' \href{http://dx.doi.org/10.1103/PhysRevD.61.072006}{{\em Phys.
  Rev. D} {\bfseries 61} (2000) 072006},
  \href{http://arxiv.org/abs/hep-ex/9907014}{{\ttfamily arXiv:hep-ex/9907014}}.

\bibitem{E391a:2006fxm}
{\bfseries E391a} Collaboration, J.~K. Ahn {\em et~al.}, ``{New limit on the
  $K^0_L \to \pi^0 \nu \overline{\nu}$ decay rate},''
  \href{http://dx.doi.org/10.1103/PhysRevD.74.051105}{{\em Phys. Rev. D}
  {\bfseries 74} (2006) 051105},
  \href{http://arxiv.org/abs/hep-ex/0607016}{{\ttfamily arXiv:hep-ex/0607016}}.
  [Erratum: Phys.Rev.D 74, 079901 (2006)].

\bibitem{E391a:2007qcj}
{\bfseries E391a} Collaboration, J.~K. Ahn {\em et~al.}, ``{Search for the
  Decay $K^0_L \to \pi^0 \nu \overline{\nu}$},''
  \href{http://dx.doi.org/10.1103/PhysRevLett.100.201802}{{\em Phys. Rev.
  Lett.} {\bfseries 100} (2008) 201802},
  \href{http://arxiv.org/abs/0712.4164}{{\ttfamily arXiv:0712.4164 [hep-ex]}}.

\bibitem{E391a:2009jdb}
{\bfseries E391a} Collaboration, J.~K. Ahn {\em et~al.}, ``{Experimental study
  of the decay $K^0_L\to \pi^0 \nu \overline{\nu}$},''
  \href{http://dx.doi.org/10.1103/PhysRevD.81.072004}{{\em Phys. Rev. D}
  {\bfseries 81} (2010) 072004},
  \href{http://arxiv.org/abs/0911.4789}{{\ttfamily arXiv:0911.4789 [hep-ex]}}.

\bibitem{KOTO:2016vwr}
{\bfseries KOTO} Collaboration, J.~K. Ahn {\em et~al.}, ``{A new search for the
  $K_{L} \to \pi^0 \nu \overline{\nu}$ and $K_{L} \to \pi^{0} X^{0}$ decays},''
  \href{http://dx.doi.org/10.1093/ptep/ptx001}{{\em PTEP} {\bfseries 2017}
  no.~2, (2017) 021C01}, \href{http://arxiv.org/abs/1609.03637}{{\ttfamily
  arXiv:1609.03637 [hep-ex]}}.

\bibitem{KOTO2015}
{\bfseries KOTO} Collaboration, J.~K. Ahn {\em et~al.}, ``{Search for the $K_L
  \!\to\! \pi^0 \nu \overline{\nu}$ and $K_L \!\to\! \pi^0 X^0$ decays at the
  J-PARC KOTO experiment},''
  \href{http://dx.doi.org/10.1103/PhysRevLett.122.021802}{{\em Phys. Rev.
  Lett.} {\bfseries 122} no.~2, (2019) 021802},
  \href{http://arxiv.org/abs/1810.09655}{{\ttfamily arXiv:1810.09655
  [hep-ex]}}.

\bibitem{KOTO2016-18}
{\bfseries KOTO} Collaboration, J.~K. Ahn {\em et~al.}, ``{Study of the $K_L
  \to \pi^0 \nu \overline{\nu}$ Decay at the J-PARC KOTO Experiment},''
  \href{http://dx.doi.org/10.1103/PhysRevLett.126.121801}{{\em Phys. Rev.
  Lett.} {\bfseries 126} no.~12, (2021) 121801},
  \href{http://arxiv.org/abs/2012.07571}{{\ttfamily arXiv:2012.07571
  [hep-ex]}}.

\bibitem{KOTO:2024zbl}
{\bfseries KOTO} Collaboration, J.~K. Ahn {\em et~al.}, ``{Search for the
  $K_{L} \to \pi^{0} \nu \overline{\nu}$ Decay at the J-PARC KOTO
  Experiment},'' \href{http://arxiv.org/abs/2411.11237}{{\ttfamily
  arXiv:2411.11237 [hep-ex]}}.

\bibitem{NA62:2021zjw}
{\bfseries NA62} Collaboration, E.~Cortina~Gil {\em et~al.}, ``{Measurement of
  the very rare $K^+ \to \pi^+ \nu \overline{\nu}$ decay},''
  \href{http://dx.doi.org/10.1007/JHEP06(2021)093}{{\em JHEP} {\bfseries 06}
  (2021) 093}, \href{http://arxiv.org/abs/2103.15389}{{\ttfamily
  arXiv:2103.15389 [hep-ex]}}.

\bibitem{KOTOproposal}
{\bfseries J-PARC E14} Collaboration, J.~Comfort {\em et~al.}, ``{Proposal for
  $K_L \rightarrow \pi^0 \nu \overline{\nu}$ Experiment at J-PARC}.''
  \url{https://j-parc.jp/researcher/Hadron/en/pac_0606/pdf/p14-Yamanaka.pdf},
  2006.

\bibitem{ref:kaon2019}
T.~Nomura, ``{A future $K^0_L \rightarrow \pi^0 \nu \overline{\nu}$ experiment
  at J-PARC},'' \href{http://dx.doi.org/10.1088/1742-6596/1526/1/012027}{{\em
  J. Phys. Conf. Ser.} {\bfseries 1526} (2020) 012027}.

\bibitem{ref:kotobl}
{\bfseries KOTO} Collaboration, T.~Shimogawa, ``{Design of the neutral K0(L)
  beamline for the KOTO experiment},''
  \href{http://dx.doi.org/10.1016/j.nima.2010.03.078}{{\em Nucl. Instrum. Meth.
  A} {\bfseries 623} (2010) 585--587}.

\bibitem{Bak:1988bq}
J.~F. Bak {\em et~al.}, ``{$e^+ e^-$ Pair Creation by 40-{GeV} to 150-{GeV}
  Photons Incident Near the $\left<110\right>$ Axis in a Germanium Crystal},''
\href{http://dx.doi.org/10.1016/0370-2693(88)91874-6}{{\em Phys. Lett.}
  {\bfseries B202} (1988) 615--619}.

\bibitem{Kimball:1985np}
J.~C. Kimball and N.~Cue, ``{QUANTUM ELECTRODYNAMICS AND CHANNELING IN
  CRYSTALS},''
\href{http://dx.doi.org/10.1016/0370-1573(85)90021-3}{{\em Phys. Rept.}
  {\bfseries 125} (1985) 69--101}.

\bibitem{Baryshevsky:1989wm}
V.~G. Baryshevsky and V.~V. Tikhomirov, ``{Synchrotron type radiation processes
  in crystals and polarization phenomena accompanying them},''
  \href{http://dx.doi.org/10.1070/PU1989v032n11ABEH002778}{{\em Sov. Phys.
  Usp.} {\bfseries 32} (1989) 1013--1032}.
[Usp. Fiz. Nauk159,529(1989)].

\bibitem{Soldani:2022ekn}
M.~Soldani {\em et~al.}, ``{Strong enhancement of electromagnetic shower
  development induced by high-energy photons in a thick oriented tungsten
  crystal},'' \href{http://dx.doi.org/10.1140/epjc/s10052-023-11247-x}{{\em
  Eur. Phys. J. C} {\bfseries 83} no.~1, (2023) 101},
  \href{http://arxiv.org/abs/2203.07163}{{\ttfamily arXiv:2203.07163
  [hep-ex]}}.

\bibitem{MATSUMURA2024169990}
T.~Matsumura, Y.~Hirayama, G.~Y. Lim, H.~Nanjo, T.~Nomura, K.~Shiomi, and
  H.~Watanabe, ``{Measurement of muon flux behind the beam dump of the J-PARC
  Hadron Experimental Facility},''
  \href{http://dx.doi.org/10.1016/j.nima.2024.169990}{{\em Nucl. Instrum. Meth.
  A} {\bfseries 1069} (2024) 169990},
  \href{http://arxiv.org/abs/2407.17868}{{\ttfamily arXiv:2407.17868
  [physics.ins-det]}}.

\bibitem{Sato:2020kpq}
K.~Sato {\em et~al.}, ``{CsI calorimeter for the J-PARC KOTO experiment},''
  \href{http://dx.doi.org/10.1016/j.nima.2020.164527}{{\em Nucl. Instrum. Meth.
  A} {\bfseries 982} (2020) 164527}.

\bibitem{Naito:2015vrz}
D.~Naito {\em et~al.}, ``{Development of a low-mass and high-efficiency charged
  particle detector},'' \href{http://dx.doi.org/10.1093/ptep/ptv191}{{\em PTEP}
  {\bfseries 2016} no.~2, (2016) 023C01},
  \href{http://arxiv.org/abs/1512.04524}{{\ttfamily arXiv:1512.04524
  [physics.ins-det]}}.

\bibitem{Kamiji:2017deh}
I.~Kamiji and K.~Nakagiri, ``{Upgrade of In-Beam Charged Particle Detector for
  the KOTO Experiment},''
  \href{http://dx.doi.org/10.1088/1742-6596/800/1/012041}{{\em J. Phys. Conf.
  Ser.} {\bfseries 800} no.~1, (2017) 012041}.

\bibitem{Maeda:2014pga}
Y.~Maeda {\em et~al.}, ``{An aerogel Cherenkov detector for multi-GeV photon
  detection with low sensitivity to neutrons},''
  \href{http://dx.doi.org/10.1093/ptep/ptv074}{{\em PTEP} {\bfseries 2015}
  no.~6, (2015) 063H01}, \href{http://arxiv.org/abs/1412.6880}{{\ttfamily
  arXiv:1412.6880 [physics.ins-det]}}.

\bibitem{Murayama:2020mcp}
R.~Murayama {\em et~al.}, ``{A new cylindrical photon-veto detector for the $KL
  \to \pi^0 \nu \bar \nu$ experiment},''
  \href{http://dx.doi.org/10.1016/j.nima.2019.163255}{{\em Nucl. Instrum. Meth.
  A} {\bfseries 953} (2020) 163255}.

\bibitem{ref:PAScal}
J.~Kim {\em et~al.}, ``{Simulation of angular resolution of a new
  electromagnetic sampling calorimeter},''
  \href{http://dx.doi.org/10.1016/j.nima.2023.168261}{{\em Nucl. Instrum. Meth.
  A} {\bfseries 1052} (2023) 168261}.

\bibitem{ref:PDG}
{\bfseries Particle Data Group} Collaboration, P.~A. Zyla {\em et~al.},
  ``{Review of Particle Physics},''
  \href{http://dx.doi.org/10.1093/ptep/ptaa104}{{\em PTEP} {\bfseries 2020}
  no.~8, (2020) 083C01}.

\bibitem{ref:KOPIO}
{\bfseries KOPIO Project} Collaboration, ``{KOPIO Conceptual Design Report},''
  April, 2005.

\bibitem{ref:Calorimeter}
R.~M. Brown and D.~J.~A. Cockerill, ``{Electromagnetic calorimetry},''
  \href{http://dx.doi.org/10.1016/j.nima.2011.03.017}{{\em Nucl. Instrum. Meth.
  A} {\bfseries 666} (2012) 47--79}.

\bibitem{ref:CsITime}
E.~Iwai {\em et~al.}, ``{Performance study of a prototype pure CsI calorimeter
  for the KOTO experiment},''
  \href{http://dx.doi.org/10.1016/j.nima.2015.02.046}{{\em Nucl. Instrum. Meth.
  A} {\bfseries 786} (2015) 135--141}.

\bibitem{Matsumura_2023}
T.~Matsumura, ``Effect of low-energy neutrons on accidental counting rate in
  the koto experiment,''
  \href{http://dx.doi.org/10.1088/1742-6596/2446/1/012044}{{\em Journal of
  Physics: Conference Series} {\bfseries 2446} no.~1, (Feb, 2023) 012044}.
  \url{https://dx.doi.org/10.1088/1742-6596/2446/1/012044}.

\bibitem{HIKE:2023ext}
{\bfseries HIKE} Collaboration, M.~U. Ashraf {\em et~al.}, ``{High Intensity
  Kaon Experiments (HIKE) at the CERN SPS Proposal for Phases 1 and 2},''
  \href{http://arxiv.org/abs/2311.08231}{{\ttfamily arXiv:2311.08231
  [hep-ex]}}.

\bibitem{Ceravolo:2022rag}
S.~Ceravolo {\em et~al.}, ``{Crilin: A CRystal calorImeter with Longitudinal
  InformatioN for a future Muon Collider},''
  \href{http://dx.doi.org/10.1088/1748-0221/17/09/P09033}{{\em JINST}
  {\bfseries 17} no.~09, (2022) P09033},
  \href{http://arxiv.org/abs/2206.05838}{{\ttfamily arXiv:2206.05838
  [physics.ins-det]}}.

\bibitem{Cantone:2023fac}
C.~Cantone {\em et~al.}, ``{Beam test, simulation, and performance evaluation
  of PbF2 and PWO-UF crystals with SiPM readout for a semi-homogeneous
  calorimeter prototype with longitudinal segmentation},''
  \href{http://dx.doi.org/10.3389/fphy.2023.1223183}{{\em Front. in Phys.}
  {\bfseries 11} (2023) 1223183},
  \href{http://arxiv.org/abs/2308.01148}{{\ttfamily arXiv:2308.01148
  [physics.ins-det]}}.

\bibitem{Cantone:2024aix}
C.~Cantone {\em et~al.}, ``{Research and Development Status for an Innovative
  Crystal Calorimeter for the Future Muon Collider},''
  \href{http://dx.doi.org/10.1109/TNS.2024.3364771}{{\em IEEE Trans. Nucl.
  Sci.} {\bfseries 71} no.~5, (2024) 1116--1123}.

\bibitem{Cantone:2024qbx}
C.~Cantone {\em et~al.}, ``{Developing an alternative calorimeter solution for
  the future Muon Collider: The Crilin design},''
  \href{http://dx.doi.org/10.1016/j.nima.2024.169973}{{\em Nucl. Instrum. Meth.
  A} {\bfseries 1069} (2024) 169973}.

\bibitem{Korzhik:2022xln}
M.~Korzhik {\em et~al.}, ``{Ultrafast PWO scintillator for future high energy
  physics instrumentation},''
  \href{http://dx.doi.org/10.1016/j.nima.2022.166781}{{\em Nucl. Instrum. Meth.
  A} {\bfseries 1034} (2022) 166781}.

\bibitem{Cemmi:2024ojv}
A.~Cemmi, B.~D'Orsi, E.~Di~Meco, I.~Di~Sarcina, E.~Diociaiuti, M.~Moulson,
  D.~Paesani, I.~Sarra, J.~Scifo, and A.~Verna, ``{The CRILIN calorimeter:
  gamma radiation resistance of crystals and SiPMs},''
  \href{http://dx.doi.org/10.1088/1748-0221/19/10/P10016}{{\em JINST}
  {\bfseries 19} no.~10, (2024) P10016},
  \href{http://arxiv.org/abs/2410.18731}{{\ttfamily arXiv:2410.18731
  [physics.ins-det]}}.

\bibitem{NA60:2022sze}
{\bfseries NA60+} Collaboration, C.~Ahdida {\em et~al.}, ``{Letter of Intent:
  the NA60+ experiment},'' \href{http://arxiv.org/abs/2212.14452}{{\ttfamily
  arXiv:2212.14452 [nucl-ex]}}.

\bibitem{The:2890181}
{\bfseries ALICE} Collaboration, ``{Technical Design report for the ALICE Inner
  Tracking System 3 - ITS3 ; A bent wafer-scale monolithic pixel detector},''
  tech. rep., CERN, Geneva, 2024.
\newblock \url{https://cds.cern.ch/record/2890181}.
\newblock Co-project Manager: Magnus Mager, magnus.mager@cern.chds.

\bibitem{Butler:2019rpu}
{\bfseries CMS} Collaboration, J.~N. Butler and T.~Tabarelli~de Fatis, ``{A MIP
  Timing Detector for the CMS Phase-2 Upgrade},''.

\bibitem{CERN-LHCC-2020-007}
{\bfseries ATLAS} Collaboration, ``{Technical Design Report: A High-Granularity
  Timing Detector for the ATLAS Phase-II Upgrade},'' tech. rep., CERN, Geneva,
  2020.
\newblock \url{https://cds.cern.ch/record/2719855}.

\bibitem{NA62:Det}
{\bfseries NA62} Collaboration, E.~C.~G. et~al., ``{The Beam and detector of
  the NA62 experiment at CERN},''
  \href{http://dx.doi.org/10.1088/1748-0221/12/05/P05025}{{\em JINST}
  {\bfseries 12} no.~05, (2017) P05025},
  \href{http://arxiv.org/abs/1703.08501}{{\ttfamily arXiv:1703.08501
  [physics.ins-det]}}.

\bibitem{LHCb:Mighty}
{\bfseries LHCb} Collaboration, H.~Schmitz, L.~Dittmann, K.~Padeken, and
  S.~Neubert, ``{Mighty Tracker -- Performance Studies of the MightyPix for
  LHCb},'' \href{http://arxiv.org/abs/2402.08428}{{\ttfamily arXiv:2402.08428
  [physics.ins-det]}}.

\bibitem{ref:PointingCalorimeter}
S.~Ritter {\em et~al.}, ``{A Pointing Electromagnetic Calorimeter for FIP
  Experiments leveraging $X\to\gamma\gamma$ decays},'' in {\em {Proc. of CALOR
  2024 (Tsukuba, Japan)}}.
\newblock 2024.

\end{thebibliography}\endgroup

\end{document}